\documentclass[11pt]{article}

\usepackage[T1]{fontenc}
\usepackage[utf8]{inputenc}
\usepackage{tgpagella}

\usepackage[authoryear]{natbib}
\usepackage{amssymb}
\usepackage{amsfonts}
\usepackage{amsmath}
\usepackage{dsfont}
\usepackage{soul}
\usepackage[nohead]{geometry}
\usepackage{graphicx}
\usepackage{amsthm}
\usepackage{color}
\usepackage{comment}
\usepackage{setspace}
\usepackage{framed}
\usepackage{enumitem}
\usepackage{todonotes}
\usepackage{tikz}
\usepackage{hyperref}
\usepackage{rotating}
\usepackage{subfigure}
\usepackage[flushleft]{threeparttable}
\usepackage{multirow}
\newcommand{\cref}[2][1]{{\textup{(\hyperref[#2]{\ref*{#2}$_{#1}$})}}}
\usepackage{xr}
\usepackage{bm}

7

\newcommand{\bt}{\beta}
\newcommand{\ld}{\lambda}

\newcommand{\beq}{\begin{eqnarray*}}
\newcommand{\eeq}{\end{eqnarray*}}

\newtheorem{lem}{Lemma}[section]

\newtheorem{assum}{Assumption}
\newtheorem{pro}{Proposition}[section]
\numberwithin{equation}{section}
\theoremstyle{definition}

\makeatletter
\def\@biblabel#1{\hspace*{-\labelsep}}
\makeatother
\begin{document}

\title{Sparse HP Filter: Finding Kinks in the COVID-19 Contact Rate\thanks{We would like to thank the editor and an anonymous referee for helpful comments. This work is in part supported by the Ministry of Education of the Republic of Korea and the National Research Foundation of Korea (NRF-2018S1A5A2A01033487),
the McMaster COVID-19 Research Fund (Stream 2),
the European Research Council (ERC-2014-CoG-646917-ROMIA) and the UK Economic and Social Research Council for research grant (ES/P008909/1) to the CeMMAP.}}
\date{\today}

\author{
Sokbae  Lee\thanks{%
 Department of Economics, Columbia University, 420 West 118th Street,  New York, NY 10027, USA.
 Centre for Microdata Methods and Practice, Institute for Fiscal Studies, 7 Ridgmount Street, London WC1E 7AE, UK.
 E-mail: \texttt{sl3841@columbia.edu}.}   \and
Yuan Liao\thanks{
 Department of Economics, Rutgers University, 75 Hamilton St., New Brunswick, NJ 08901, USA. Email:
	\texttt{yuan.liao@rutgers.edu}.}
\and Myung Hwan Seo\thanks{%
Correspondence. Department of Economics, Seoul National University, 1 Gwanak-ro, Gwanak-gu, Seoul 08826, Korea. E-mail:
\texttt{myunghseo@snu.ac.kr}.}   \and
Youngki  Shin\thanks{%
Department of Economics, McMaster University, 1280 Main St.\ W.,\ Hamilton, ON L8S 4L8, Canada. Email:
\texttt{shiny11@mcmaster.ca}.}
}

\maketitle

\begin{abstract}
\noindent
In this paper, we estimate the time-varying COVID-19 contact rate of a Susceptible-Infected-Recovered (SIR) model.
Our measurement of the contact rate is constructed using data on actively infected, recovered  and deceased cases.
We propose a new trend filtering method that is a variant of the Hodrick-Prescott (HP) filter, constrained by the number of possible kinks.  We term it the \emph{sparse HP filter} and apply it to daily data from five countries: Canada, China, South Korea, the UK and the US. Our new method yields the kinks that are well aligned with actual events in each country.
We find that the sparse HP filter provides a fewer kinks than the $\ell_1$ trend filter, while both methods fitting data equally well.
Theoretically, we establish risk consistency of both the sparse HP and $\ell_1$ trend filters.
Ultimately, we propose to use  time-varying \emph{contact  growth rates}  to document and monitor outbreaks of COVID-19.
\end{abstract}

\thispagestyle{empty}



\onehalfspacing

\newpage
\setcounter{page}{1}
\pagenumbering{arabic}

\section{Introduction}

Since March 2020, there has been a meteoric rise in economic research on COVID-19.
New research outputs have been appearing on the daily and weekly basis at an unprecedented level.\footnote{The major outlets for economists are: arXiv working papers, NBER working papers, and CEPR's new working paper series called ``Covid Economics: Vetted and Real-Time Papers'' among others.}
To sample a few,
 \citet{Ng:NBER} quantified the macroeconomic impact of COVID-19 by using data on costly and deadly disasters in recent US history;
\citet{Manski:2020:JoE} and \citet{Manski:NBER} applied the principle of partial identification to the infection rate and antibody tests, respectively;
\citet{CKS:2020} used the US state-level data to study determinants of social distancing behavior.

Across a wide spectrum of research, there is a rapidly emerging strand of literature based on
a Susceptible-Infected-Recovered (SIR) model and its variants \citep[e.g.,][for a review of the SIR and related models]{Hethcote:2000}.
Many economists have embraced the SIR-type models  as new tools to study the COVID-19 pandemic.
\cite{Avery:NBER} provided a review of the SIR models for economists, calling for new research in economics.
A variety of economic models and policy simulations have been built on the SIR-type models.
See \cite{Acemoglu:NBER},
\cite{Alvarez:NBER},
\cite{Atkeson:NBER},
\cite{Eichenbaum:NBER},
\cite{Pindyck:NBER},
\cite{Stock:NBER},
\cite{kim2020estimating},
and
\cite{Toda}
among many others.

One of central parameters in the SIR-type models is
the contact rate, typically denoted by $\beta$.\footnote{It is also called the transmission rate by \cite{Stock:NBER}.}
 It measures
``the average number of adequate contacts (i.e., contacts sufficient for transmission) of a person per unit time''
\citep{Hethcote:2000}.
The contact number $\beta/\gamma$ is the product between $\beta$ and the average infectious period, denoted by $1/\gamma$;  the contact number is interpreted as
``the average number of adequate contacts of a typical infective during the infectious period'' \citep{Hethcote:2000}.

The goal of this paper is to estimate the  time-varying COVID-19 contact rate, say $\beta_t$.
In canonical SIR models, $\beta$ is a time-constant parameter.
However, it may vary over time due to multiple factors.
For example, as pointed by \cite{Stock:NBER}, self-isolation, social distancing and lockdown may reduce $\beta$.
To estimate a SIR-type model,
\citet{FVC:ver2}
allowed for a time-varying contact rate to reflect behavioral and policy-induced changes associated with social distancing.
In particular, they estimated $\beta_t$ using data on deaths at city, state and country levels.
Their main focus was to simulate  future outcomes  for many cities, states and countries.

Researchers have also adopted nonlinear time-series models from the econometric toolbox.
For example, \cite{Li:Linton} analyzed  the daily data on the number of new cases and the number
of new deaths  with a quadratic
time trend model in logs. Their main purpose was to estimate the peak of the pandemic.
\cite{Liu2020} studied the density forecasts of the  daily number of active infections
for a panel of countries/regions. They modeled the growth rate of active infections as autoregressive fluctuations around a  piecewise linear trend with a single break.
\cite{Hartl:et:al} used a linear trend model in logs with a trend break to fit German confirmed cases.
\cite{Harvey:Kattuman} used a  Gompertz model with a time-varying trend to fit and forecast German and UK new cases and deaths.

In this paper, we aim to synthesize  the time-varying contact rate with nonparametric time series modeling.
Especially, we build a new nonparametric regression model for $\beta_t$ that allows for a piecewise linear trend with multiple kinks at unknown dates.
We analyze daily data from Johns Hopkins University Center for Systems Science and Engineering \citep[][JHU CSSE]{JHU}
and suggest a particular transformation of data that can be regarded as a noisy measurement of time-varying $\beta_t$.
Our measurement of $\beta_t$, which is constructed from daily data on confirmed, recovered and deceased cases, is different from that of \citet{FVC:ver2} who used only death data.
We believe both measurements are complements to each other.
However,
the SIR model  is at best a first-order approximation to the real world;
a raw series of $\beta_t$  would be too noisy to draw on inferences regarding the underlying contact rate.
In fact, the raw series exhibits high degrees of skewness and time-varying volatility even after the log transformation.

To extract the time-varying signal from the noisy measurements, we
consider nonparametric trend filters that produce possibly multiple kinks in $\beta_t$ where the kinks are induced by government policies and
changes in individual behavior. 
A natural candidate method that yields the kinks is $\ell_1$ trend filtering \citep[e.g.,][]{kim2009}.
However, $\ell_1$ trend filtering is akin to LASSO; hence, it may have a problem of producing too many kinks, just like LASSO selects too many covariates. In view of this concern, we propose a novel filtering method by adding a constraint on the maximum number of kinks to the popular \citet{HP} (HP) filter. It turns out that this method produces a smaller number of the kink points than $\ell_1$ trend filtering when
both methods fit data equally well. In view of that, we call our new method  the \emph{sparse HP filter}.
We find that the estimated kinks  are well aligned with actual events in each country.
To document and monitor outbreaks of COVID-19, we
propose to use  piecewise constant \emph{contact  growth rates} using the piecewise linear trend estimates from the sparse HP filter. They provide not only an informative summary of past outbreaks but also a useful surveillance measure.

The remainder of the paper is organized as follows.
In Section~\ref{sec:model}, we describe a simple time series model of the time-varying contact rate.
In Section~\ref{sec:filtering}, we introduce two classes of filtering methods.
In Section~\ref{sec:first-look}, we have a first look at the US data, as a benchmark country.
In Section~\ref{sec:empirical-results}, we present empirical results for five countries: Canada, China, South Korea,
the UK and the US.
In Section~\ref{sec:theory}, we establish  risk consistency of both the sparse HP and $\ell_1$ trend filters.
Section~\ref{sec:conclusions} concludes and appendices include additional materials.
The replication R codes for the empirical results are available at \url{https://github.com/yshin12/sparseHP}.
Finally, we add the caveat that the empirical analysis in the paper was carried out in mid-June using daily observations up to June 8th. As a result, some remarks and analysis might be out of sync with the COVID-19 pandemic in real time.

\section{A Time Series Model of the COVID-19 Contact Rate}\label{sec:model}

In this section, we develop a time-series model of the contact rate.
Our model specification is inspired by the classical SIR model which has been adopted by many economists in
the current coronavirus pandemic.

We start with a discrete version of the SIR model, augmented with deaths, adopted from \cite{Pindyck:NBER}:
\begin{align}\label{eq:SIR}
\begin{split}
\Delta I_{t} & =\beta S_{t-1} I_{t-1} - \gamma I_{t-1}, \\
\Delta D_{t} & = \gamma_d I_{t-1}, \\
\Delta R_{t} & = \gamma_r I_{t-1}, \\
1 &= S_t + I_t + D_t + R_t, \\
\gamma &= \gamma_r + \gamma_d,
\end{split}
\end{align}
where the (initial) population size is normalized to be 1,
$S_t$ is the proportion of the population that is susceptible,
$I_t$ the fraction infected,
$D_t$ the proportion that have died,
and
$R_t$ the fraction that have recovered.
 The parameter $\gamma = \gamma_r + \gamma_d$ governs the rate at which
 infectives transfer to the state of being deceased or recovered.


 In the emerging economics literature on  COVID-19, the contact rate $\beta$ is viewed as the parameter that can be affected
 by changes in individual behavior and government policies through social distancing and lockdown.
 We follow this literature and let $\beta = \beta_t$ be time-varying.

Let $C_t$ be the proportion of confirmed cases, that is $C_t = I_t + R_t + D_t$.
In words, the confirmed cases consist of actively infected, recovered  and deceased cases.
Use the equations in \eqref{eq:SIR} to obtain
\begin{align}
\beta_t = Y_t := \frac{\Delta C_{t}}{I_{t-1} S_{t-1}}.  \label{eq:infected:stoc:coint}
\end{align}
Assume that we have daily data on $\Delta C_t$, $\Delta R_t$ and $\Delta D_t$.
From these, we can construct
cumulative $C_t$, $R_t$ and $D_t$.
 Then $S_t = 1 - C_t$ and $I_t = C_t - R_t - D_t$.
This means that we can obtain time series of $\beta_t$ from $Y_t$. We formally assume this in the following.

\begin{assum}[Data]\label{assum:data}
 For each $t$, we observe $(C_t, R_t, D_t)$.
 \end{assum}

By Assumption~\ref{assum:data}, we can  construct $Y_t = \Delta C_{t}/(I_{t-1} S_{t-1})$.
Assumption~\ref{assum:data} is a key assumption in the paper.
We use daily data from JHU CSSE
and they are subject to measurement errors, which could bias our estimates.
In Appendix A, we show that the time series model given in this section is robust to some degree of under-reporting of confirmed cases.
However, our estimates are likely to be biased if the underreporting is time-varying. For example, this could happen because testing capacity in many countries has expanded over the time period. Nonetheless, we believe that our measurement of $Y_t$ primarily captures the genuine underlying trend of $\beta_t$.
Moreover, because the SIR model in \eqref{eq:SIR} is at best a first-order approximation,
a raw series of  $Y_t$ would be too noisy to be used as the actual series of the underlying contact rate
$\beta_t$.
In other words, $\beta_t \neq Y_t$ in actual data and  it would be natural to include an error term in $Y_t$. Because $\beta_t$ has to be positive, we adopt
a multiplicative error structure and make the following assumption.

 \begin{assum}[Time-Varying Signal plus Regression Error]\label{assum:beta_t}
 For each $t$, the unobserved random variable $\beta_t$ satisfies
 \begin{align*}
\log Y_t = \log \beta_t + u_t,
\end{align*}
where  the error term $u_t$ has the following properties:
\begin{enumerate}
\item $\mathbb{E}[ u_t | \mathcal{F}_{t-1} ] = 0$,  where $\mathcal{F}_{t-1}$ is the natural filtration at time $t-1$,
\item $\mathbb{E}[ u_t^2 | \mathcal{F}_{t-1} ] = \sigma_t^2 > 0$ for some time-varying conditional variance  $\sigma_t^2$.
\end{enumerate}
 \end{assum}

 Define
 \begin{align}\label{def:y}
  y_t := \log (\Delta C_t) - \log(I_{t-1}) - \log S_{t-1}.
 \end{align}
  Under Assumption~\ref{assum:beta_t}, \eqref{eq:infected:stoc:coint} can be rewritten as
 \begin{align}\label{main-eq-est}
y_t = \log \beta_t + u_t,
\end{align}
The time-varying parameter $\log \beta_t$ would not be  identified without further restrictions.
Because it is likely to be affected by government policies and cannot change too rapidly, we will assume that it follows
a piecewise trend:

\begin{assum}[Piecewise Trend]\label{assum:f_t}
 The time-varying parameter $f_{0,t} := \log \beta_t$  follows a piecewise trend with at most $\kappa$
  kinks, where
  the set of kinks is defined by
  $\{t=1,...,T:  f_{0,t}-f_{0,t-1}\neq f_{0,t+1}-f_{0,t}\}$
  and
  the locations of kinks are unknown.
 \end{assum}

The main goal of this paper is to estimate $\log \beta_t$ and its kinks under
Assumptions~\ref{assum:data}, \ref{assum:beta_t} and \ref{assum:f_t}.

\section{Filtering the COVID-19 Contact Rate}\label{sec:filtering}

We consider two different classes of trend filtering methods to produce piecewise estimators  of
$f_{0,t} := \log \beta_t$.
The first class is based on $\ell_1$ trend filtering, which   has become popular recently. See, e.g.,
\cite{kim2009},
\cite{tibshirani2014},
and
\cite{wang2016trend}
among others.

The starting point of the second class is the  HP filter, which  has been popular in macroeconomics
and
has been frequently used  to separate trend from cycle.
The standard convention in the literature is to set  $\lambda = 1600$ for quarterly time series.
For example,
\citet{Ravn:02}
suggested a method for adjusting the HP filter for the frequency of observations;
\citet{deJong:16} and \citet{Explicit:HP} established some representation results;
\citet{Hamilton:2018} provided criticism on the HP filter;
\citet{phillips2019boosting} advocated a boosted version of the HP filter via
$L_2$-boosting \citep{bHP} that  can detect multiple structural breaks.
We view that the kinks might be more suitable than the breaks for modelling $\beta_t$ using daily data.
It is unlikely that in a few days, the degree of contagion of COVID-19 would be diminished with an abrupt  jump by social distancing and lockdown.
The original HP filter cannot produce any kink just as ridge regression does not select any variable.
We build the sparse HP filter by drawing on the recent literature that uses an $\ell_0$-constraint or -penalty
\citep[see, e.g.][]{bertsimas2016,chen2018,chen2018arXiv,Huang:2018}.

\subsection{$\ell_1$ Trend Filtering}

In $\ell_1$ trend filtering, the trend estimate $f_t$ is a minimizer of
\begin{align}\label{L1filter}
 \sum_{t=1}^T (y_t - f_t)^2 + \lambda \sum_{t=2}^{T-1} | f_{t-1} - 2 f_t + f_{t+1} |,
\end{align}
which is related to \citet{HP} filtering; the latter is the minimizer of
\begin{align}\label{HPfilter:def}
 \sum_{t=1}^T (y_t - f_t)^2 + \lambda \sum_{t=2}^{T-1} ( f_{t-1} - 2 f_t + f_{t+1} )^2.
\end{align}
In this paper, the main interest is to find the kinks in the trend. For that purpose,  $\ell_1$ trend filtering is
more suitable than the HP filtering.
The main difficulty of using \eqref{L1filter} is the choice of $\lambda$.
This is especially challenging
since the time series behavior of $y_t$ is largely unknown.

The $\ell_1$ trend filter is akin to LASSO. In view of an analogy to square-root LASSO \citep{sqrt-lasso}, it might be useful to consider a
  square-root variant of \eqref{L1filter}:
\begin{align}\label{L1filter-sqrt}
 \left( \sum_{t=1}^T (y_t - f_t)^2 \right)^{1/2} + \lambda \sum_{t=2}^{T-1} | f_{t-1} - 2 f_t + f_{t+1} |.
\end{align}
We will call \eqref{L1filter-sqrt} \emph{square-root $\ell_1$ trend filtering}.
Both \eqref{L1filter} and \eqref{L1filter-sqrt} can be solved via convex optimization software, e.g., \textbf{CVXR} \citep{CVXR}.

\subsection{Sparse Hodrick-Prescott Trend Filtering}\label{sec:SHP}

As an alternative to $\ell_1$ trend filtering, we may exploit Assumption~\ref{assum:f_t} and consider an $\ell_0$-constrained  version of trend flitering:
\begin{align}\label{L0-constrained}
\begin{split}
 &\sum_{t=1}^T (y_t - f_t)^2 \\
 & \text{ subject to } \\
& \sum_{t=2}^{T-1}  1\{ f_t  - f_{t-1} \neq f_{t+1} - f_t \} \leq \kappa.
\end{split}
\end{align}
The formulation in \eqref{L0-constrained} is related to the method called best subset selection \citep[see, e.g.][]{bertsimas2016,chen2018}.
It requires only the input of $\kappa$. However,
because of the nature of the $\ell_0$-(pseudo)norm, it would not work well if the signal-to-noise ratio (SNR) is low
\citep{hastie2017extended,mazumder2017subset}. This is likely to be a concern for our measurement of the log contact rate.

To regularize the best subset selection procedure, it has been suggested in the literature that \eqref{L0-constrained} can be combined with
$\ell_1$ or $\ell_2$ penalization \citep{bertsimas2020,mazumder2017subset}.
We adopt \citet{bertsimas2020} and propose
 an $\ell_0$-constrained version of the Hodrick-Prescott filter:
\begin{align}\label{sparseHP}
\begin{split}
& \sum_{t=1}^T (y_t - f_t)^2
+ \lambda \sum_{t=2}^{T-1} ( f_{t-1} - 2 f_t + f_{t+1} )^2\\
&\text{subject to } \\
&\sum_{t=2}^{T-1}
1\{ f_t  - f_{t-1} \neq f_{t+1} - f_t \} \leq \kappa.
\end{split}
\end{align}
As in \eqref{L0-constrained},
the tuning parameter $\kappa$ controls how many kinks are allowed for. Thus, we have a direct control of the resulting segments of different slopes.
The $\ell_2$ penalty term is useful to deal with the low SNR problem with the COVID 19 data.
We will call \eqref{sparseHP} \emph{sparse HP trend filtering}.

Problem \eqref{sparseHP} can be solved by mixed integer quadratic programming (MIQP).
Rewrite the objective function in \eqref{sparseHP} as
\begin{align*}
 \sum_{t=1}^T (y_t - f_t)^2 + \lambda \sum_{t=2}^{T-1} ( f_{t-1} - 2 f_t + f_{t+1} )^2
\end{align*}
subject to $z_t \in \{0, 1\},  t=2,\ldots,T-1$,
$\underline{f} \leq f_t \leq \overline{f}$,
$\sum_{t=2}^{T-1} z_t \le \kappa$,
and
\begin{align*}
&- M z_t \leq f_{t-1} - 2 f_t + f_{t+1} \leq M z_t, \; t=2,\ldots,T-1.
\end{align*}
This is called a big-M formulation that requires that
\[
\max_t | f_{t-1} - 2 f_t + f_{t+1} | \leq M.
\]
We need to choose the auxiliary parameters $\underline{f}$, $\overline{f}$ and $M$.
We set $\underline{f} = \min y_t$ and $\overline{f} = \max y_t$.
One simple practical method for choosing $M$ is to set
\begin{align}\label{def:M}
M =  \max_{t=2,\ldots,T-1} |y_{t-1} - 2 y_t + y_{t+1}|.
\end{align}

To implement the proposed method, it is simpler to write the MIQP problem above
in matrix notation.
Let $\bm{y}$ denote the $(T \times 1)$ vector of $y_t$'s
and $\bm{1}$ a vector of 1's whose dimension may vary. We solve
\begin{align}\label{MIQP:formulation}
\min_{\bm{f}, \bm{z}}
\left[
(\bm{y}- \bm{f})^\top (\bm{y}- \bm{f})
+ \lambda \bm{f}^\top \bm{D}^\top \bm{D} \bm{f}
\right]
\end{align}
subject to  $\bm{z} \in \{0, 1\}^{T-2}$,
$\underline{f} \bm{1}  \leq \bm{f} \leq \overline{f} \bm{1}$,
$\bm{1}^\top \bm{z} \leq \kappa$,
$-M \bm{z} \leq \bm{D} \bm{f} \leq M \bm{z}$,
where
$\bm{D}$ is the $(T-2) \times T$ second-order difference matrix such that
$$
\bm{D} =
\left[
\begin{matrix} 1 &  -2  &  1      &        &        &    &   \\
                 &   1  & -2      &  1     &        &    &   \\
                 &      &  \ddots & \ddots & \ddots &    &   \\
				 &      &         & 1      & -2     &  1 &   \\
				 &      &         &        & 1      & -2 &  1	\\
\end{matrix}
\right]
$$
with entries not shown above being zero.
Let $\widehat{\bm{f}}$ and $\widehat{\bm{z}}$ denote the resulting maximizers.
It is straightforward to see that $\widehat{\bm{f}}$ also solves  \eqref{sparseHP}.
Therefore,
$\widehat{\bm{f}}$ is the $(T \times 1)$ vector of trend estimates
and $\widehat{K} := \{ t =2,\ldots,T-1: \widehat{z}_t = 1 \}$
is the index set of estimated kinks.
The MIQP problem can be solved via modern mixed integer programming
software, e.g., \textbf{Gurobi}. Because the sample size for $y_t$ is typically less than 100, the computational speed of MIQP is
fast enough to carry out cross-validation to select tuning parameters.
We summarize the equivalence between the original and MIQP formulation in the following proposition.

\begin{pro}
Define
\begin{align*}
\begin{split}
\mathbb{F}(\kappa) := \{ \bm{f} = (f_1,\ldots,f_T): &  \min_t f_t \geq \underline{f},  \max_t f_t \leq \overline{f},
 \max_t | f_{t-1} - 2 f_t + f_{t+1} | \leq M, \\
 & \sum_{t=2}^{T-1}
1\{ f_t  - f_{t-1} \neq f_{t+1} - f_t \} \leq \kappa
\}.
\end{split}
\end{align*}
Let $\widehat{\bm{f}}_{\textrm{SHP}} : \{ \widehat{f}_t: t=1,\ldots, T\}$ denote a solution to
\begin{align*}
\min_{\bm{f} \in \mathbb{F}(\kappa)}
S_T (\bm{f}, \lambda) :=
(\bm{y}- \bm{f})^\top (\bm{y}- \bm{f})
+ \lambda \bm{f}^\top \bm{D}^\top \bm{D} \bm{f}.
\end{align*}
Let $\bm{\widehat{f}}_{\textrm{MIQP}}$ denote a solution to \eqref{MIQP:formulation}.
Then, both
$\widehat{\bm{f}}_{\textrm{SHP}}$ and $\bm{\widehat{f}}_{\textrm{MIQP}}$ are equivalent in the sense that
$
\widehat{\bm{f}}_{\textrm{SHP}} \in \mathbb{F}(\kappa),
$
$
\bm{\widehat{f}}_{\textrm{MIQP}} \in \mathbb{F}(\kappa),
$
and
$
S_T (\widehat{\bm{f}}_{\textrm{SHP}}, \lambda)
=
S_T (\bm{\widehat{f}}_{\textrm{MIQP}}, \lambda).
$
\end{pro}

\subsection{Selection of Tuning Parameters}\label{sec:loocv}

We first consider the sparse HP filter. There are two tuning parameters: $\lambda$ and $\kappa$.
It is likely that there will be an initial stage of coronavirus spread, followed by lockdown or social distancing.
Even without any policy intervention, it will come down since many people will voluntarily select into self-isolation
and there is a chance of herd immunity.
Hence, the minimum $\kappa$ is at least 1. 
If $\kappa$ is too large, it is difficult to interpret the resulting kinks.
 In view of these, we set the possible values  $\kappa \in \mathcal{K} = \{2,3,4\}$.
For each pair of $(\kappa,\lambda)$, let $\widehat{\bm{f}}_{-s}(\kappa,\lambda)$ denote the leave-one-out estimator of $\bm{f}_s$. That is,
it is the sparse HP filter estimate by solving:
\begin{align}\label{sparseHP:lov}
\begin{split}
& \sum_{t=1, t \neq s}^T (y_t - f_t)^2
+ \lambda \sum_{t=2}^{T-1} ( f_{t-1} - 2 f_t + f_{t+1} )^2\\
&\text{subject to } \\
&\sum_{t=2}^{T-1}
1\{ f_t  - f_{t-1} \neq f_{t+1} - f_t \} \leq \kappa.
\end{split}
\end{align}
The only departure from \eqref{sparseHP} is that we replace
the fidelity term
$\sum_{t=1}^T (y_t - f_t)^2$
with
$\sum_{t=1, t \neq s}^T (y_t - f_t)^2$.
We choose the optimal $(\kappa,\lambda)$ by
\begin{align}\label{lovcv}
\min_{(\kappa,\lambda) \in \mathcal{K} \times \mathcal{L}}
\sum_{t=1}^T  \left\{ y_t - \widehat{\bm{f}}_{-t}(\kappa,\lambda) \right\}^2,
\end{align}
where $\mathcal{L}$ is the set for possible values of $\lambda$.
We view $\lambda$ as an auxiliary tuning parameter that mitigates the low SNR problem.
Hence, we take $\mathcal{L}$ to be in the range of relatively smaller values than the typical values used for the HP filter.
In the numerical work, we let $\Lambda$ to a grid of  equi-spaced points in the $\log_2$-scale.

We now turn to the HP, $\ell_1$  and square-root $\ell_1$ trend filters.
For each filter, we  choose $\lambda$ such that the fidelity term
$\sum_{t=1}^T (y_t - f_t)^2$ is the same as that of the sparse HP filter.
In this way, we can compare different methods holding the same level of fitting the data.
Alternatively, we may choose $\lambda$ by leave-one-out cross validation for each filtering method. However, in that case, it would be more difficult to make a comparison
across different methods. Since our main focus is to find the kinks in the contact rate, we will fine-tune all the filters to have the same level of $\sum_{t=1}^T (y_t - f_t)^2$
based on the sparse HP filter's cross validation result.

\section{A First Look at the Time-Varying Contact Rate}\label{sec:first-look}


As a benchmark, we have a first look at  the US data.
The dataset is  obtained via R package \textbf{coronavirus} \citep{Krispin},
 which provides
a daily summary of COVID-19 cases from Johns Hopkins University Center for Systems Science and Engineering
\citep{JHU}.
Following \citet{Liu2020}, we set the first date of the analysis to begin when the number of cumulative cases reaches 100
(that is, March 4 for the US).
To smooth data minimally, we take $Y_t$ in \eqref{eq:infected:stoc:coint}
to be a three-day simple moving average: that is,
$Y_t = (\breve{Y}_{t}+\breve{Y}_{t-1} + \breve{Y}_{t-2})/3$, where
$\breve{Y}_{t}$ is the daily observation of $Y_t$ constructed from the dataset.\footnote{\citet{Liu2020} used one-sided three-day rolling averages;
\citet{FVC:ver2} took 5-day centered moving averages.}
Then, we take the log to obtain $y_t = \log Y_t$.

\begin{figure}[htbp]
\begin{center}
\caption{US Data}\label{data-US}
\vskip10pt
\begin{tabular}{cc}
\includegraphics[scale=0.45]{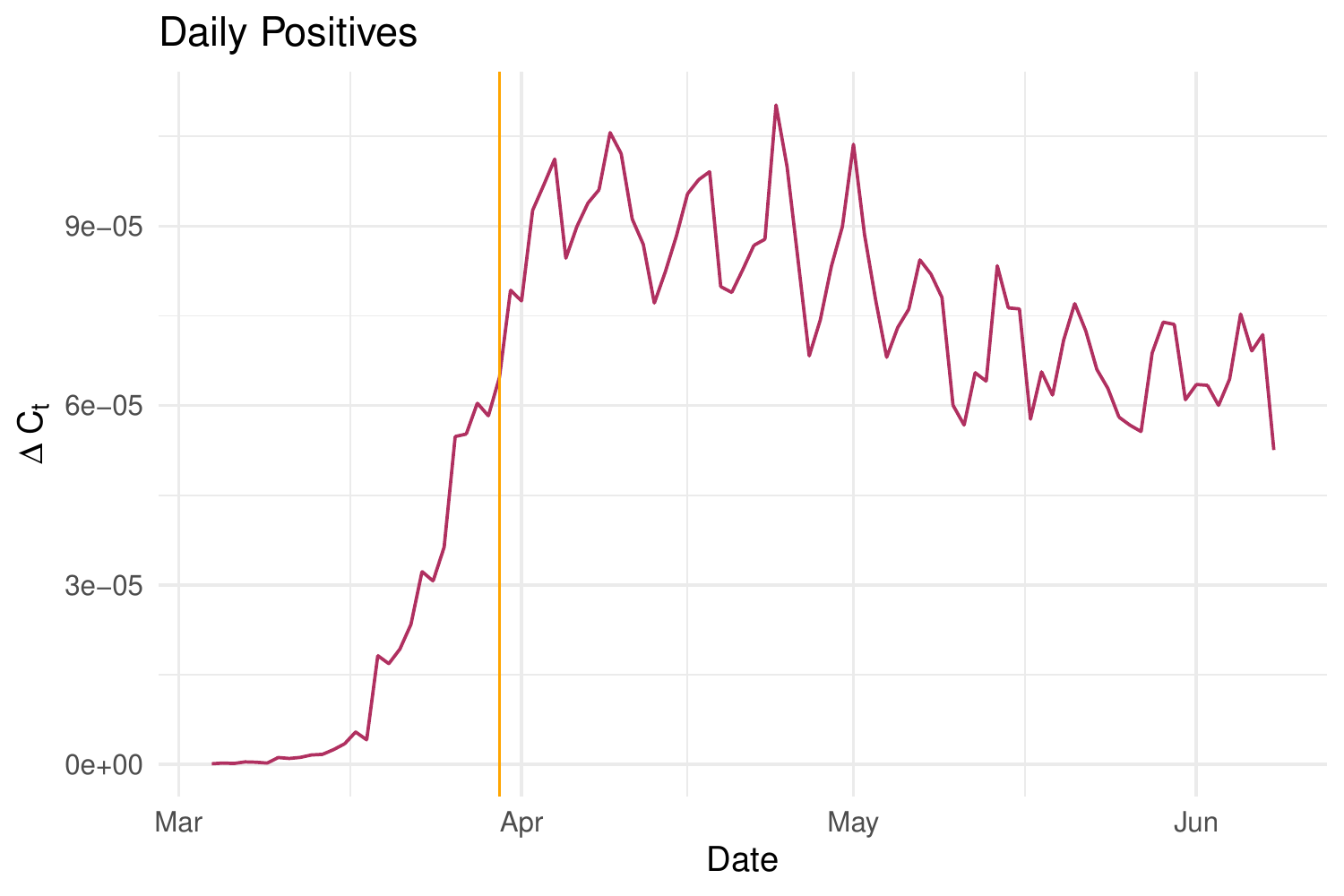} &
\includegraphics[scale=0.45]{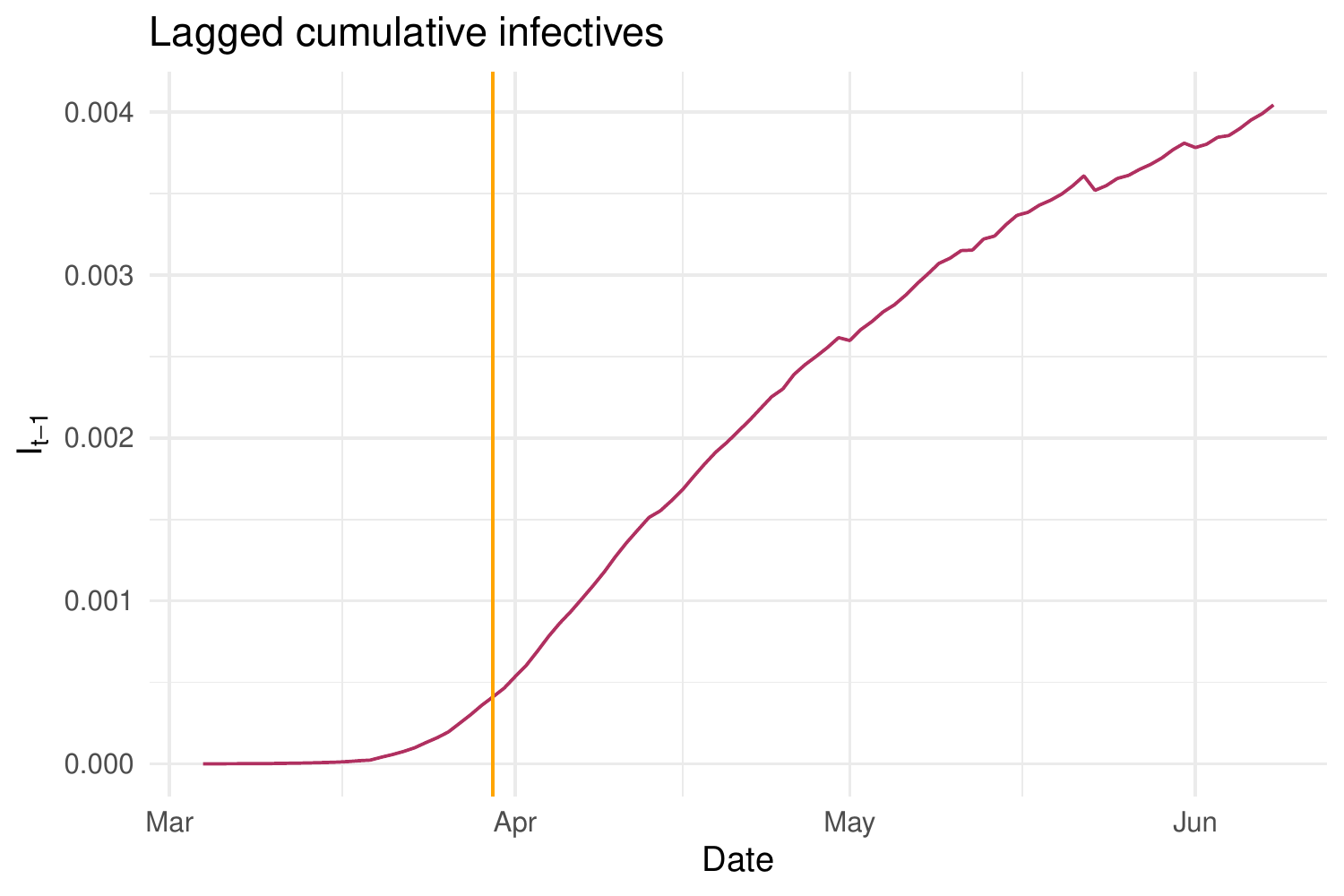} \\
\includegraphics[scale=0.45]{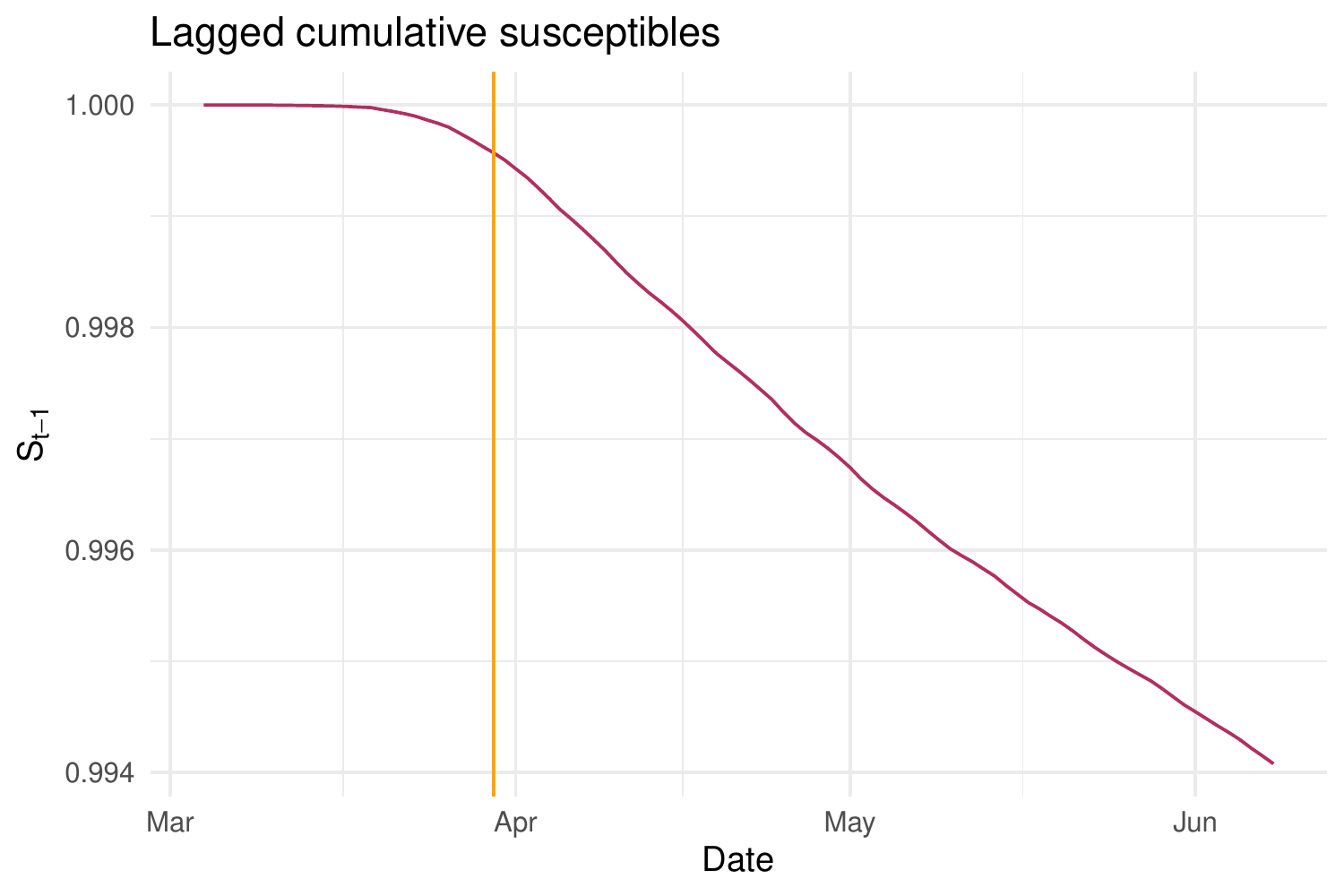} &
\includegraphics[scale=0.45]{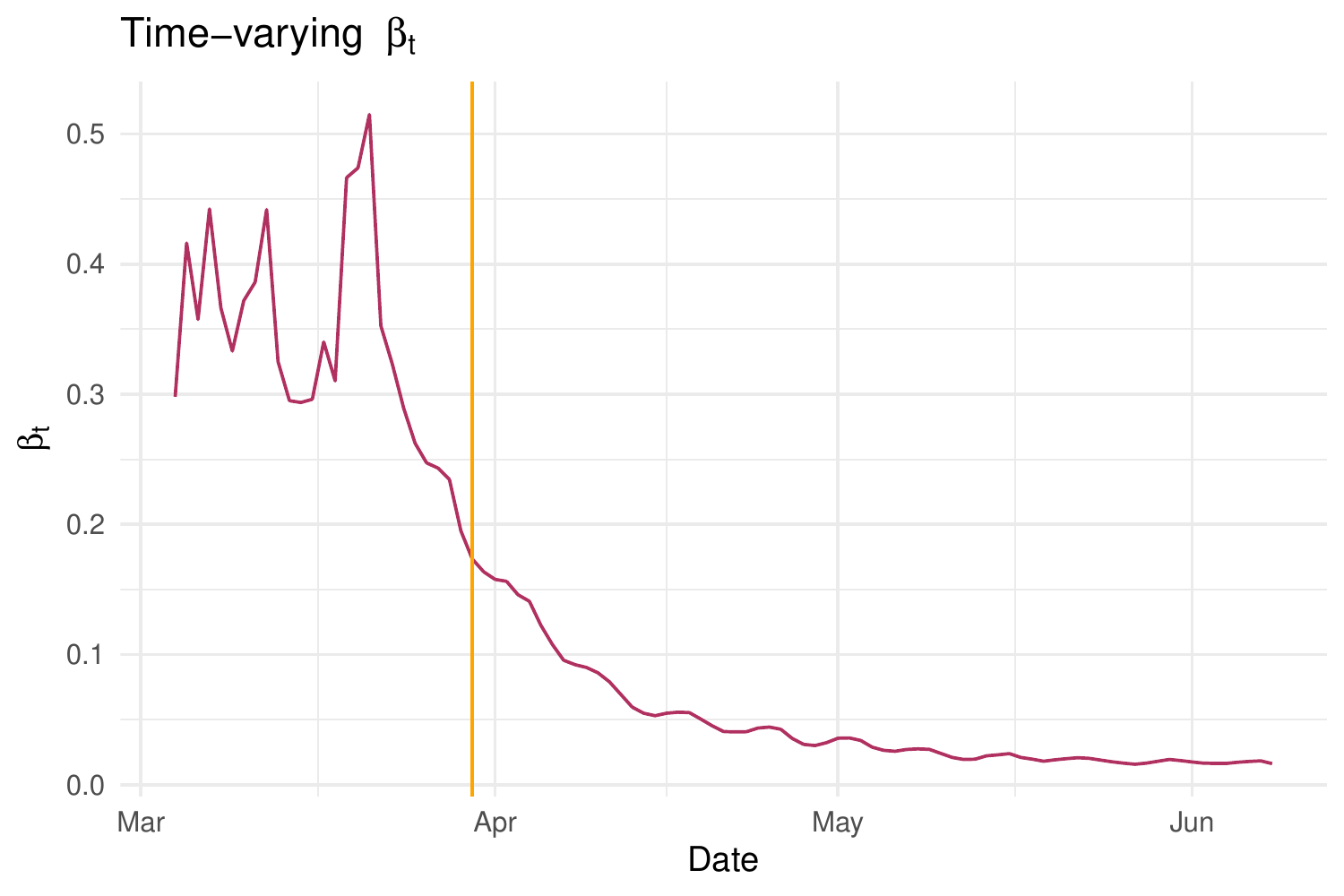}
\end{tabular}
\end{center}
\noindent \footnotesize{Note: The orange vertical line denotes the lockdown date, March 30. The population size is normalized to be 1. }
\end{figure}

Figure~\ref{data-US} has four panels.
The top-left panel shows the fraction of daily positives,
the top-right panel the fraction of lagged cumulative infectives,
the bottom-left panel  the fraction of lagged cumulative susceptibles,
and the bottom-right $Y_t = \Delta C_{t}/(I_{t-1} S_{t-1})$.
In the US, statewide stay-at-home
 orders started in California on March 20
and extended to 30 states by March 30 \citep{NYtimes}.
The  inserted vertical line in the figure corresponds to March 30, which we will call the ``lockdown'' date for simplicity, although
there was no lockdown at the national level.
As a noisy measurement of $\beta_t$, $Y_t$ shows enormous skewness and fluctuations especially in the beginning of the study period. This indicates that the signal-to-noise ratio is high and is time-varying as well.
This pattern of the data has motivated Assumption~\ref{assum:beta_t}.
Because $S_{t-1}$ is virtually one throughout the analysis period (0.994 on June 8, which is the last date of the sample),
$Y_t \approx  \Delta C_{t}/I_{t-1}$, which is daily positives divided by the lagged infectives.

\begin{figure}[htp]
\begin{center}
\caption{$\log Y_t$ as a raw time series of $\log \beta_t$ and parametric fitting}\label{data-US-log-beta}
\vskip10pt
\begin{tabular}{cc}
\includegraphics[scale=0.45]{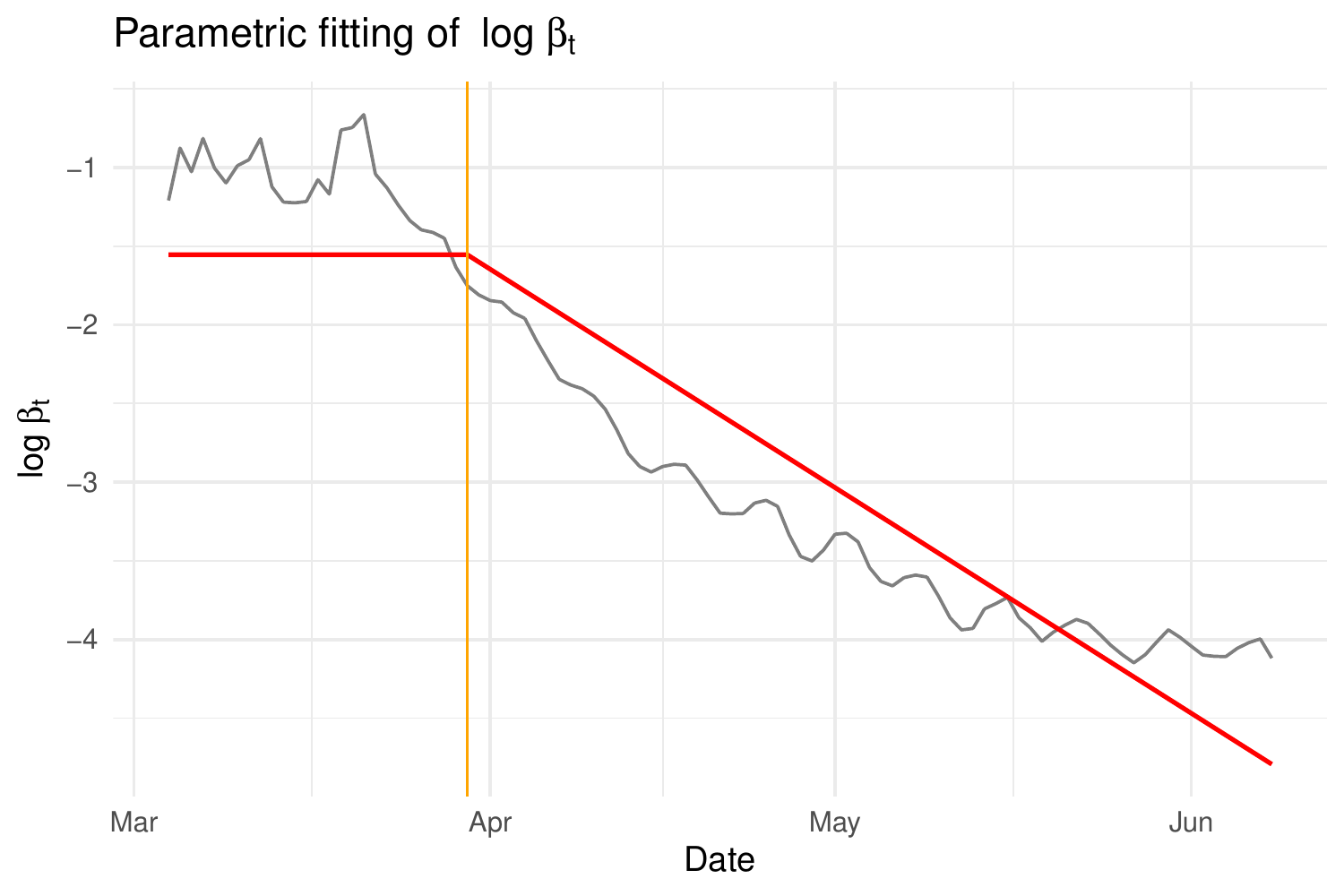} &
\includegraphics[scale=0.45]{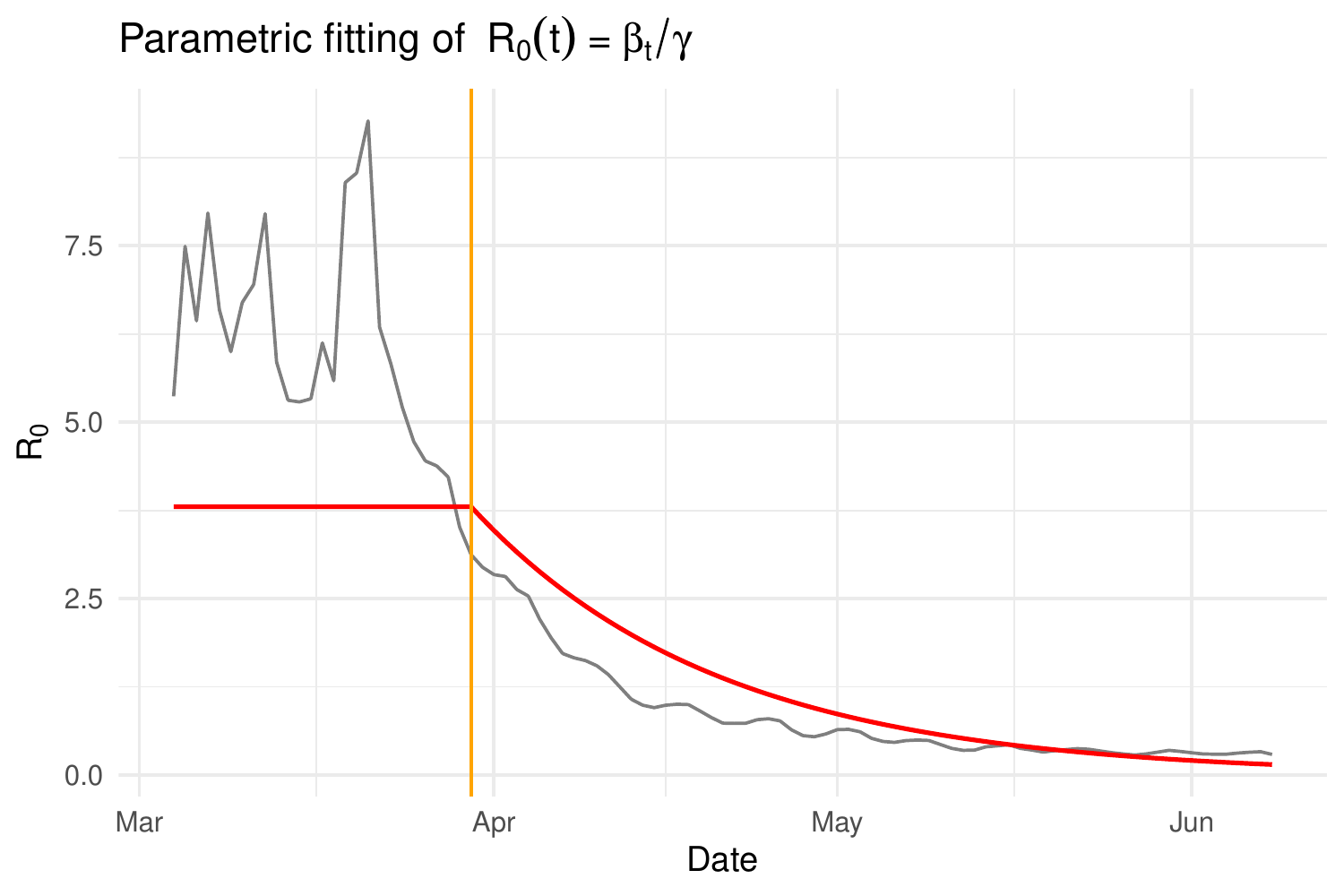} \\
\includegraphics[scale=0.45]{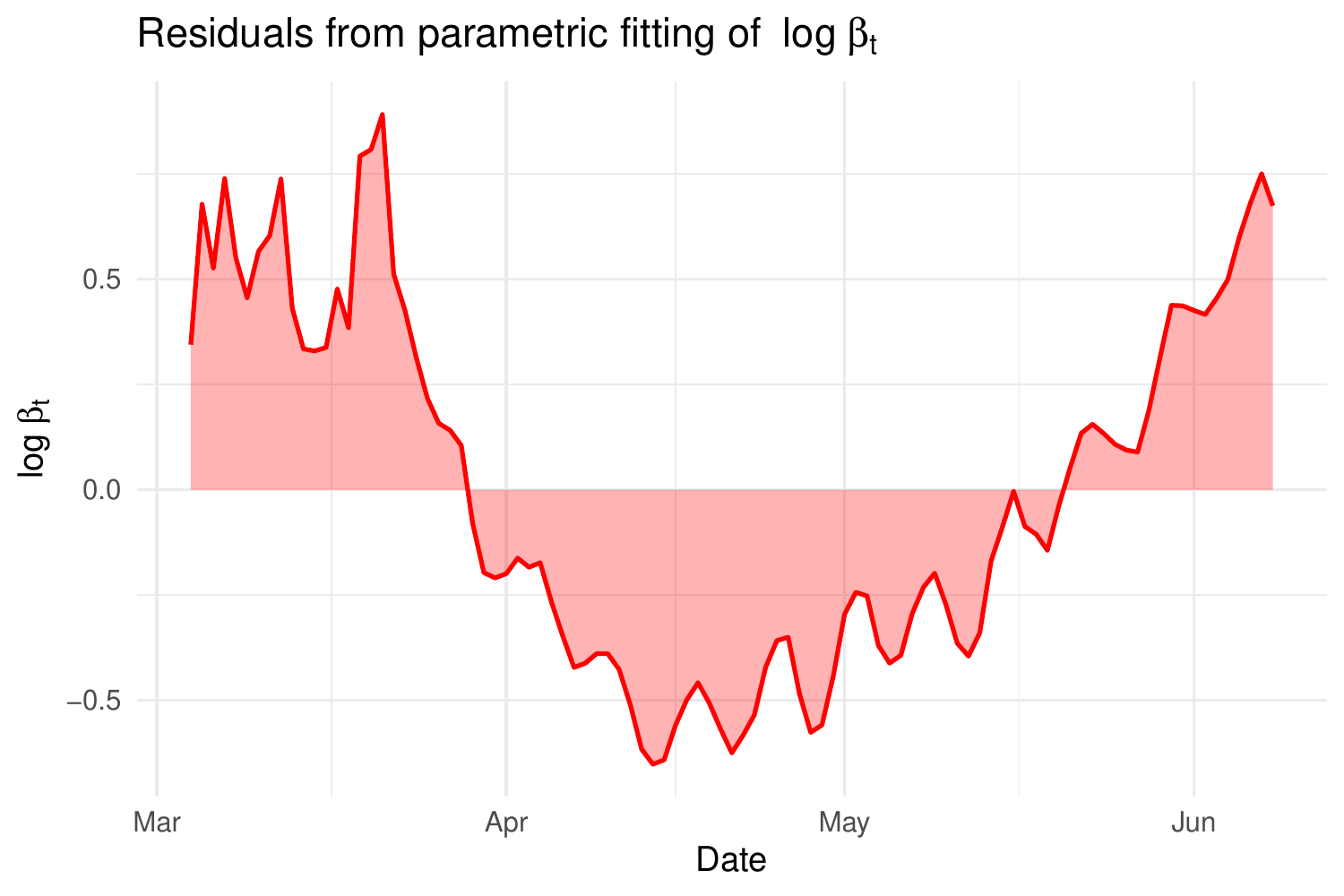}  &
\includegraphics[scale=0.45]{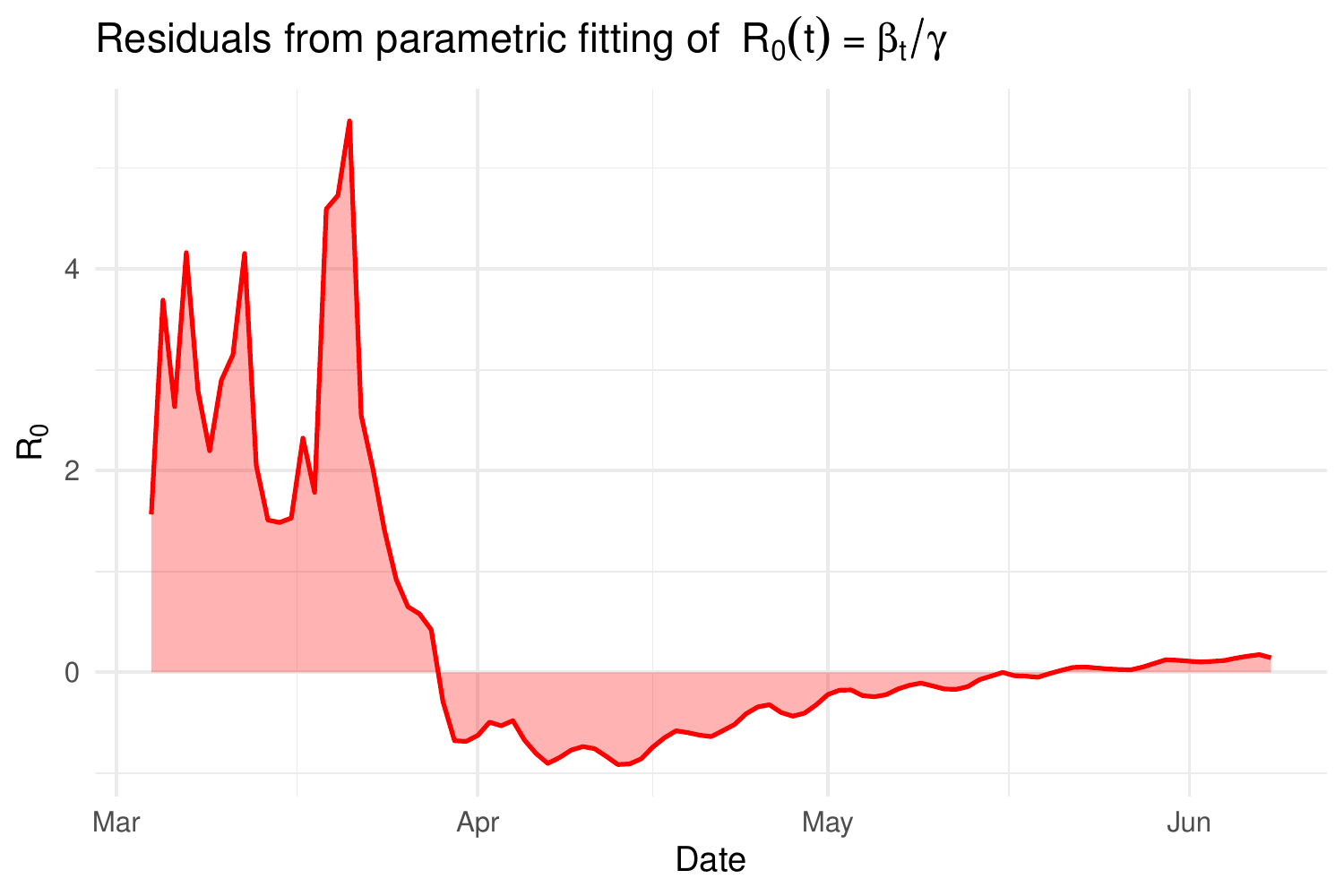}
\end{tabular}
\end{center}
\noindent \footnotesize{Note: The orange vertical line denotes the lockdown date, March 30.}
\end{figure}

Figure~\ref{data-US-log-beta} shows the raw data along with parametric fitting.
The top-left panel shows the logarithm of $Y_t$, which still exhibits some degree of skewness and time-varying variance.
The fitted regression line is based on the following parametric regression model:
\begin{align}\label{regmodel1}
y_t
= \alpha_0 + \alpha_1 (t-t_0) 1(t > t_0) + \varepsilon_t,
\end{align}
where $t_0$ is March 30.
The simple idea behind \eqref{regmodel1}
is that an initial, time-constant contact rate began to diminish over time after a majority of US states imposed stay-at-home orders.

In simple SIR models, the contact number $\beta/\gamma$ is identical to the basic reproduction number denoted by $R_0$, which is viewed as a key threshold quantity in the sense that
``an infection can get started in a fully susceptible population if and only if $R_0 > 1$
for many deterministic epidemiology models'' \citep{Hethcote:2000}.
Since $\beta_t$ is time-varying in our framework, we may define a time-varying basic reproduction number by
$R_0(t) := \beta_t/\gamma$.

The top-right panel shows the estimates of time-varying $R_0(t)$:\footnote{The formula given in \eqref{time-varying-R}
is valid if errors are  homoskedastic, which is unlikely to be true in actual data. However, we present \eqref{time-varying-R} here because it is simpler. Our main analysis focuses on estimation of the kinks based on $y_t$, not on estimating $R_0(t)$. We use the latter mainly to appreciate the magnitude of the kinks.}
\begin{align}\label{time-varying-R}
\widehat R_0(t) := \exp[\widehat \alpha_0 + \widehat \alpha_1 (t-t_0) 1(t > t_0) ]/\gamma,
\end{align}
where  $\gamma = 1/18$ is taken from \citet{Acemoglu:NBER}.
This corresponds to 18 days of the average infectious period.
The parametric estimates of $R_0(t)$  started above 4 and
reached $0.15$ at the end of the sample period.

 The left-bottom panel shows the residual plot in terms of $y_t$
and the right-bottom panel the residual plot in terms of $R_0(t)$.
In both panels, the estimated residuals seem to be biased and show autocorrelation.
Especially, the positive values of residuals at the end of the sample period is worrisome
because the resulting prediction  would be too optimistic.


\section{Estimation Results}\label{sec:empirical-results}

In this section, we present estimation results for five countries: Canada, China, South Korea, the UK and the US.
These countries are not meant to be a random sample of the world; they are selected based on our familiarity with them
so that we can interpret the estimated kinks with  narratives.
We look at the US as a benchmark country and provide a detailed analysis in Section~\ref{section:US}.
A condensed version of the estimation results for other countries are provided  in Section~\ref{section:RoW}.

\subsection{Benchmark: the US}\label{section:US}

\begin{figure}[thp]
\begin{center}
\caption{Sparse HP Filtering: Leave-One-Out Cross Validation}\label{data-US-SHP-CV}
\vskip10pt
\begin{tabular}{c}
\includegraphics[scale=0.55]{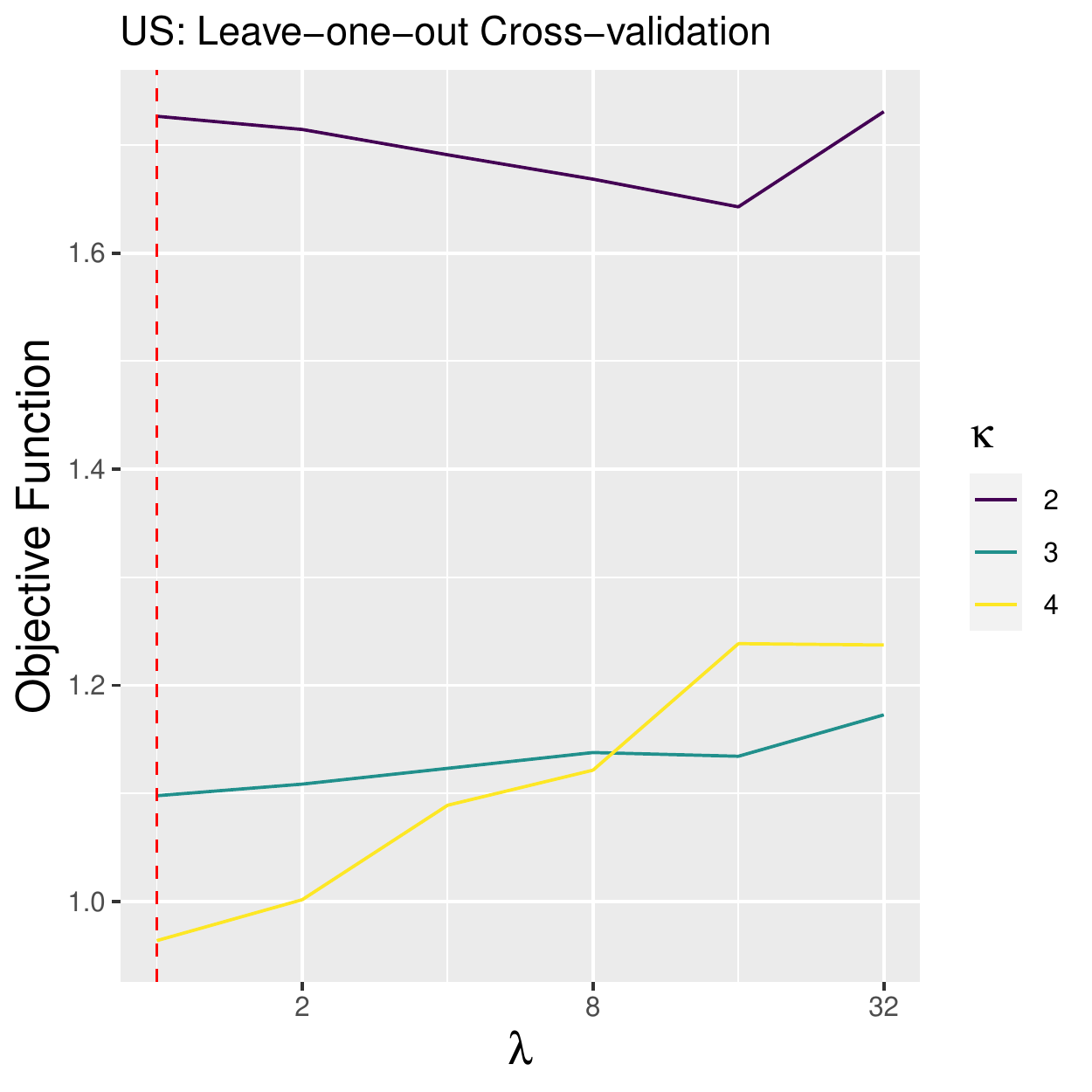} \\
\end{tabular}
\end{center}
\noindent \footnotesize{Note: The red vertical line denotes the minimizer $(\widehat{\kappa},\widehat{\ld})=(4,1)$ of the cross-validation objective function. The x-axis is represented by the $\log_2$ scale.}
\end{figure}

Figure~\ref{data-US-SHP-CV} summarizes the results of leave-one-out cross validation (LOOCV) as described in
Section~\ref{sec:loocv}. The range of tuning parameters were:
$\kappa\in\{2,3,4\}$ and $\ld=\{2^{0}, 2^{1}, \ldots, 2^{5}\}$.
We can see that the choice of $\kappa$ seems to matter more than that of $\lambda$. Clearly,
$\kappa = 2$ provides the worst result and $\kappa = 3$ and $\kappa = 4$ are relatively similar.
The LOOCV criterion function was minimized at $(\widehat{\kappa},\widehat{\ld})=(4,1)$.

\begin{figure}[thp]
\begin{center}
\caption{Sparse HP Filtering}\label{data-US-SHP}
\vskip10pt
\begin{tabular}{cc}
\includegraphics[scale=0.45]{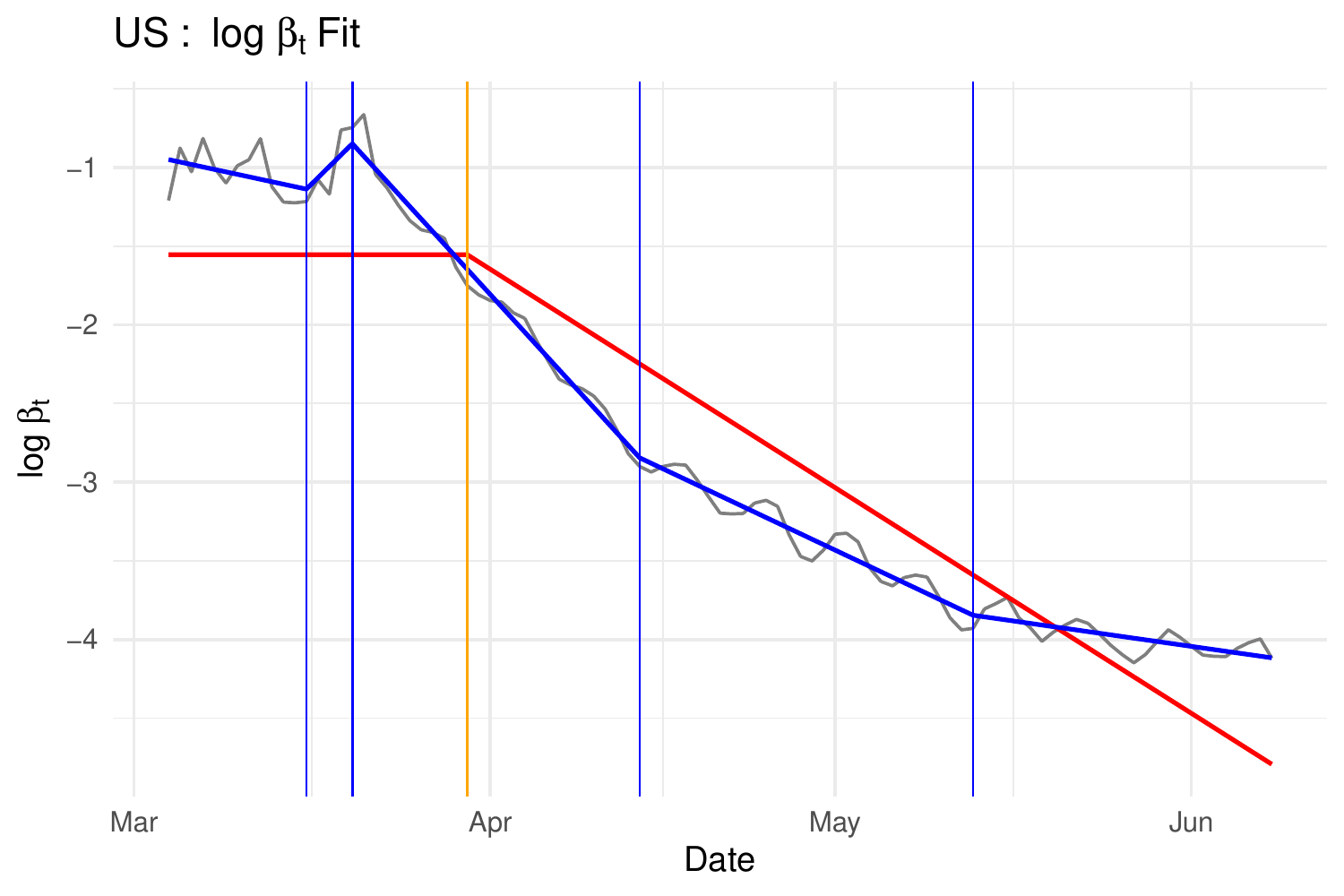} &
\includegraphics[scale=0.45]{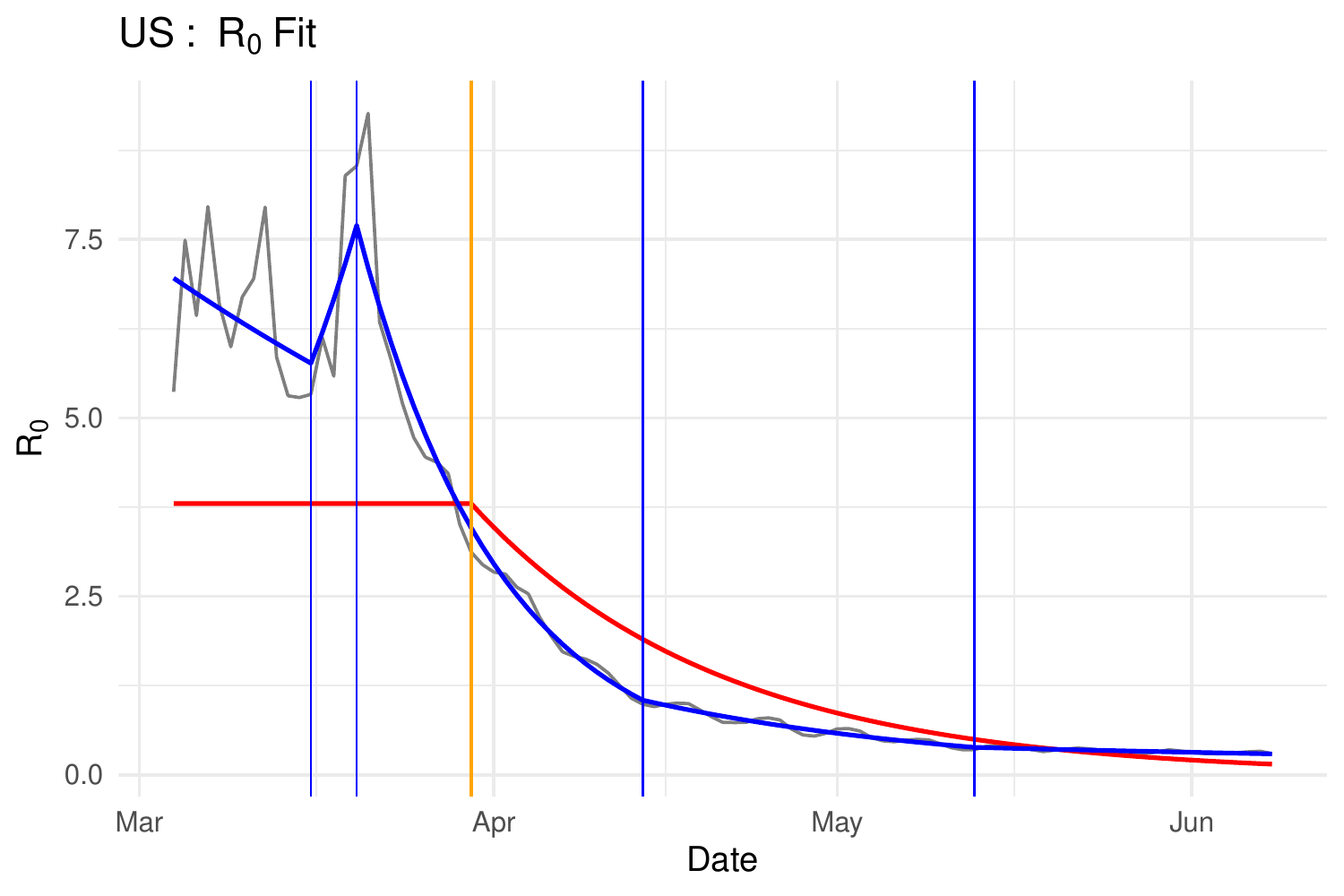} \\
\includegraphics[scale=0.45]{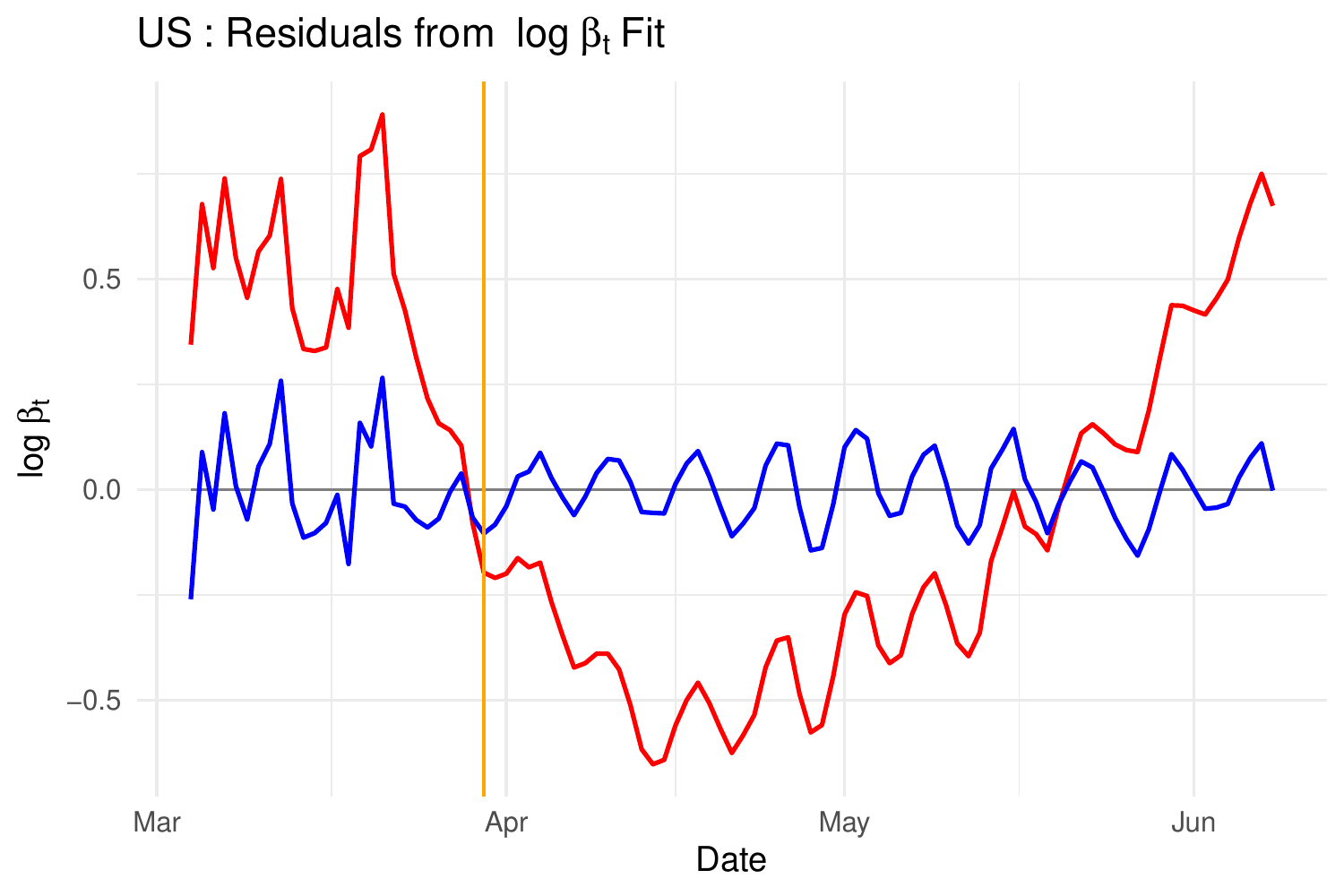}  &
\includegraphics[scale=0.45]{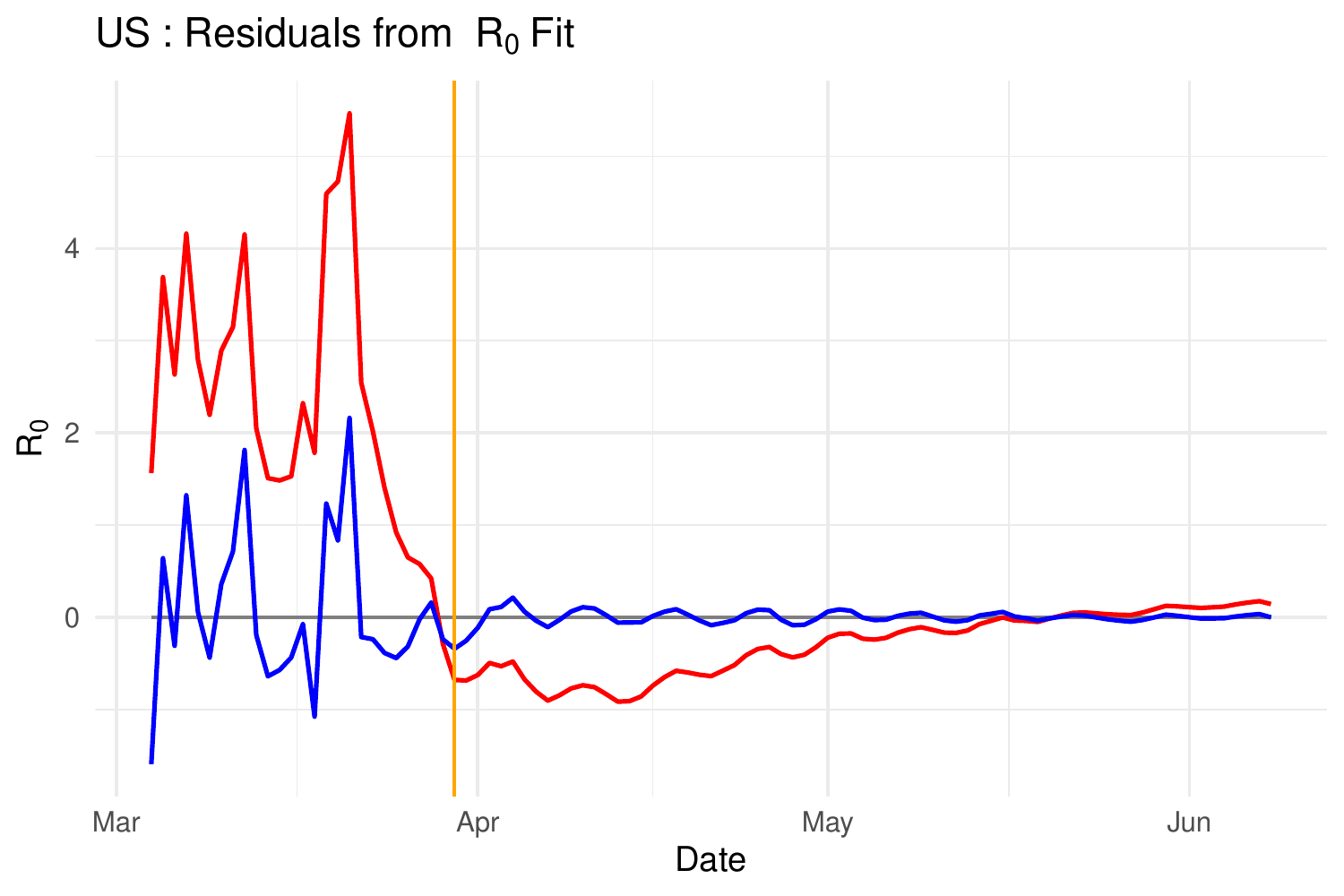} \\
\multicolumn{2}{c}{\includegraphics[scale=1]{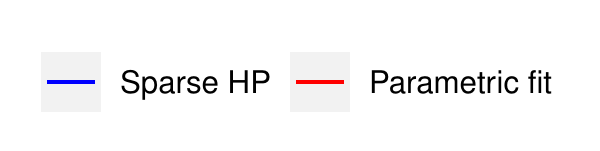}}
\end{tabular}
\end{center}
\noindent \footnotesize{Note:
The grey curve in each panel represents the data $y_t$.
Sparse HP filtering solves \eqref{sparseHP} and the parametric fit uses the linear regression \eqref{regmodel1}.
The estimated kinks denoted by blue vertical lines are:
March 16, March 20, April 14, and May 13.
The orange vertical line denotes the lockdown date, March 30.}
\end{figure}

Based on the tuning parameter selection in Figure~\ref{data-US-SHP-CV},
we show  estimation results for the sparse HP filter in Figure~\ref{data-US-SHP}.
The structure of Figure~\ref{data-US-SHP} is similar to that of Figure~\ref{data-US-log-beta}.
The top-left panel shows estimates of the sparse HP filter along with the raw series of $y_t$ and
the parametric estimates shown in Figure~\ref{data-US-log-beta}.
The top-right panel displays counterparts in terms of $R_0(t)$.
The bottom panels exhibit residual plots for the $\log \beta_t$ and  $R_0(t)$ scales.
The trend estimates from the sparse HP filter fit the data much better than the simple parametric estimates.
The estimated kink dates are: March 16, March 20, April 14, and May 13. There are five periods based on them.

\begin{enumerate}
\item March 4 - March 16: this period corresponds to the initial epidemic stage;
\item March 16 - March 20:  the contact rate was peaked at the end of this period;
\item March 20 - April 14: a sharp decrease of the contact rate is striking;
\item April 14 - May 13: the contact rate decreased but less steeply;
\item May 13 - June 8: it continued to go down but its slope got more flattened.

\end{enumerate}
To provide narratives on these dates,
President Trump declared a national emergency  on March 13;
The Centers for Disease Control and Prevention (CDC)
recommended no gatherings of 50 or more people on March 15;
New York City's public schools system announced that it would close on March 16;
and California started  stay-at-home orders on March 20 \citep{NYtimes:timeline,NYtimes}.
These events indicate that the second period was indeed the peak of the COVID-19 epidemic in the US.
The impact of social distancing and stay-at-home orders across a majority of states is clearly visible in the third period.
The fourth and fifth periods include state reopening:
for example, stay-at-home order expired in Georgia and Texas on April 30; in Florida on May 4;
in Massachusetts on May 18; in New York on May 28 \citep{NYtimes:reopening}.
In short,
unlike the parametric model with a single kink, the nonparametric trend estimates detect multiple changes in the slopes and provide kink dates,
which are well aligned with the actual events.

\begin{figure}[htbp]
\begin{center}
\caption{Selection of $\lambda$}\label{lambda:other-filters}
\vskip10pt
\begin{tabular}{ccc}
\includegraphics[scale=0.45]{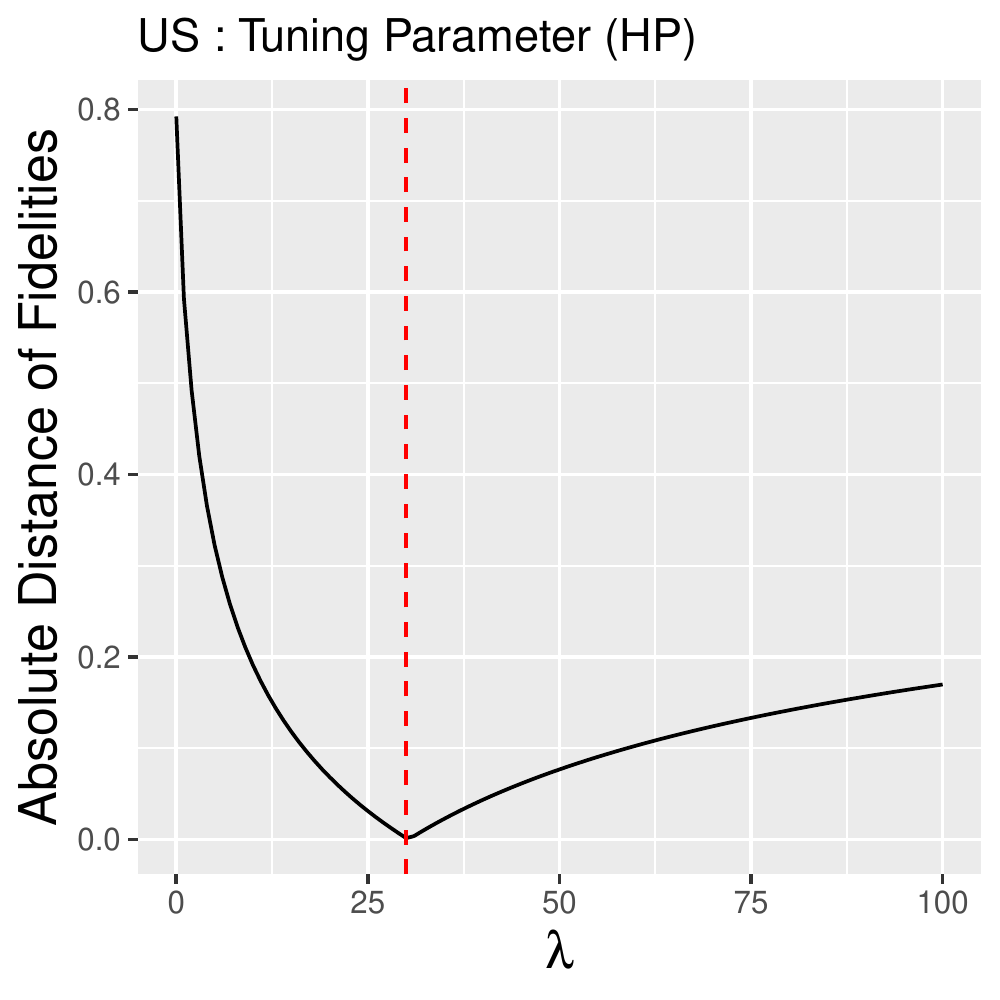} &
\includegraphics[scale=0.45]{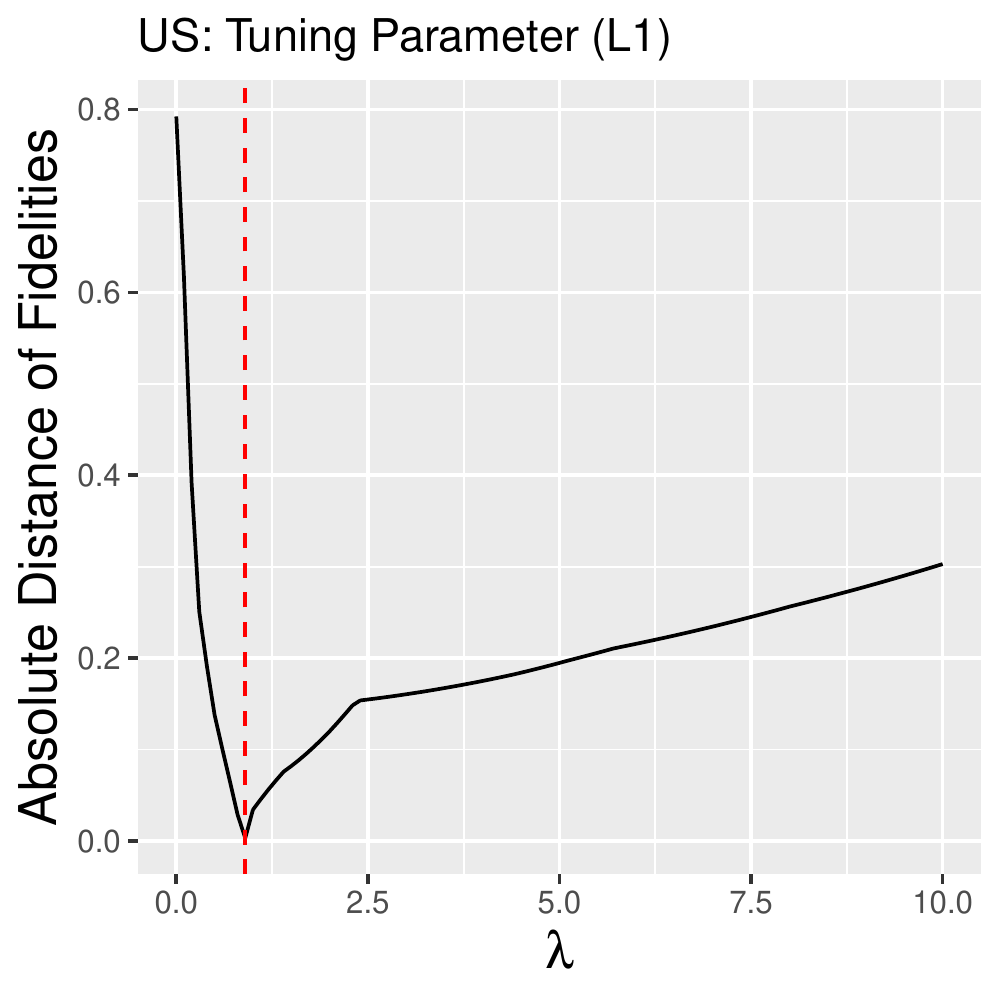} &
\includegraphics[scale=0.45]{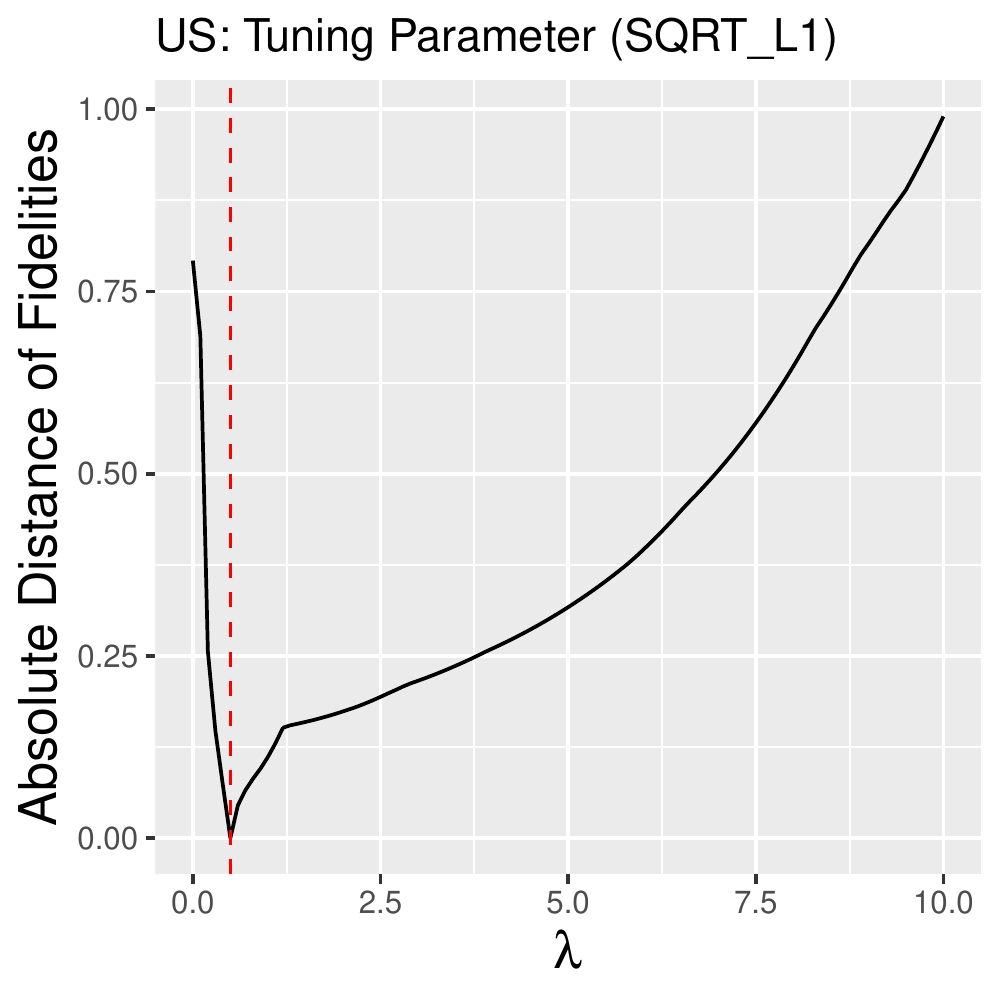}
\end{tabular}
\end{center}
\noindent \footnotesize{Note: The thunning parameter $\ld$ is chosen by minimizing the distance between two fidelities as described in Section 3.3. The selected tuning parameters for HP, $\ell_1$, and square-root $\ell_1$ are as 30, 0.9, and 0.5, repectively. }
\end{figure}

\begin{figure}[htbp]
\begin{center}
\caption{HP Filtering}\label{data-US-HP}
\vskip10pt
\begin{tabular}{cc}
\includegraphics[scale=0.5]{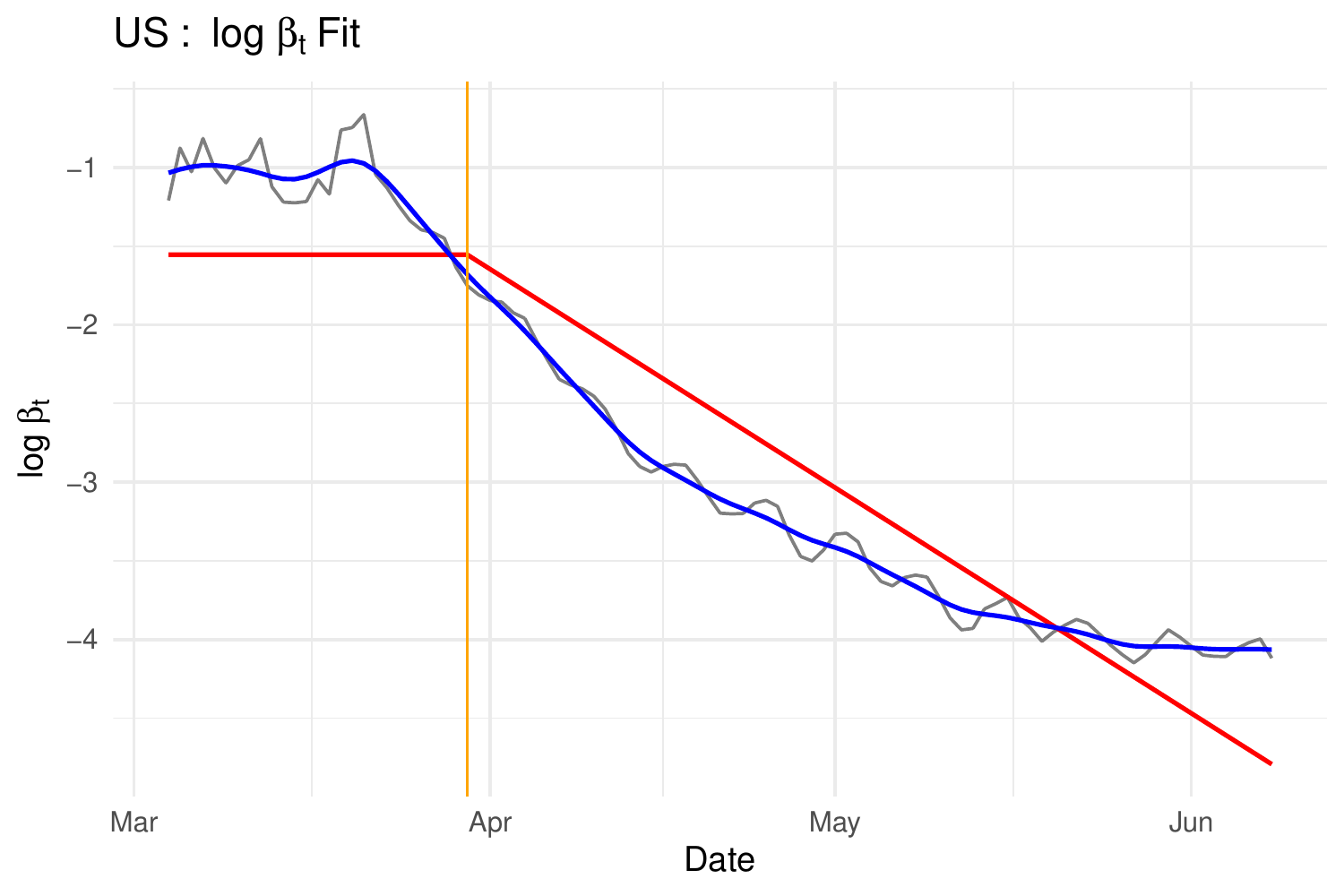} &
\includegraphics[scale=0.5]{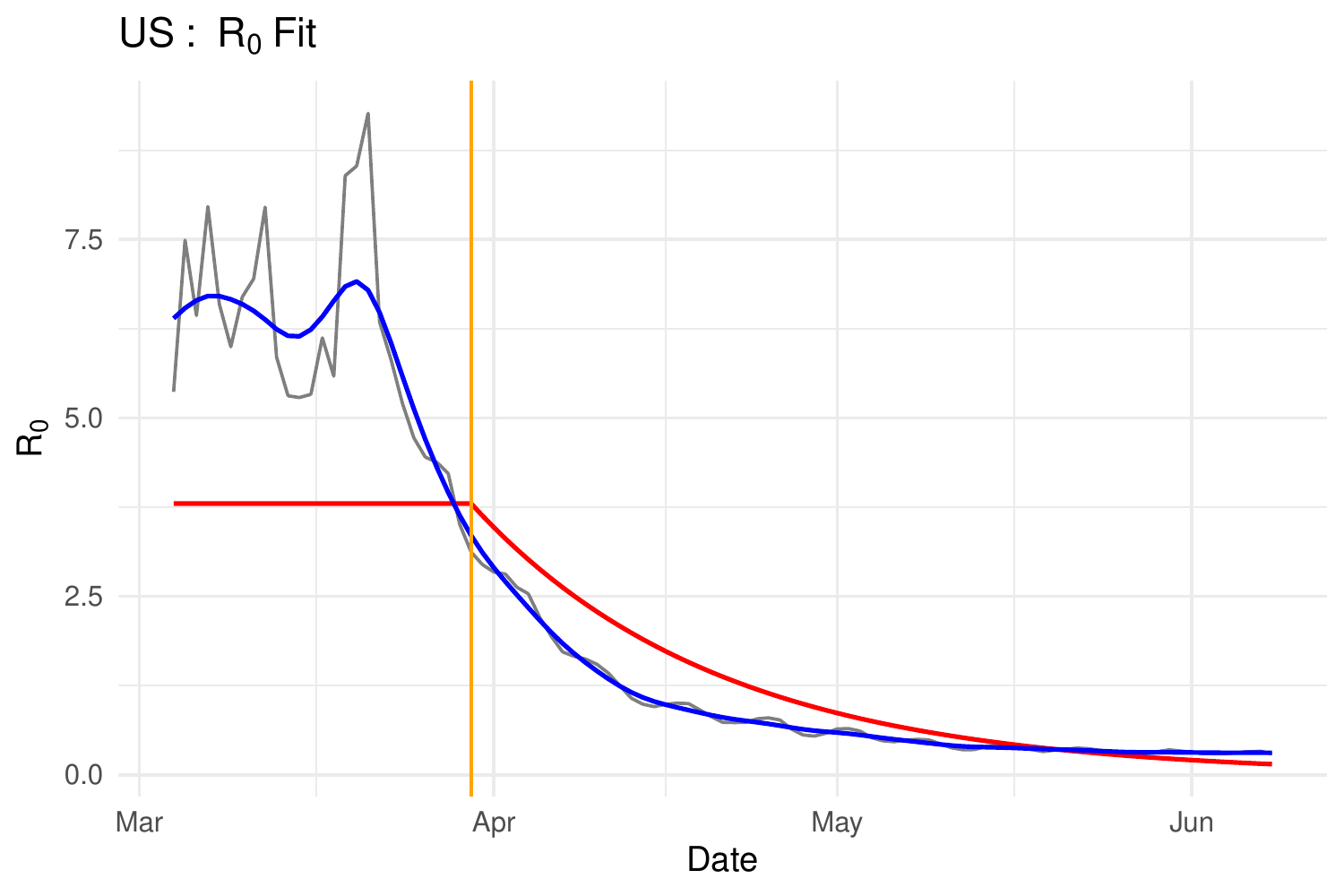} \\
\includegraphics[scale=0.5]{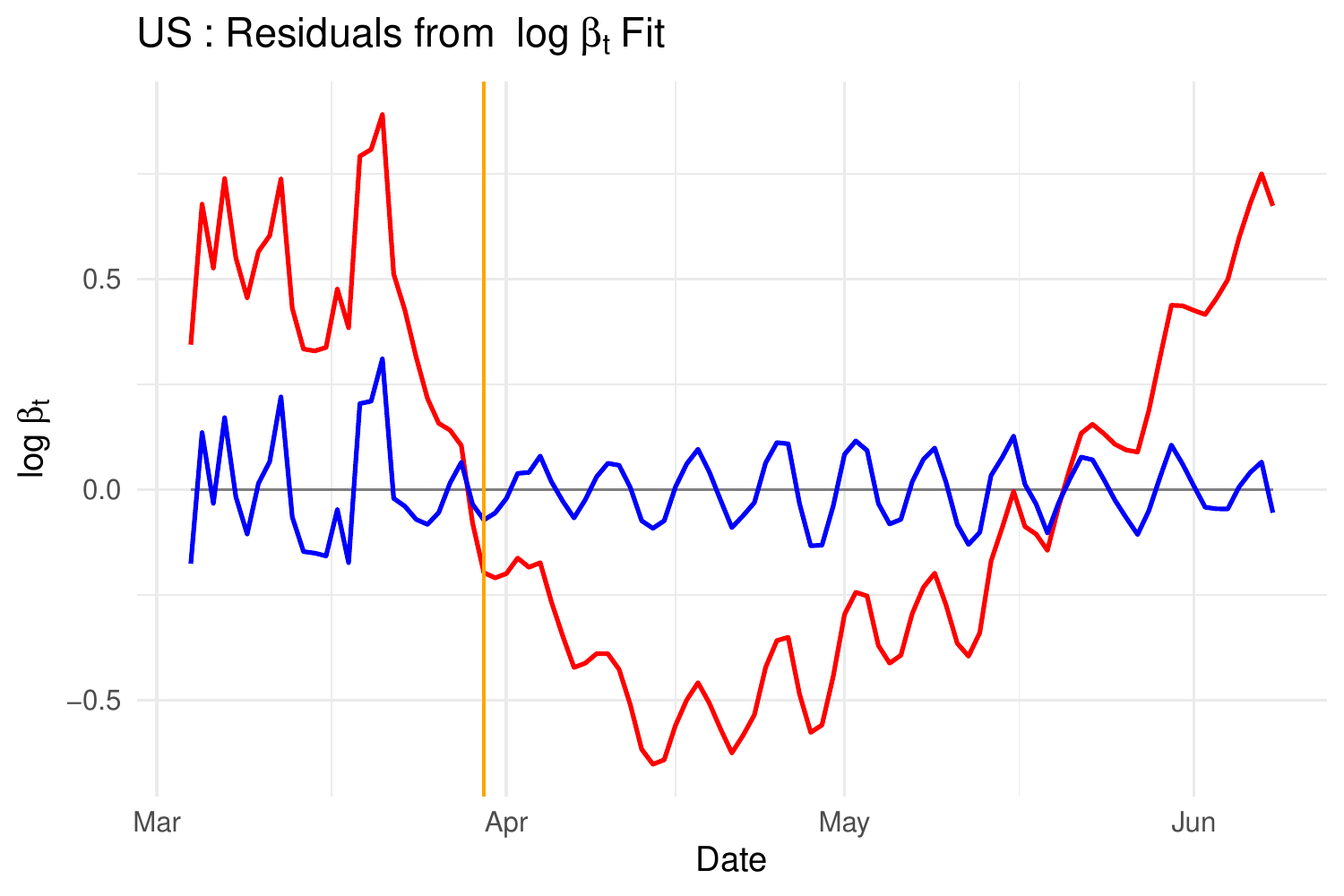}  &
\includegraphics[scale=0.5]{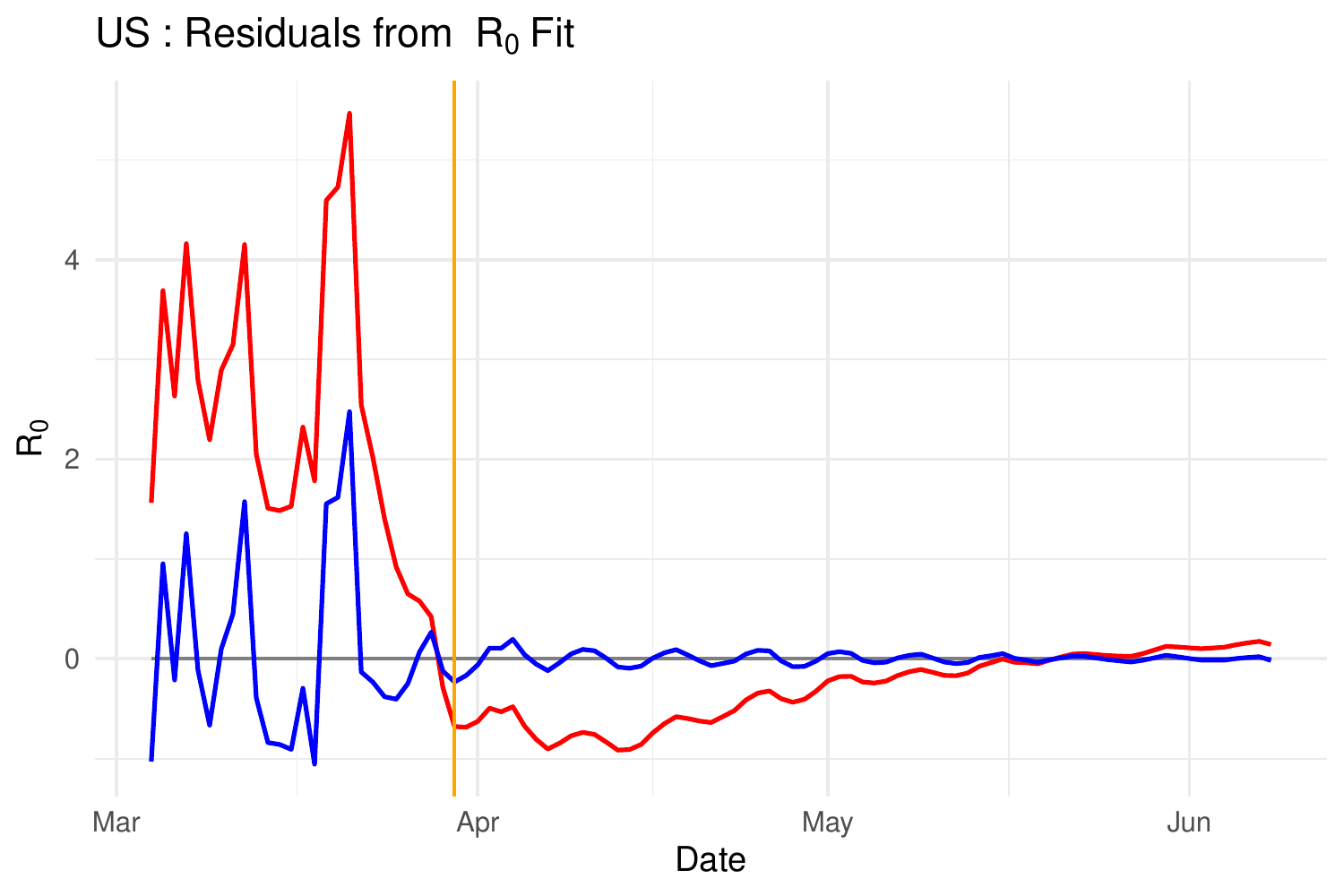} \\
\multicolumn{2}{c}{\includegraphics[scale=1]{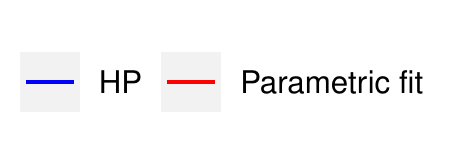}}
\end{tabular}
\end{center}
\noindent \footnotesize{Note:
HP filtering solves \eqref{HPfilter:def}.
The orange vertical line denotes the lockdown date, March 30.}
\end{figure}

We now turn to different filtering methods. In Figure~\ref{lambda:other-filters}, we show selection of $\lambda$
for the HP, $\ell_1$ and square-root $\ell_1$ filters. As explained in Section~\ref{sec:loocv},
the penalization parameter $\lambda$ is chosen to ensure that all different methods have the same level of fitting the data.
Figure~\ref{data-US-HP} shows the estimation results for the HP filter.
The HP trend estimates trace data pretty well after late March, as clear in residual plots. However, there is no kink
in the estimates due to the nature of the $\ell_2$ penalty term in the HP filter.
The tuning parameter was $\lambda = 30$, which is 30 times as large as the one used in the sparse HP filter.
This is because for the HP filter, $\lambda$ is the main tuning parameter; however,  for the sparse HP filter, $\lambda$ plays a minor role of regularizing the $\ell_0$ constrained method.

\begin{figure}[htbp]
\begin{center}
\caption{$\ell_1$ Filtering}\label{data-US-L1}
\vskip10pt
\begin{tabular}{cc}
\includegraphics[scale=0.5]{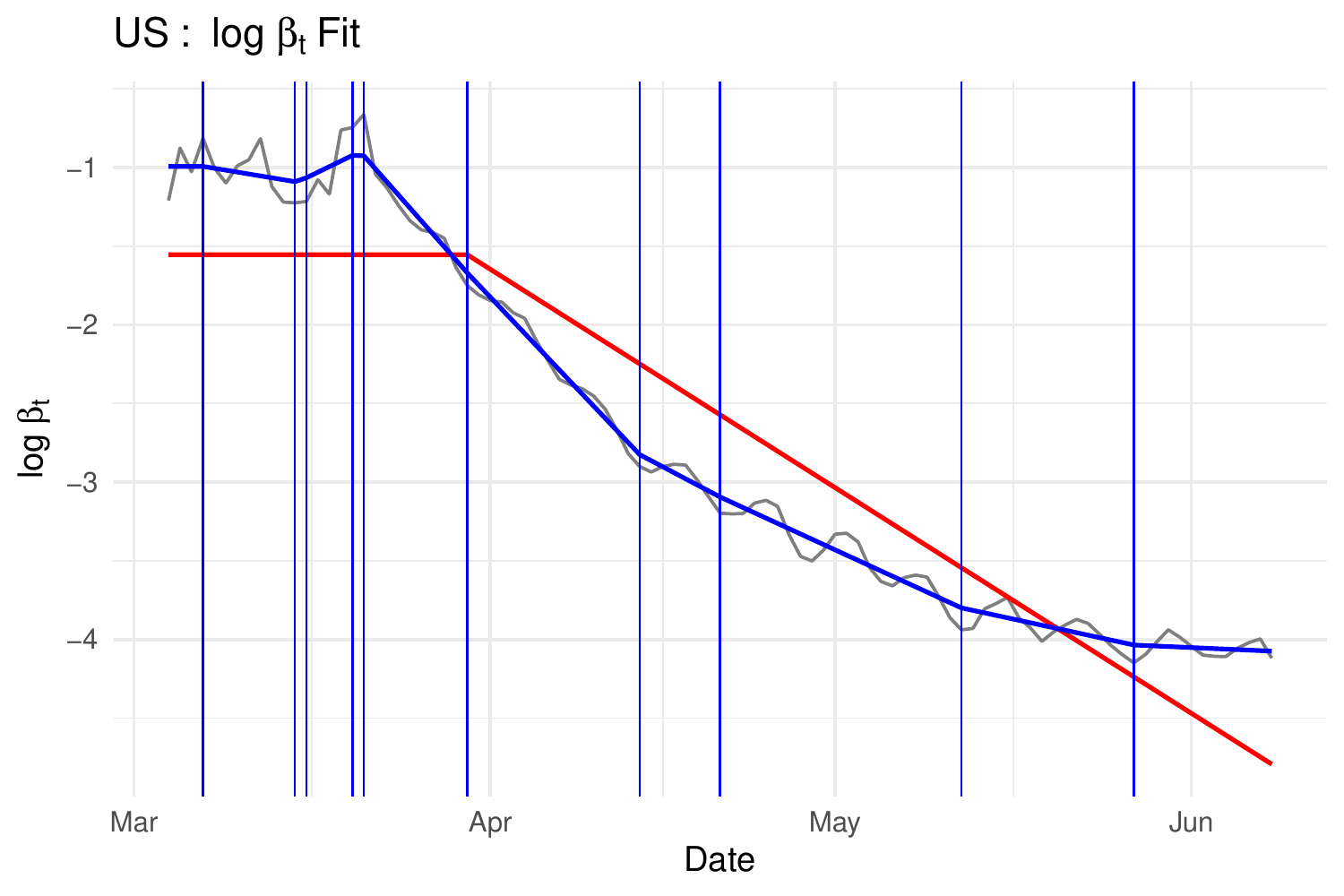} &
\includegraphics[scale=0.5]{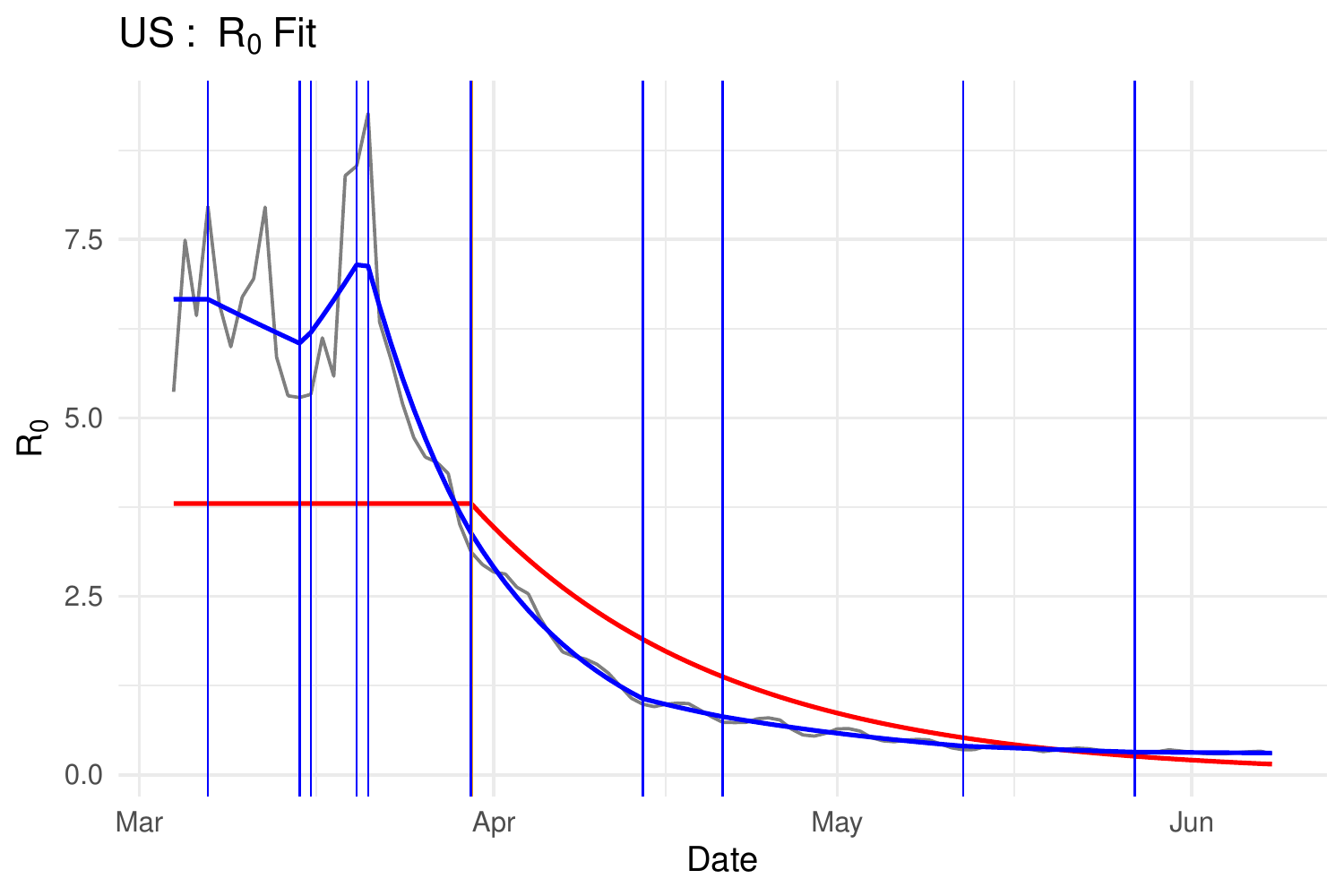} \\
\includegraphics[scale=0.5]{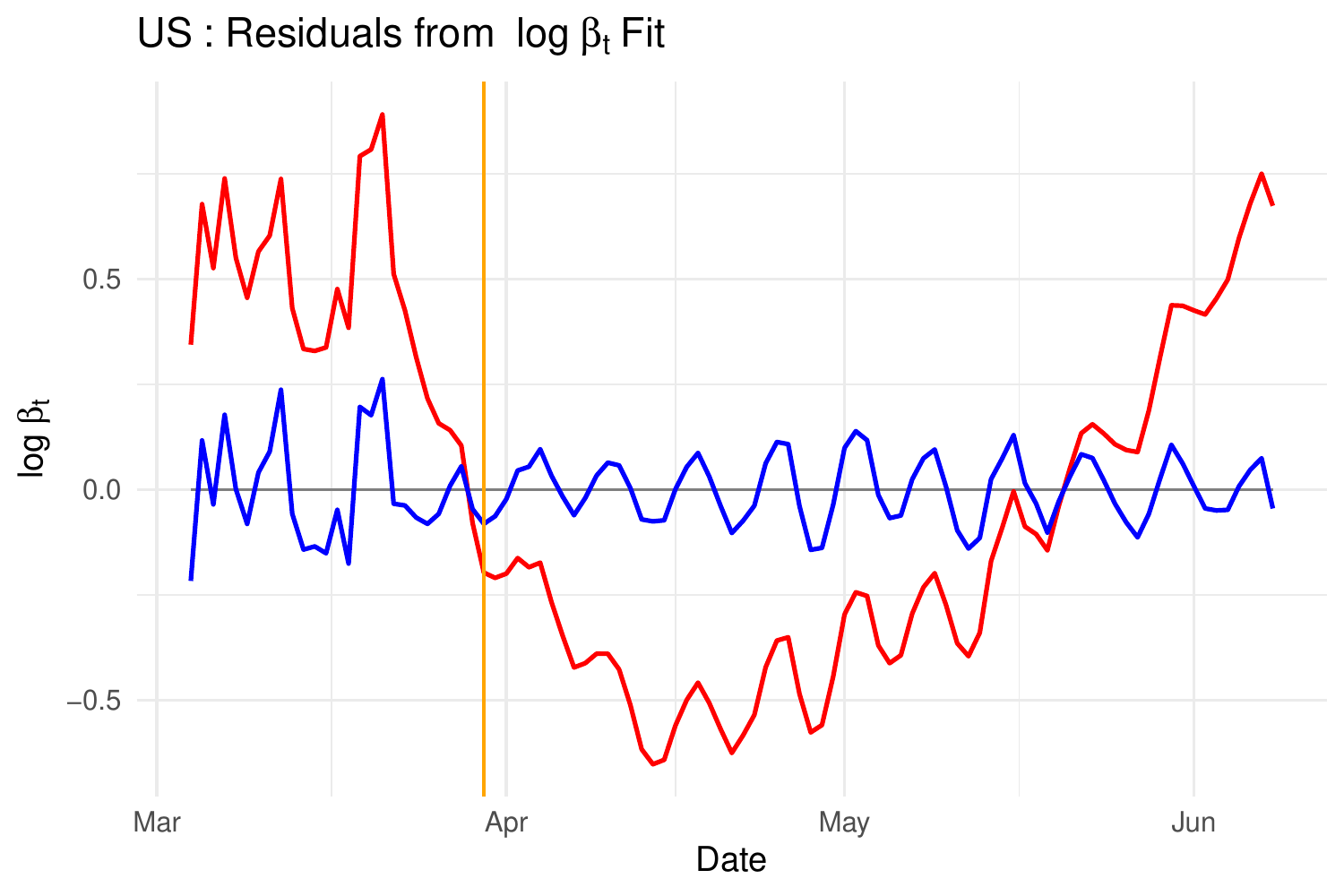}  &
\includegraphics[scale=0.5]{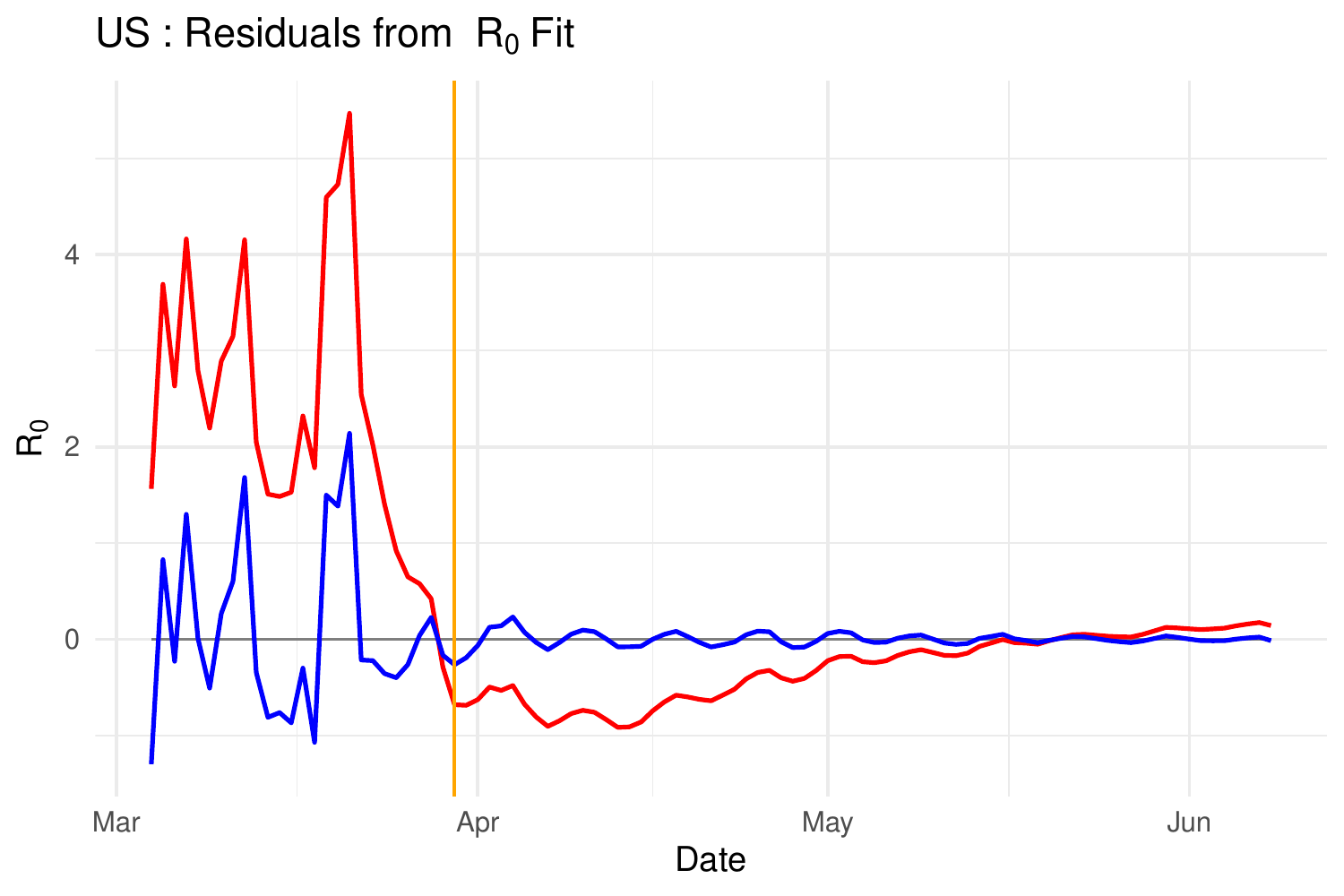} \\
\multicolumn{2}{c}{\includegraphics[scale=1]{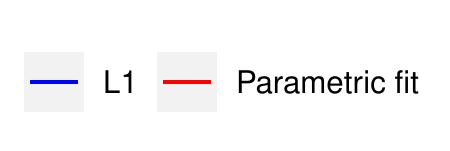}}
\end{tabular}
\end{center}
\noindent \footnotesize{Note:
$\ell_1$ filtering solves \eqref{L1filter}.
The estimated kinks denoted by blue vertical lines are: March 7, March 15, March 16, March 20, March 21, March 30, April 14, April 21, May 12, and May 27. The orange vertical line denotes the lockdown date, March 30.
The $\ell_1$-filtering kink dates are calculated by any $t$ such that $|\Delta^2 \log \hat{\beta}_t| > \eta$, where $\eta=10^{-6}$ is an effective zero.
}

\end{figure}

\begin{figure}[htbp]
\begin{center}
\caption{Square-root $\ell_1$ Filtering}\label{data-US-SQRT_L1}
\vskip10pt
\begin{tabular}{cc}
\includegraphics[scale=0.5]{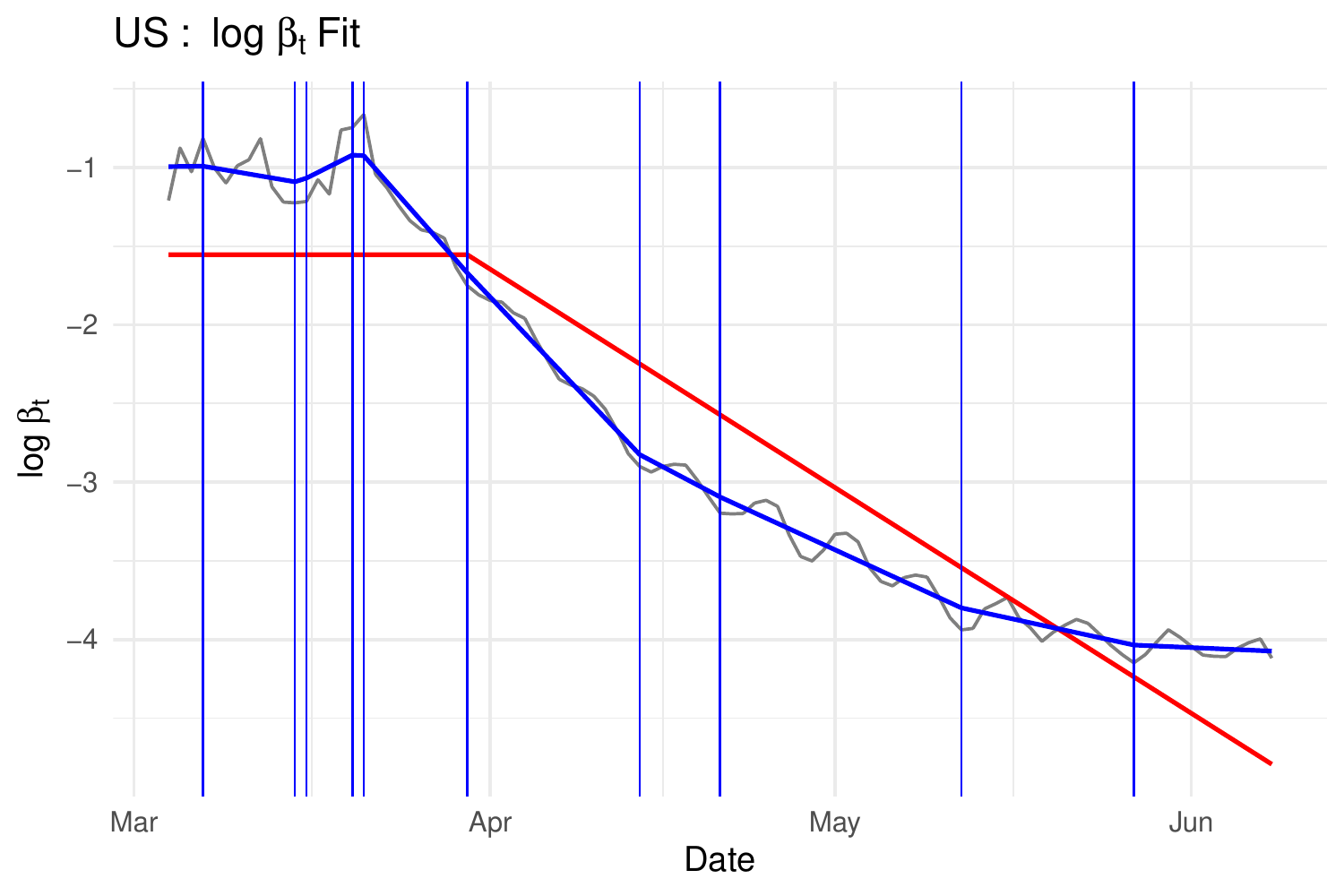} &
\includegraphics[scale=0.5]{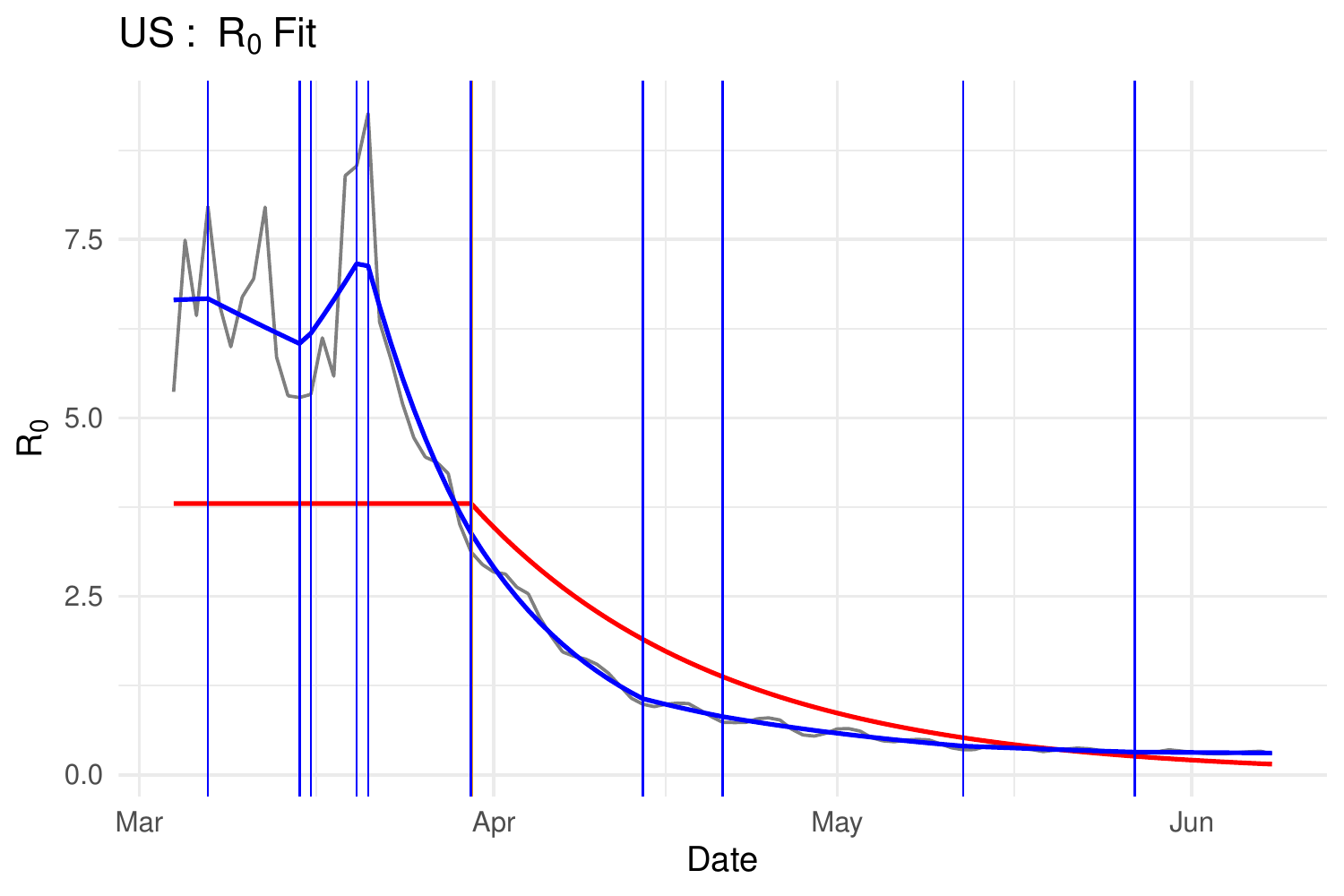} \\
\includegraphics[scale=0.5]{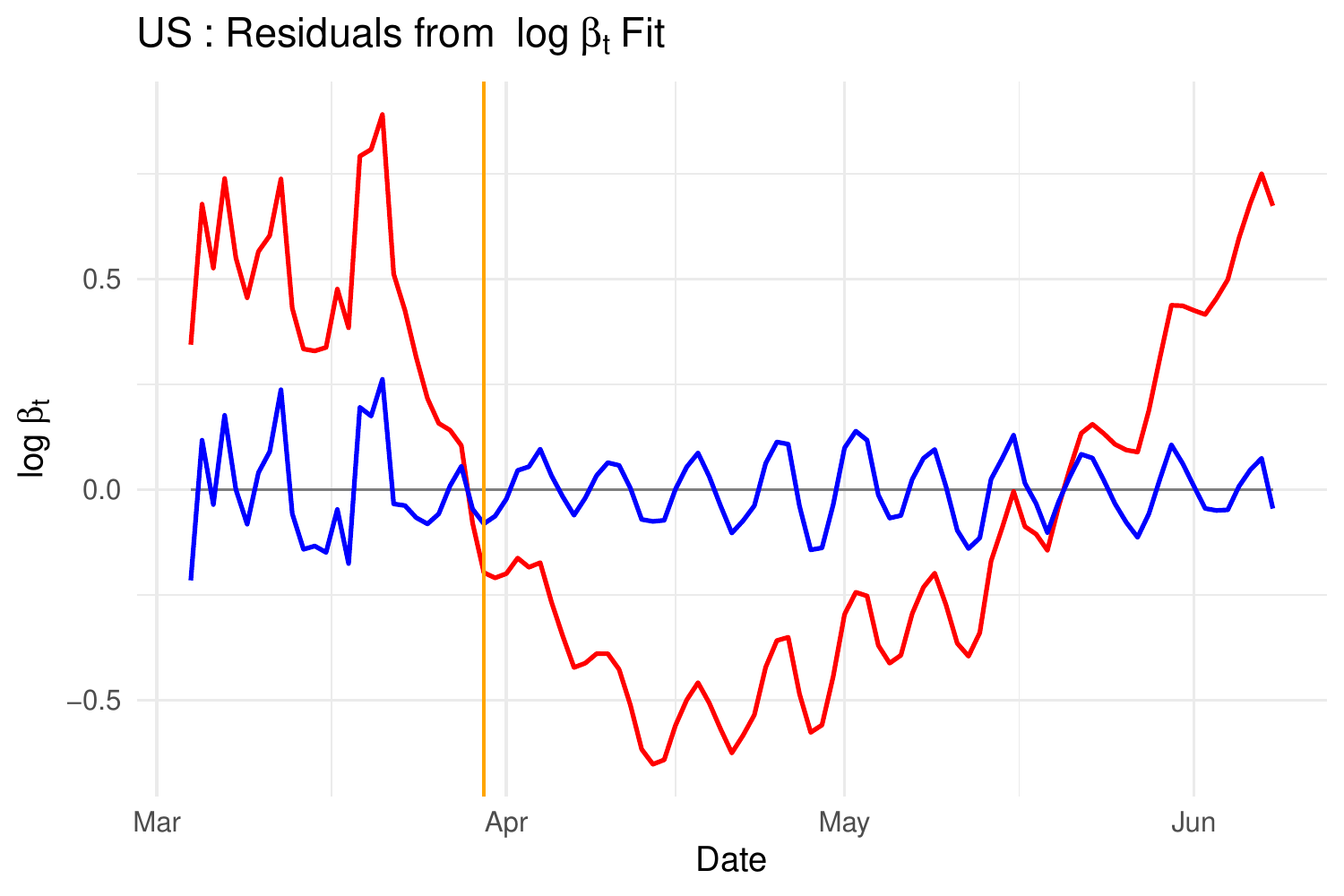}  &
\includegraphics[scale=0.5]{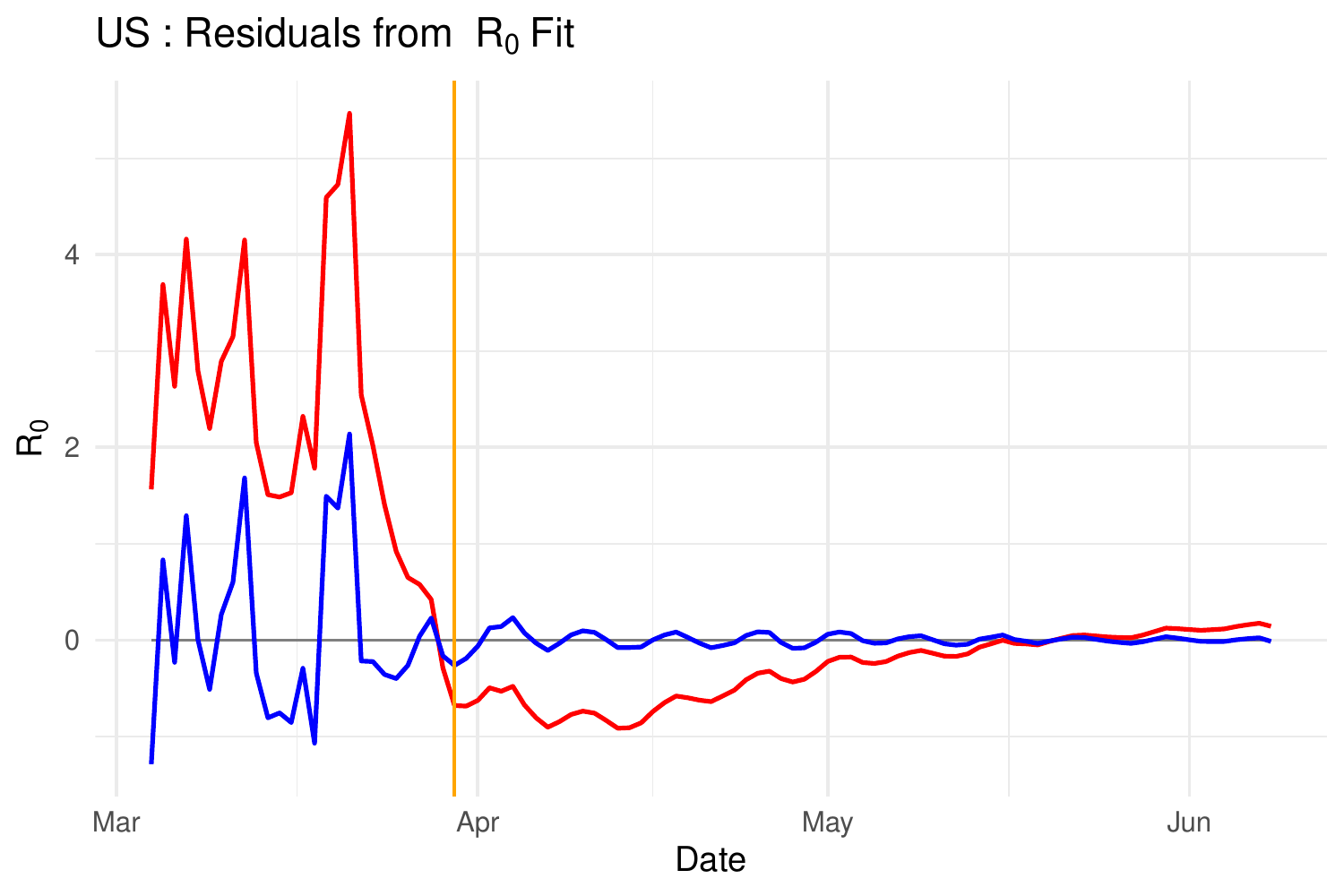}\\
\multicolumn{2}{c}{\includegraphics[scale=1]{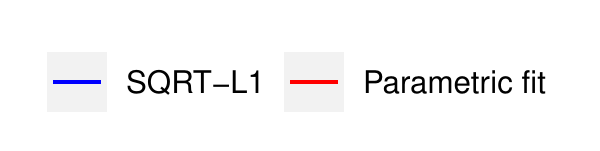}}
\end{tabular}
\end{center}
\noindent \footnotesize{Note:
Square-root $\ell_1$ filtering solves \eqref{L1filter-sqrt}.
 The estimated kinks denoted by blue vertical lines are: March 7, March 15, March 16, March 20, March 21, March 30, April 14, April 21, May 12, and May 27.
The orange vertical line denotes the lockdown date, March 30.}
\end{figure}

\begin{figure}[htbp]
\begin{center}
\caption{Sparse HP and $\ell_1$ Filtering for the US}\label{fig:US-filter}
\vskip10pt
\begin{tabular}{cc}
\includegraphics[scale=0.5]{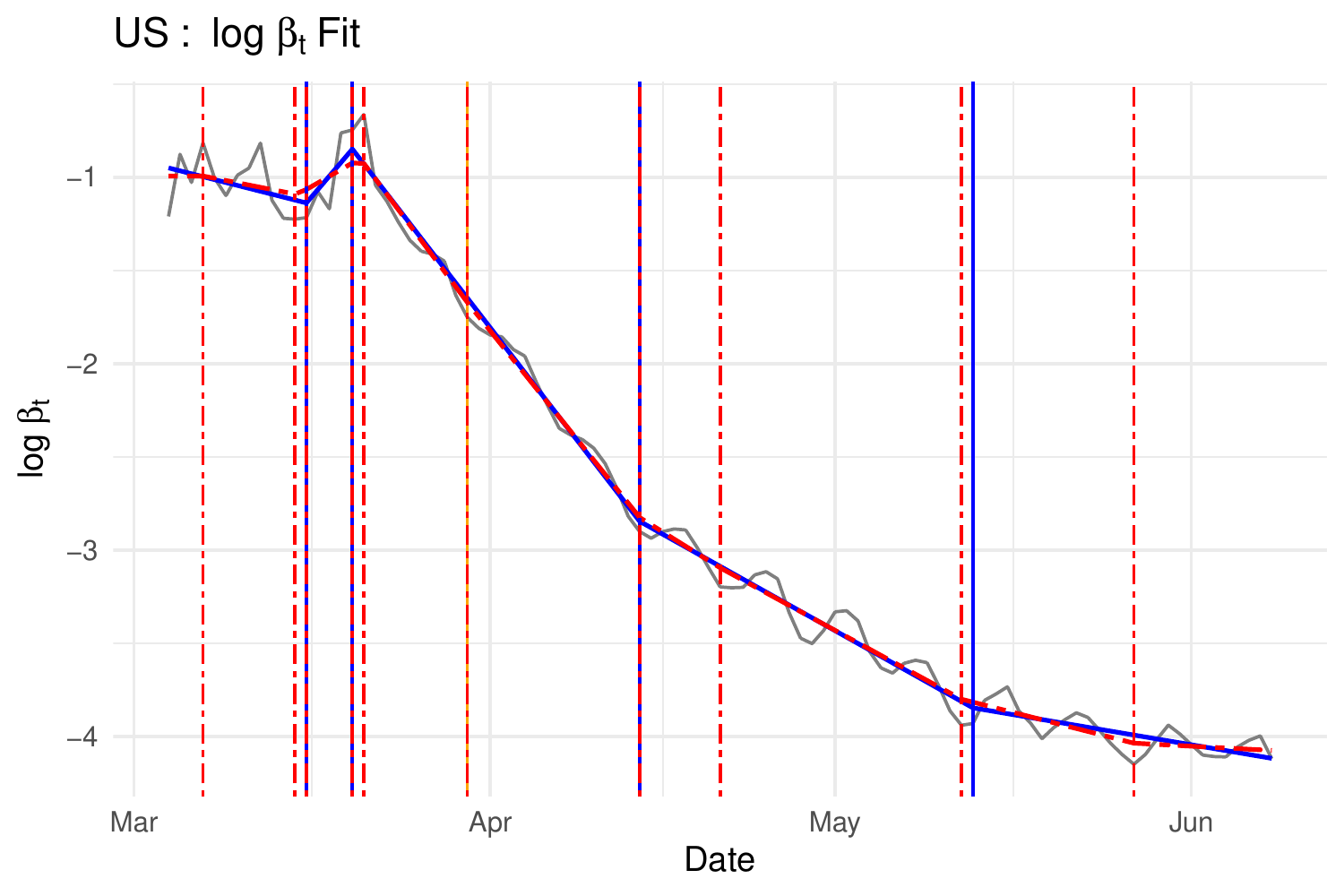} &
\includegraphics[scale=0.5]{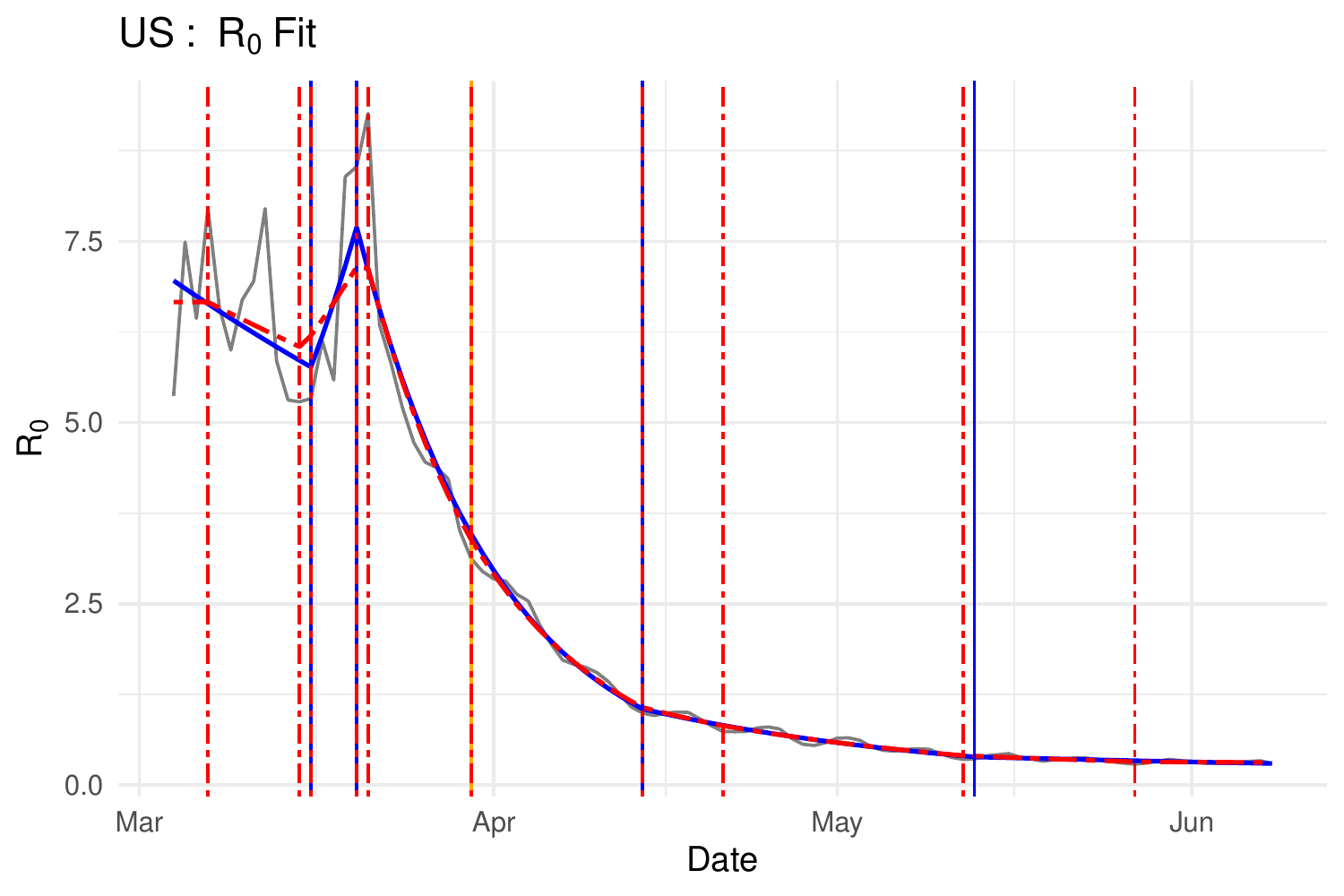} \\
\includegraphics[scale=0.5]{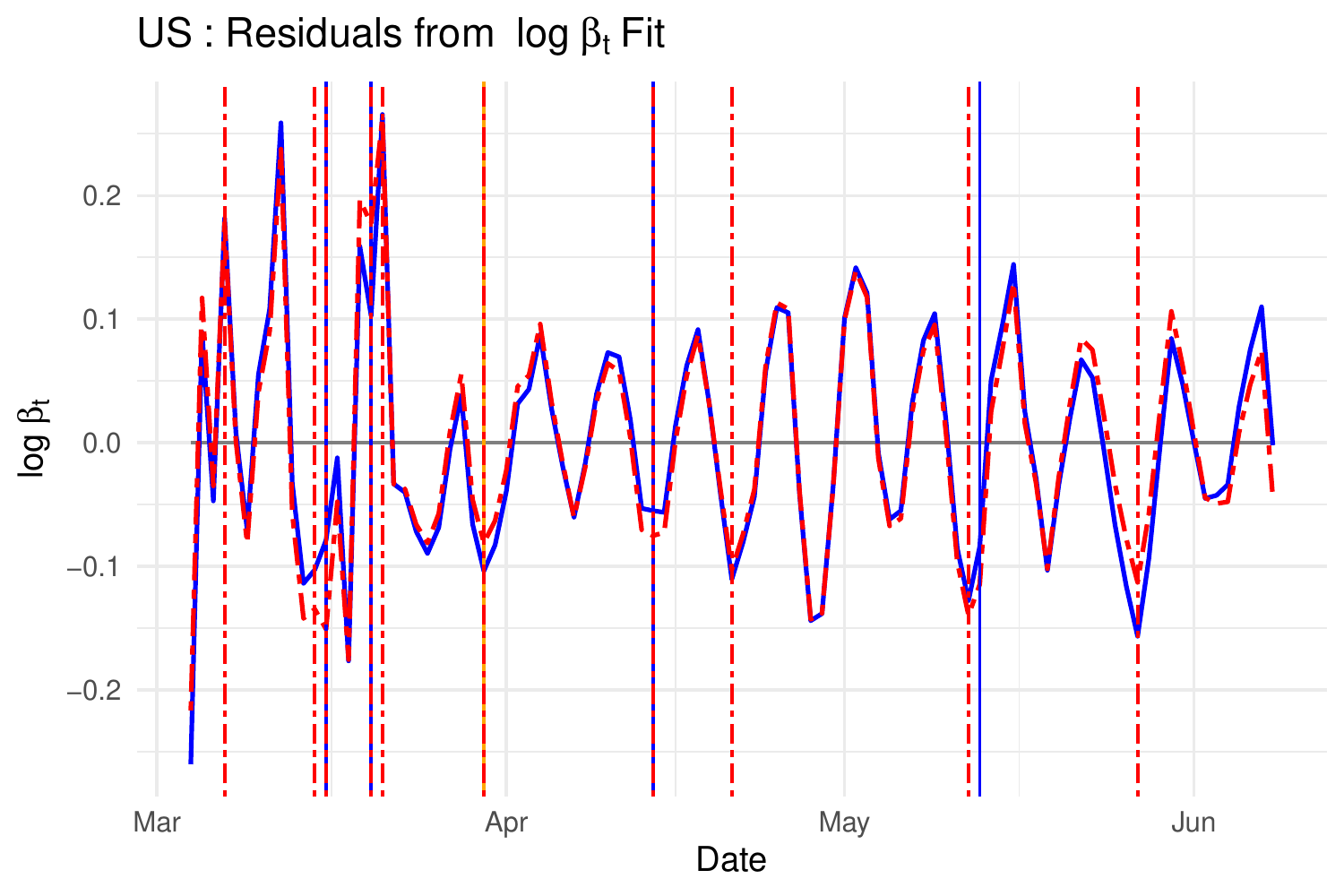}  &
\includegraphics[scale=0.5]{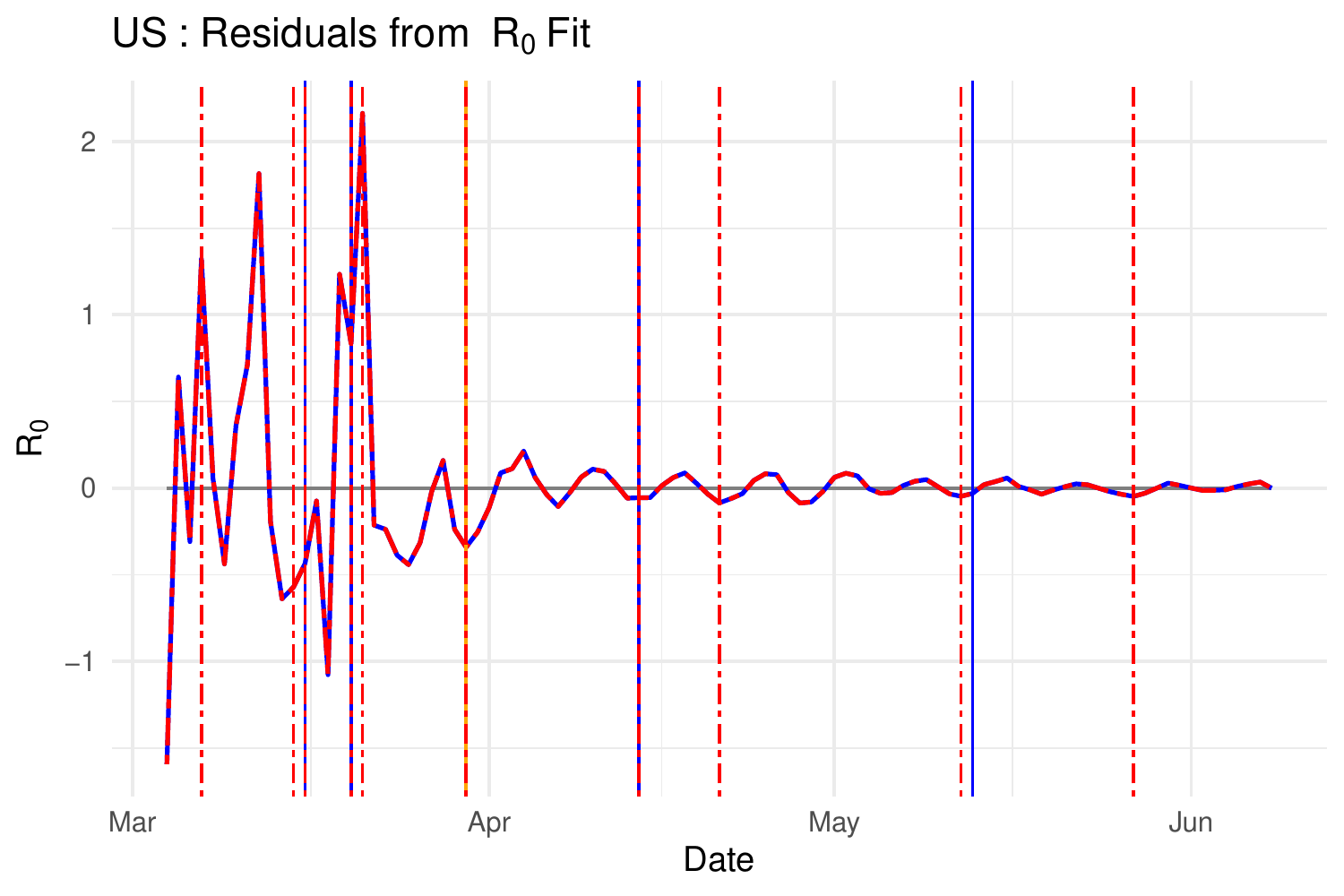} \\
\multicolumn{2}{c}{\includegraphics[scale=1]{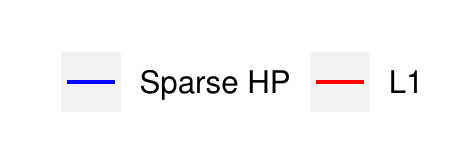}}
\end{tabular}
\end{center}
\noindent \footnotesize{Note: The Sparse HP kinks (blue) are:
March 16, March 20, April 14, and May 13.
The $\ell_1$ kinks (red) are:
March 7, March 15, March 16, March 20, March 21, March 30, April 14, April 21, May 12, and May 27.
The orange vertical line denotes the lockdown date, March 30.}
\end{figure}

In Figure~\ref{data-US-L1}, we plot estimation results using $\ell_1$ trend filtering.
The results look similar to those in Figure~\ref{data-US-HP}, but there are now 10 kink points:
March 7, March 15, March 16, March 20, March 21, March 30, April 14, April 21, May 12, May 27.
They are dates $t$ such that $|\Delta^2 \log \widehat{\beta}_t| > \eta$, 
where $\Delta^2$ is the double difference operator and
$\eta=10^{-6}$ is an effective zero.\footnote{The results are robust to the size of the effective zero and do not change even if we set $\eta=10^{-3}$. Gurobi used for the Sparse HP filtering also imposes some effective zeros in various constraints. We use the default values of them. For example, the integer tolerance level and the general feasibility tolerance level are $10^{-5}$ and $10^{-6}$, respectively.}
The tuning parameter $\ld=0.9$ was chosen by minimizing the distance between the fidelity of $\ell_1$  and that of the Sparse HP.
Recall that the sparse HP filter produces the kinks on
March 16, March 20, April 14, and May 13.
In other words, the $\ell_1$ filter estimates 6 more kinks than the sparse HP filter when both fit the data equally well.
It is unlikely that two adjacent dates (March 15-16 and March 20-21) correspond to two different regimes in the time-varying contact rate.
This suggests that the $\ell_1$ filter may over-estimate the number of kinks.
Figure~\ref{data-US-SQRT_L1} shows estimation results for the square-root $\ell_1$ trend filters.
The chosen $\ld=0.5$ was smaller than that of the $\ell_1$ trend filter due to the change in the scale of the fidelity term;
however, the trend estimates look very similar and the estimated kinks are identical between
the $\ell_1$ and square-root $\ell_1$ trend filters.
In Figure~\ref{fig:US-filter}, we plot the sparse HP filter estimates along with $\ell_1$ filter estimates.
Both methods have produced very similar trend estimates, but the number of kinks is substantially different:
only 4 kinks for the sparse HP filter but 10 kinks for the $\ell_1$ filter.


\subsection{Other Countries: Canada, China, South Korea and the UK}\label{section:RoW}

In this section, we provide condensed estimation results for other countries.
We focus on the sparse HP and $\ell_1$ filters whose tuning parameters are chosen as in the previous section.
Appendices \ref{appendix:SHP} and \ref{appendix:L1} contain the details of the selection of tuning parameters.

\begin{figure}[htbp]
\begin{center}
\caption{Sparse HP and $\ell_1$ Filtering for Canada}\label{fig:Canada-filter}
\vskip10pt
\begin{tabular}{cc}
\includegraphics[scale=0.5]{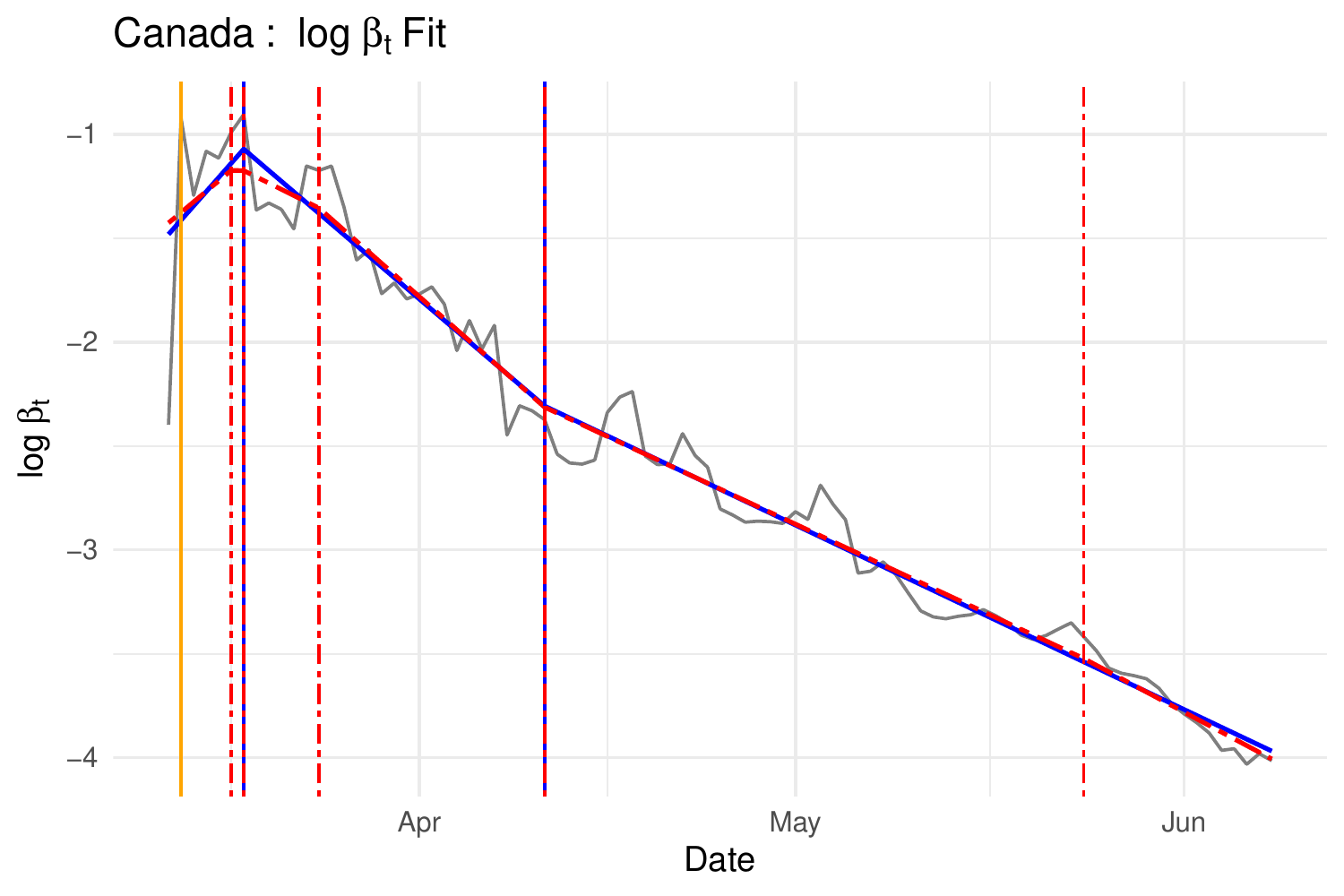} &
\includegraphics[scale=0.5]{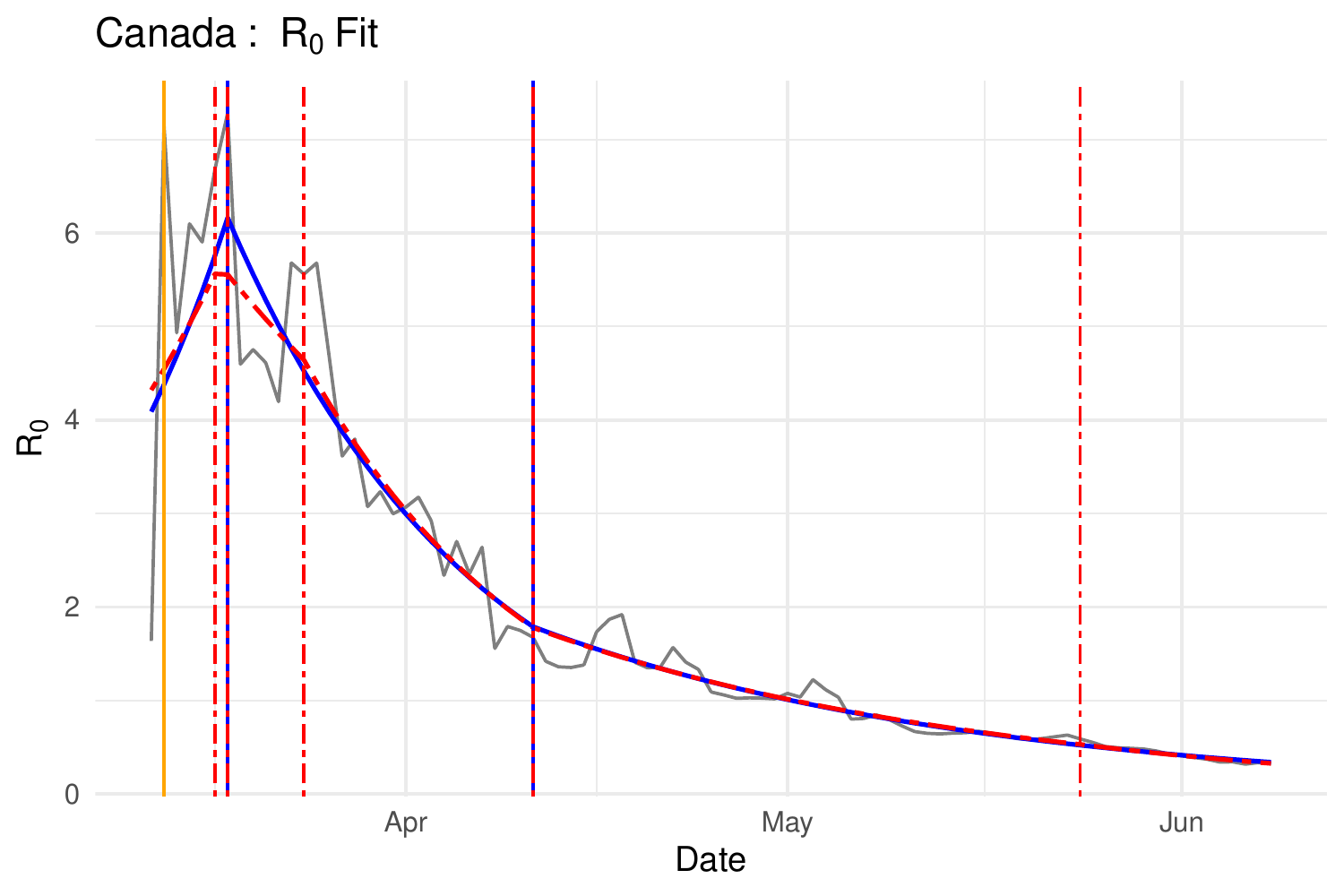} \\
\includegraphics[scale=0.5]{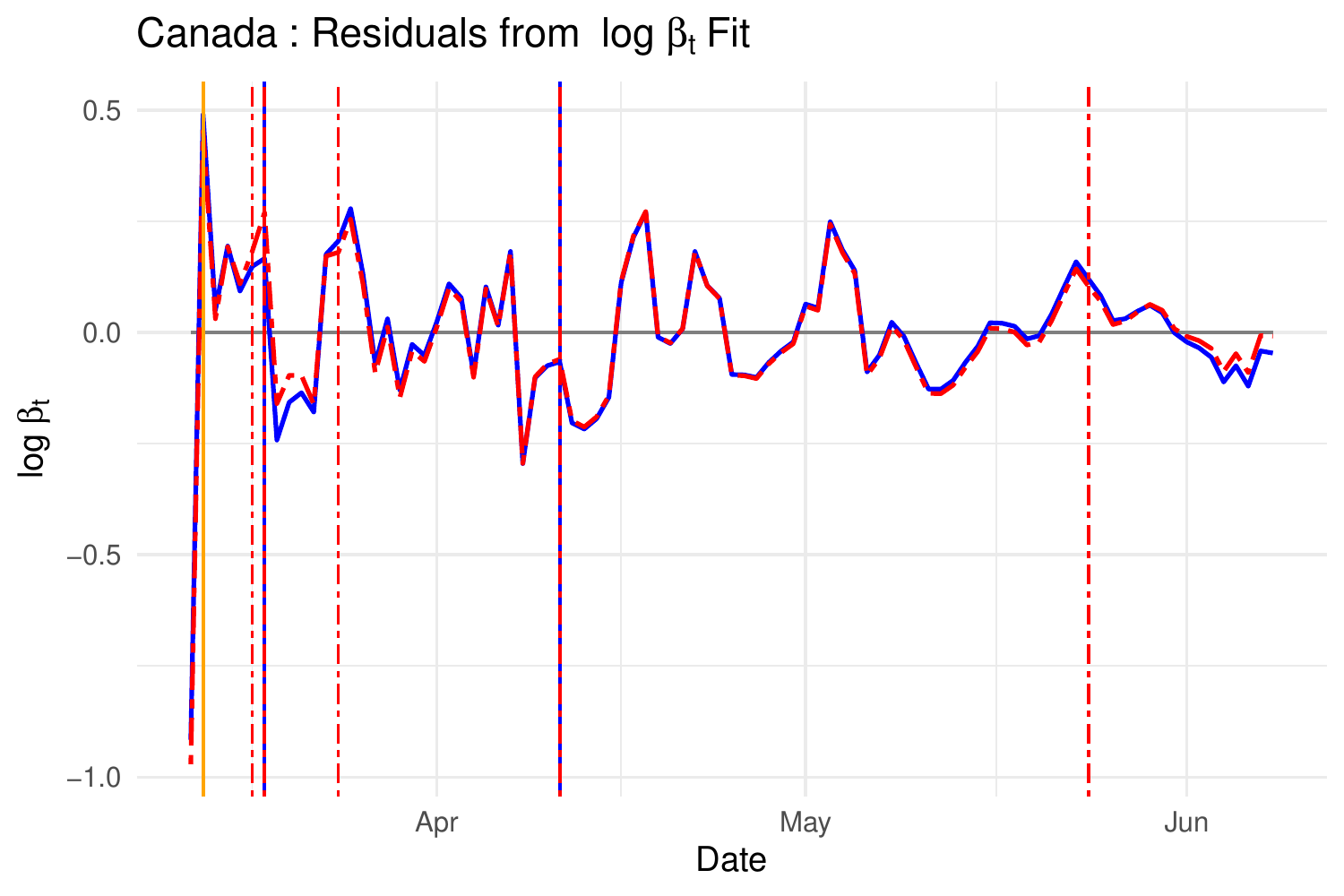}  &
\includegraphics[scale=0.5]{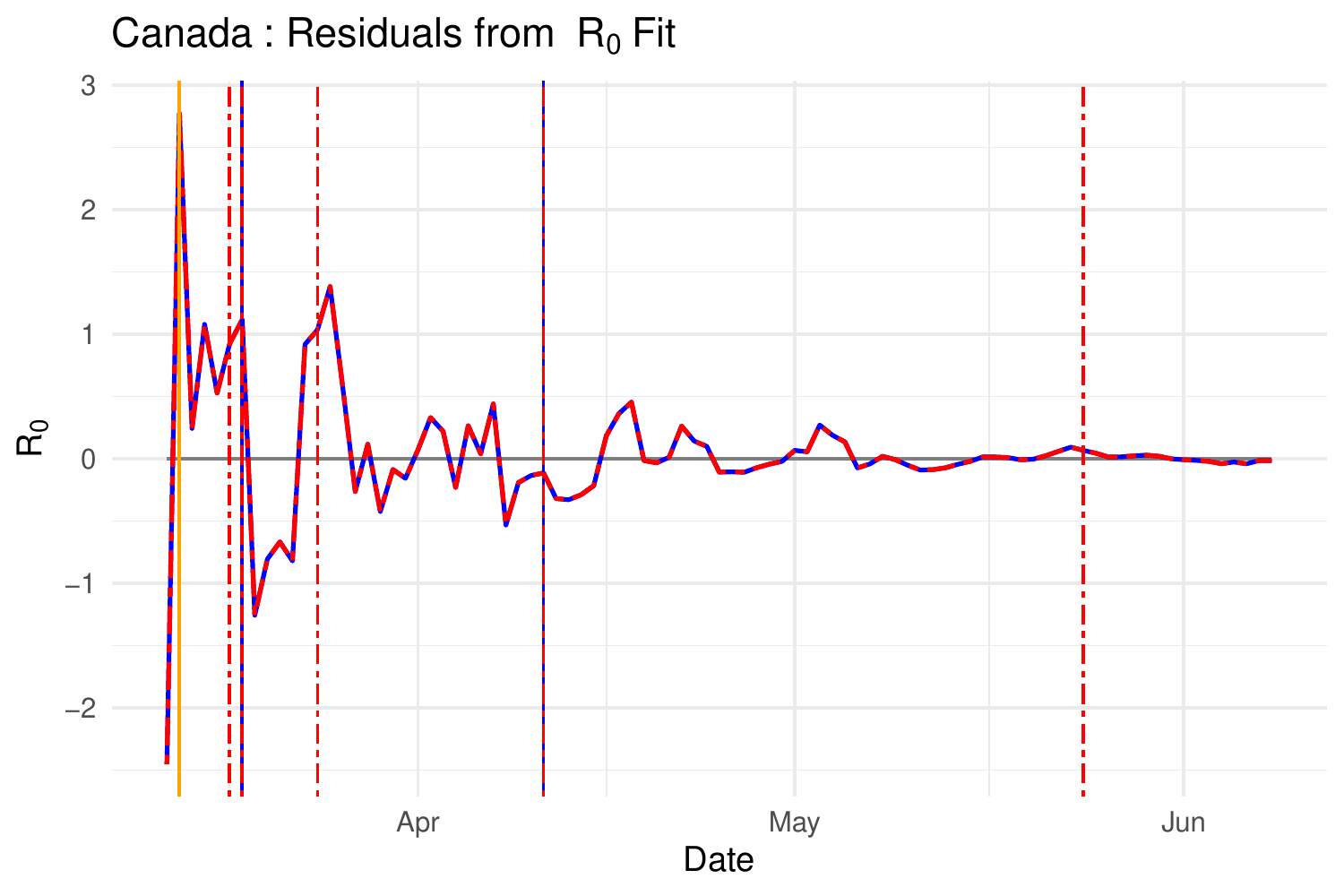}
\end{tabular}
\end{center}
\noindent \footnotesize{Note: The Sparse HP kinks (blue) are:
March 18 and April 11.
The $\ell_1$ kinks (red) are:
March 17, March 18, March 24, April 11, and May 24.
The orange vertical line denotes the lockdown date, March 13.}

\end{figure}

Figure~\ref{fig:Canada-filter} shows the empirical results of Canada. The estimated kink dates are:
March 18 and April 11. Based on them, we can classify observations into three periods:

\begin{enumerate}
\item March 6 - March 18: This is an initial period of the epidemic in Canada. The contact rate was peaked at the end of this period. Several lockdown measures started to be imposed.
\item March 18 - April 11: We observe a sharp decrease in the contact rate in this period. Additional measures were imposed.
\item April 11 - June 8: The contact rate decreased but less steeply.
\end{enumerate}


Quebec and Ontario are the two provinces hardest hit by COVID-19. In Quebec, daycares, public schools, and universities are closed on March 13 followed by non-essential businesses and public gathering places on March 15. Montreal declared state of emergency on March 27 \citep{CTV-Quebec:timeline}. Similarly, all public schools in Ontario are closed on March 12. The state of emergency was announced in Ontario on March 17 and ordered to close all non-essential businesses on March 23 \citep{Global-Ontario:timeline}. We set the lockdown date in Canada on March 13 as other provincial governments as well as the federal government started to recommend the social distancing measures strongly along with the cancellation of various events on the date \citep{CBC:March13}.
These tight lockdown and social distancing measures seemed to contribute the sharp decline of the contact rate in the second period. Both governments started to announce the plans to lift the lockdown measures at the end of April, which corresponds to the third period. Lockdown fatigue would also cause the slower decrease of the contact rate. In sum, a series of social distancing measures have been effective to decrease the contact rate but with some lags. The sparse HP filtering separates these periods reasonably well. However, the $\ell_1$ filtering overfits the model with 5 kinks.

\begin{figure}[htbp]
\begin{center}
\caption{Sparse HP and $\ell_1$ Filtering for China}\label{fig:China-filter}
\vskip10pt
\begin{tabular}{cc}
\includegraphics[scale=0.5]{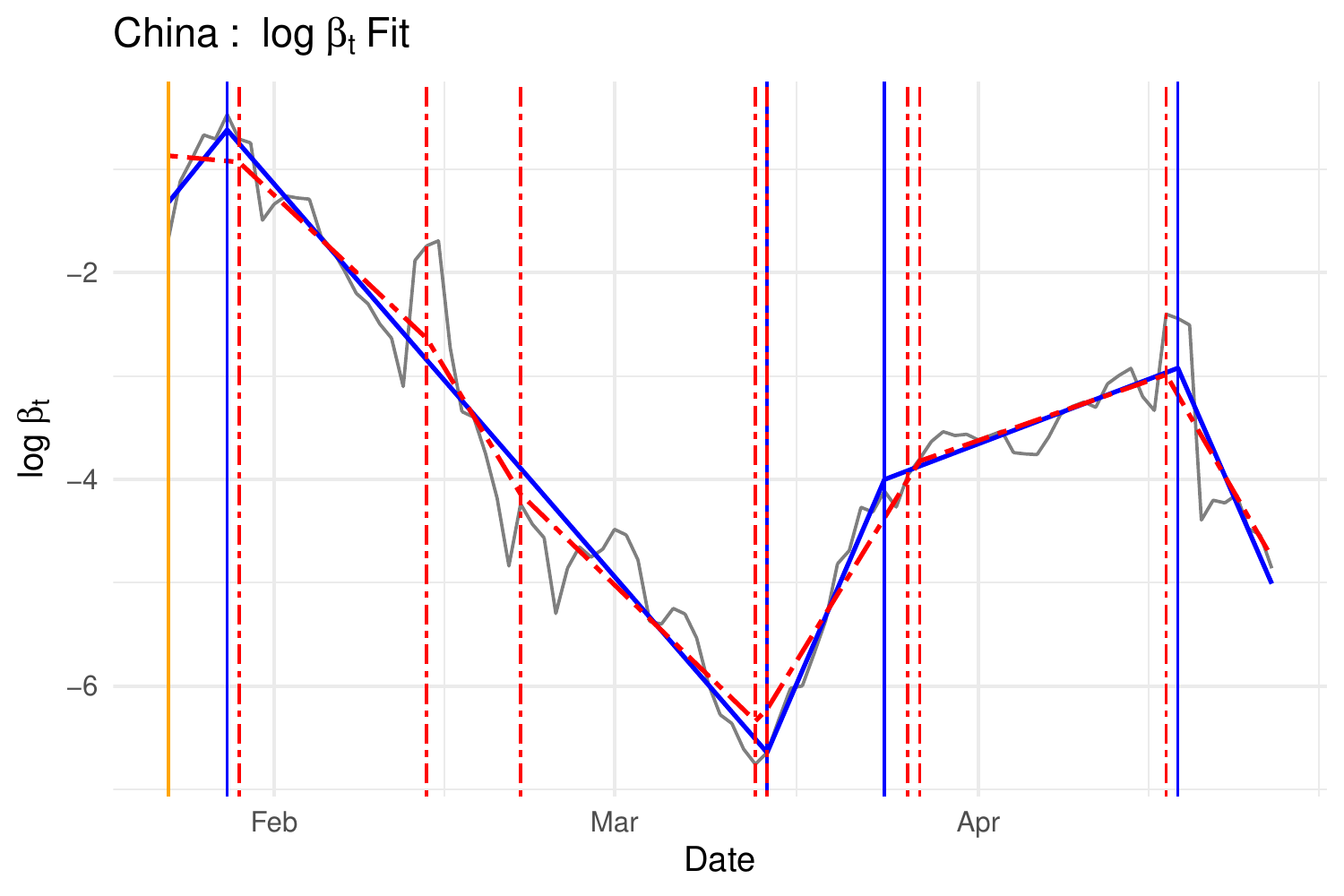} &
\includegraphics[scale=0.5]{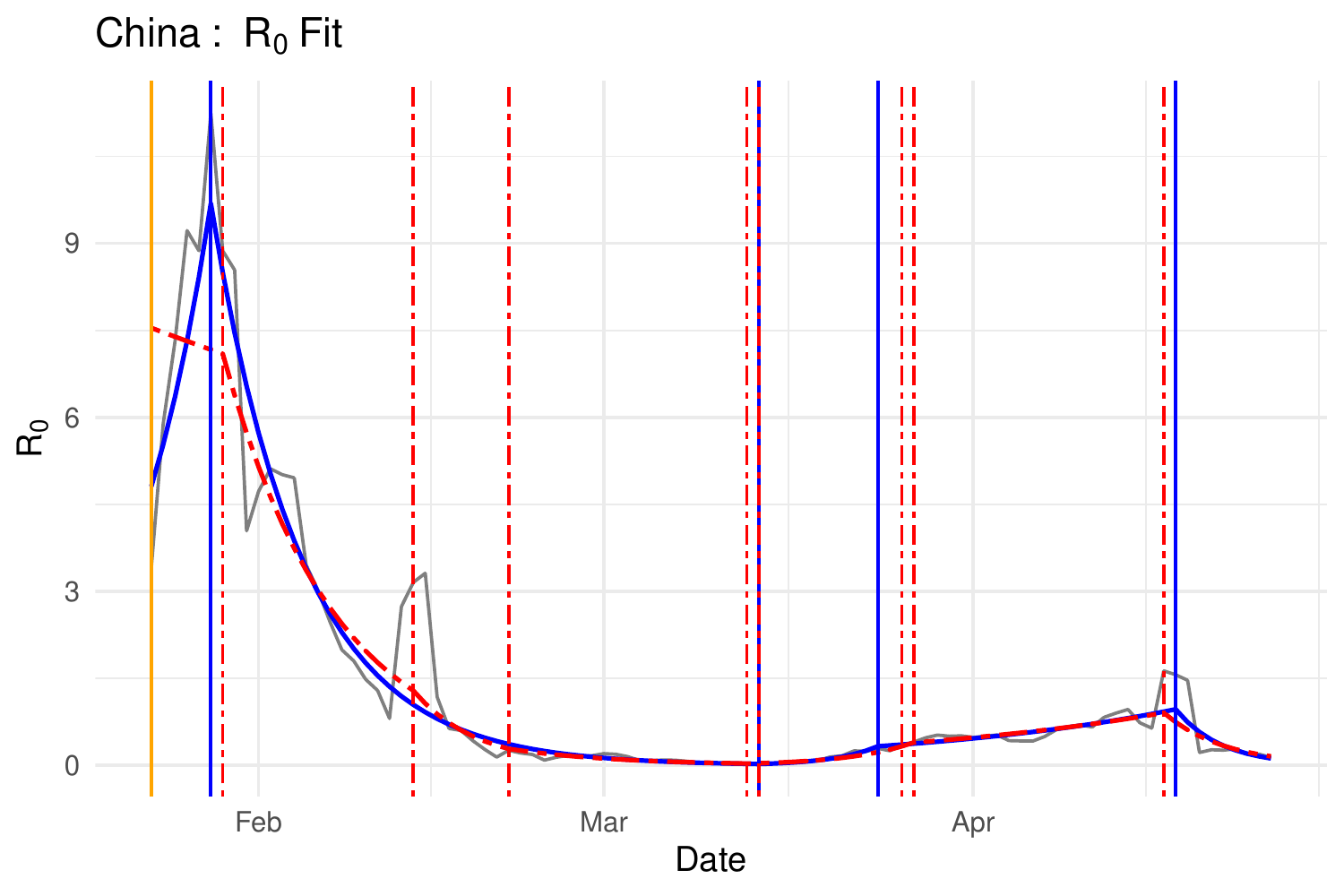} \\
\includegraphics[scale=0.5]{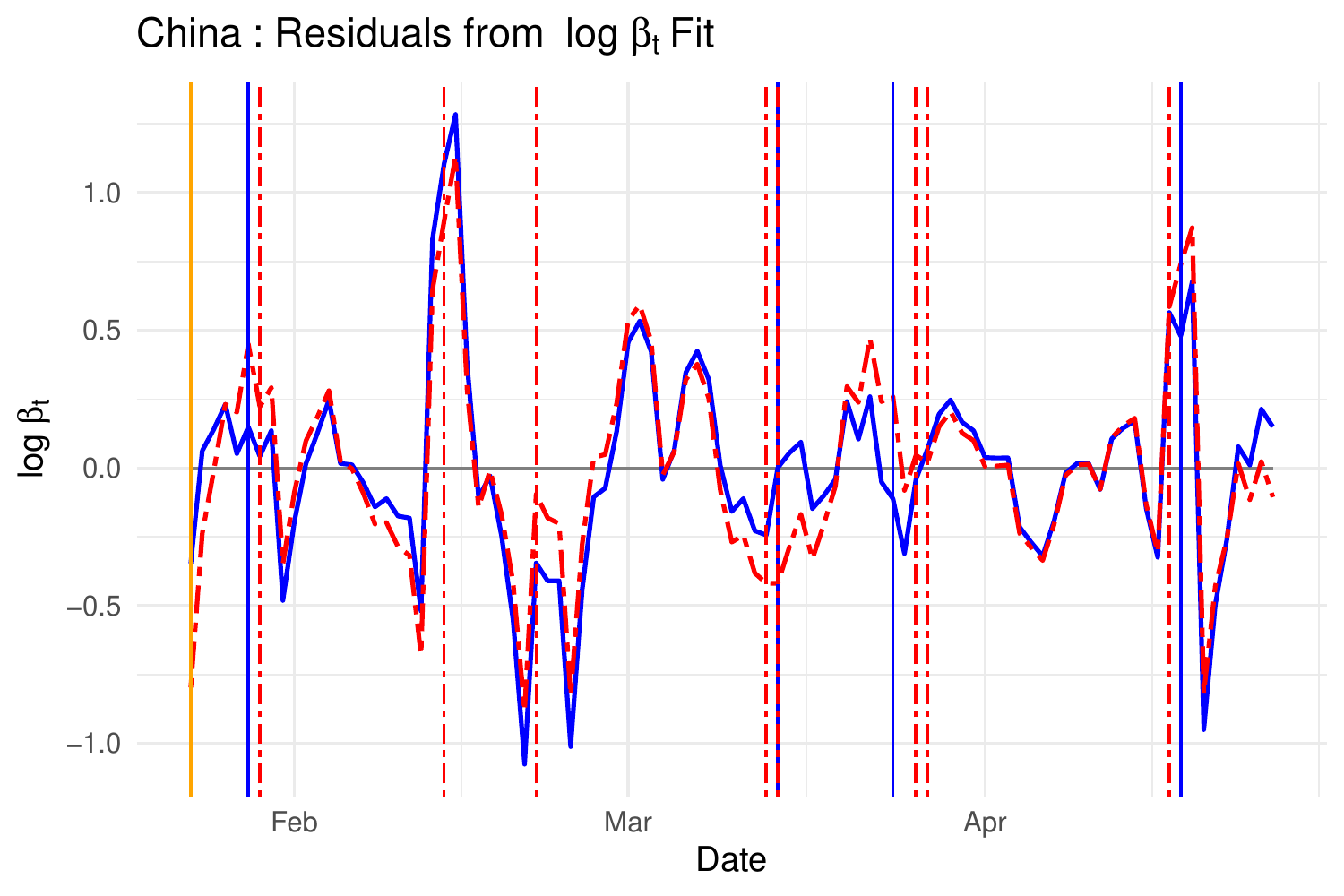}  &
\includegraphics[scale=0.5]{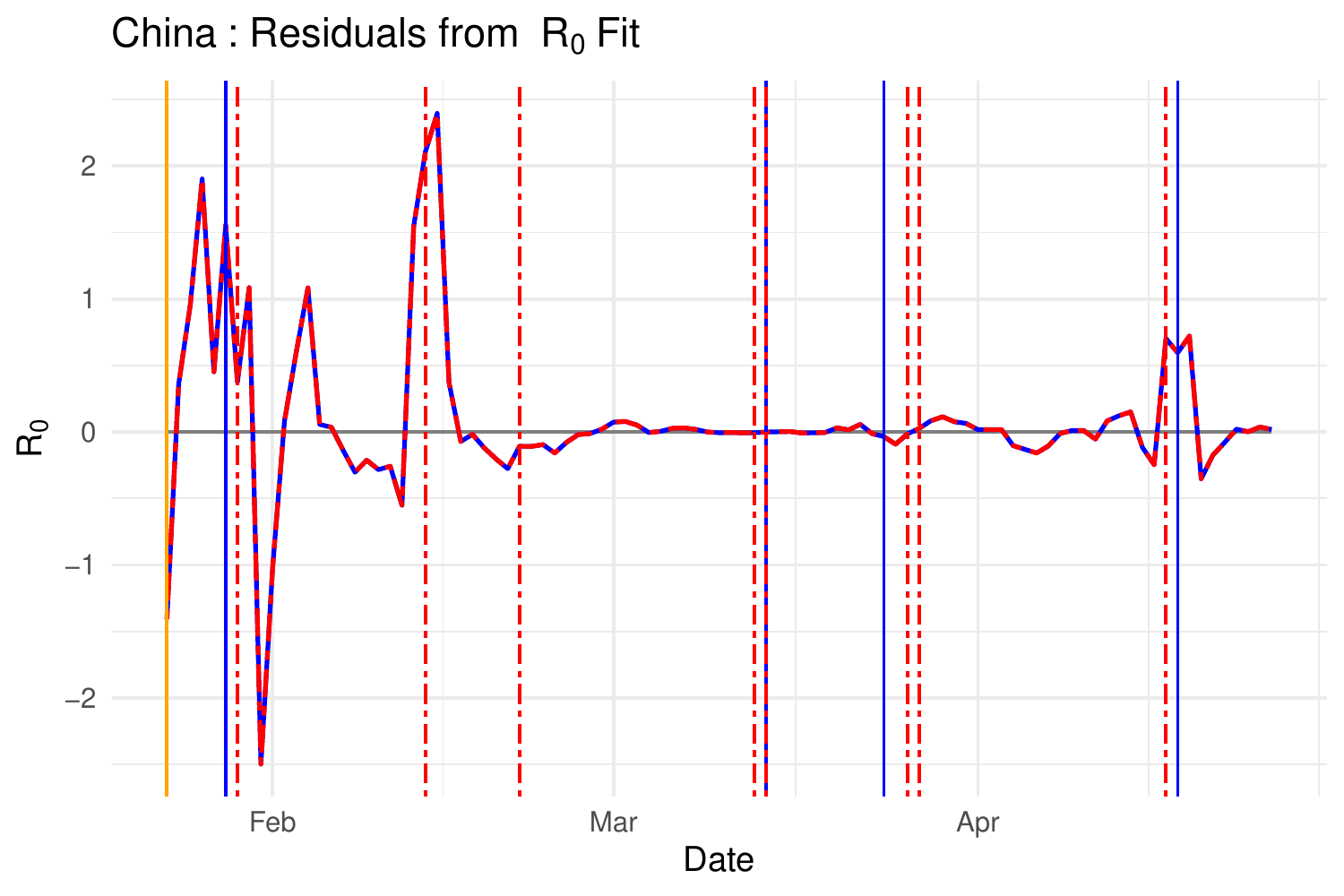}
\end{tabular}
\end{center}
\noindent
\footnotesize{Note: The Sparse HP kinks (blue) are:
January 28, March 14, March 24, and April 18.
The $\ell_1$ kinks (red) are:
January 29, February 14, February 22, March 13, March 14, March 26, March 27, and April 17.
The orange vertical line denotes the lockdown date, January 23.}
\end{figure}

Figure~\ref{fig:China-filter} shows the results for China. Since the pandemic is almost over in China, we use the data censored on April 26th when the 3-day-average of newly confirmed cases is less than 10.
The estimated kink dates are:
 January 28, March 14, March 24, and April 18.
 Based on them, we can classify observations into five periods:

\begin{enumerate}
\item January 23 - January 28: This is an initial period of the epidemic in China. Since the official confirmation of the novel coronavirus on December 31, 2019, the confirmed cases had increased rapidly. President Xi presided and issued instructions on the epidemic control on January 20.
 The travel ban on Wuhan was imposed on January 23, 2020 in the period of the Lunar New Year holidays \citep{NYtimes:timeline}. We set this date as the lockdown date.

\item January 28 - March 14: The contact rate shows a sharp decrease during this period.     The Lunar New Year holiday was  extended to February 2 across the country. China's National Health Commission (NHC)  imposed social distancing measures on January 26.  By January 29, all 31 provinces in China upgraded the public health emergency response to the most serious level.  By early February, nationwide strict social distancing policies were in place.



\item March 14 - March 24: This period shows a V-turn of the contact rate in terms of the $\log \beta_t$ scale. It also shows an upward trending in the $R_0(t)$ scale but the level is lower than that in early February. The mass quarantine of Wuhan was partially lifted on March 19 \citep{Bloomberg:ChinaToLift}.  Most provinces    downgraded their public health emergency response level, where factories and stores started to reopen in this period.

\item March 24 - April 18:  The contact rate still increased but at a lower rate. It started to decrease again at the end of this period. We can see a slight increase in $R_0(t)$. The mass quarantine of Wuhan was lifted more and the travel to other provinces was allowed on April 8 \citep{Bloomberg:ChinaToLift}.

\item April 18 - April 26: The contact rate went down quickly and was flattened at a low level. The last hospitalized Covid-19 patient in Wuhan was discharged on April 26 \citep{XinhuaFight}.
\end{enumerate}

\begin{figure}[htbp]
\begin{center}
\caption{Sparse HP and $\ell_1$ Filtering for South Korea}\label{fig:Korea-filter}
\vskip10pt
\begin{tabular}{cc}
\includegraphics[scale=0.5]{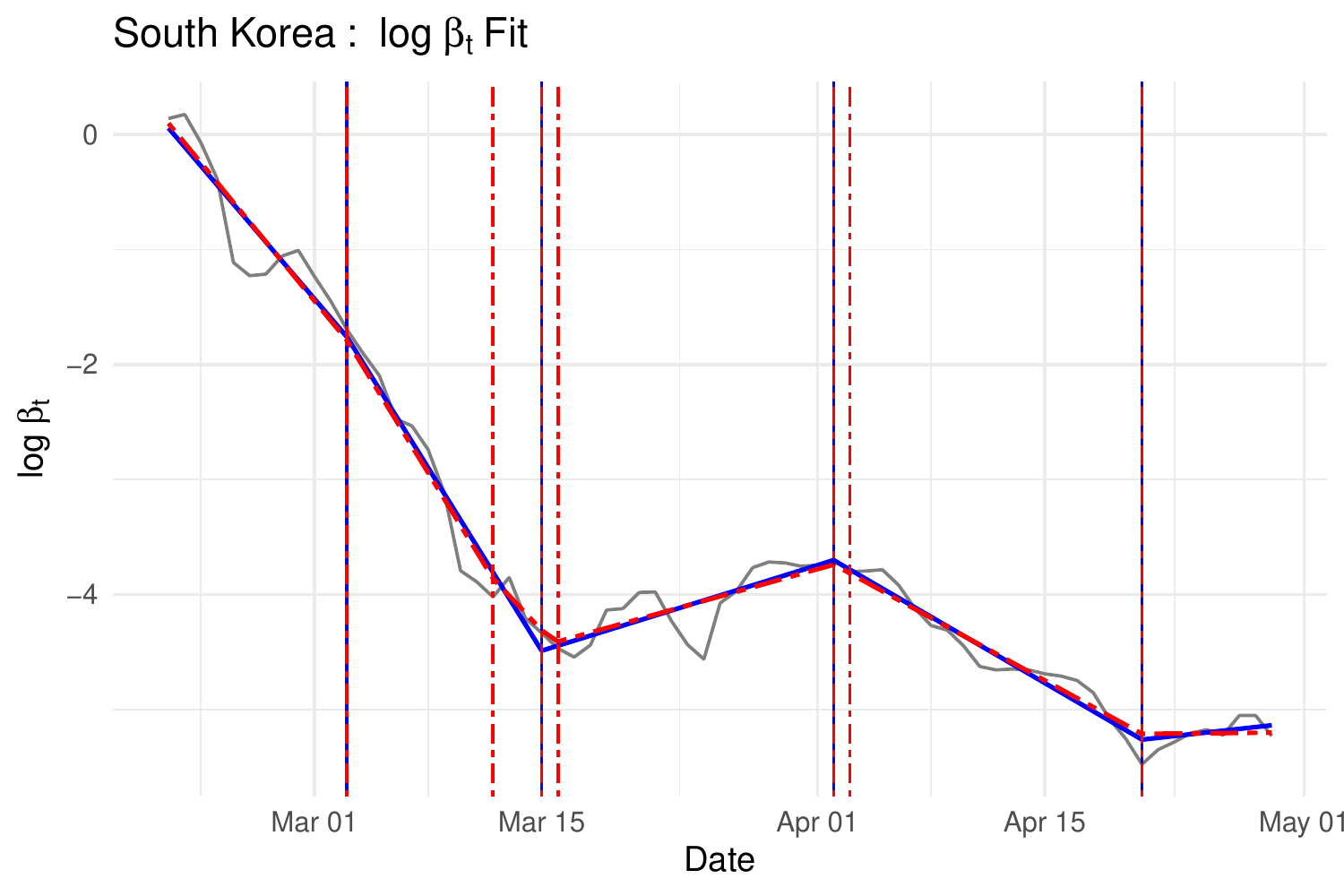} &
\includegraphics[scale=0.5]{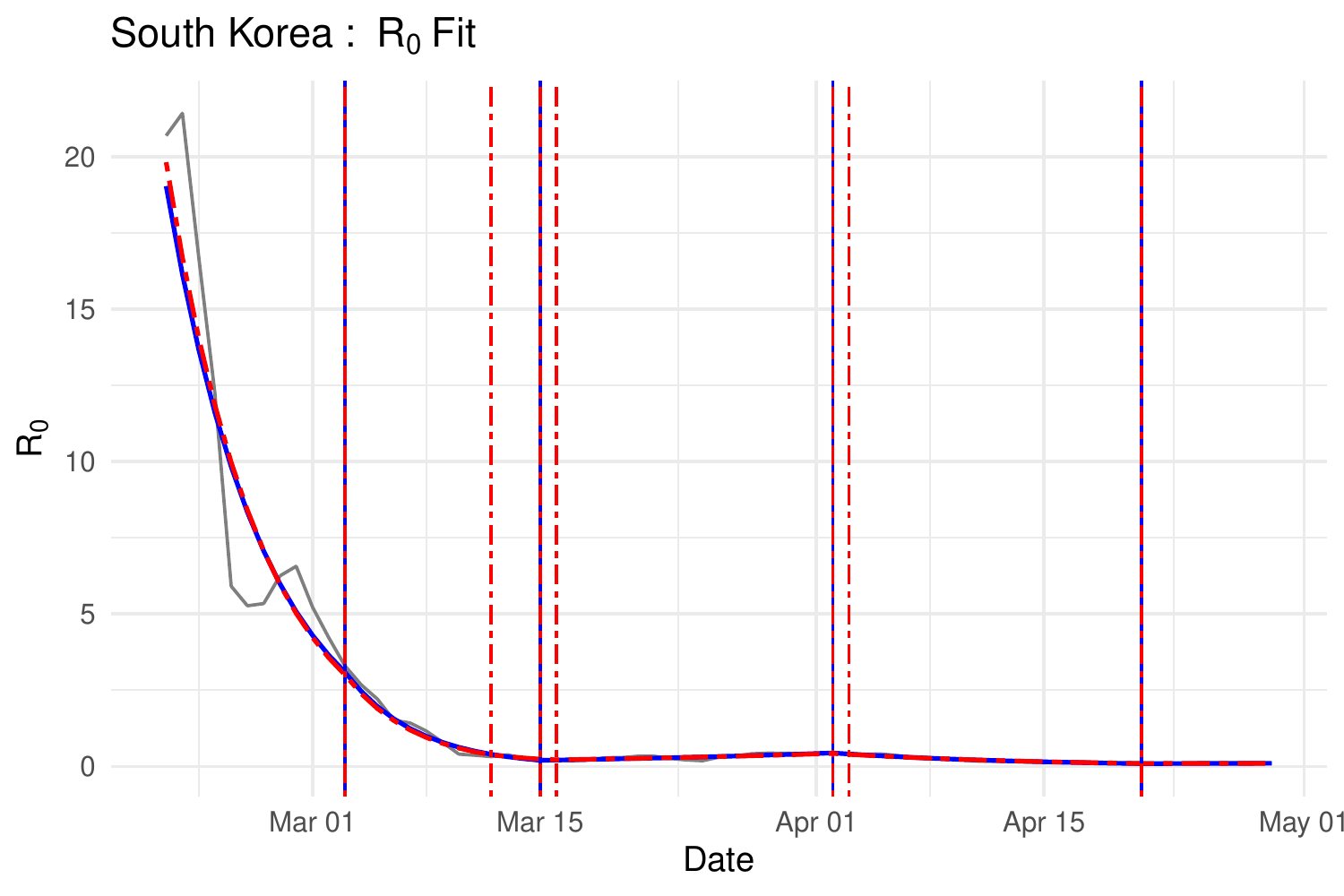} \\
\includegraphics[scale=0.5]{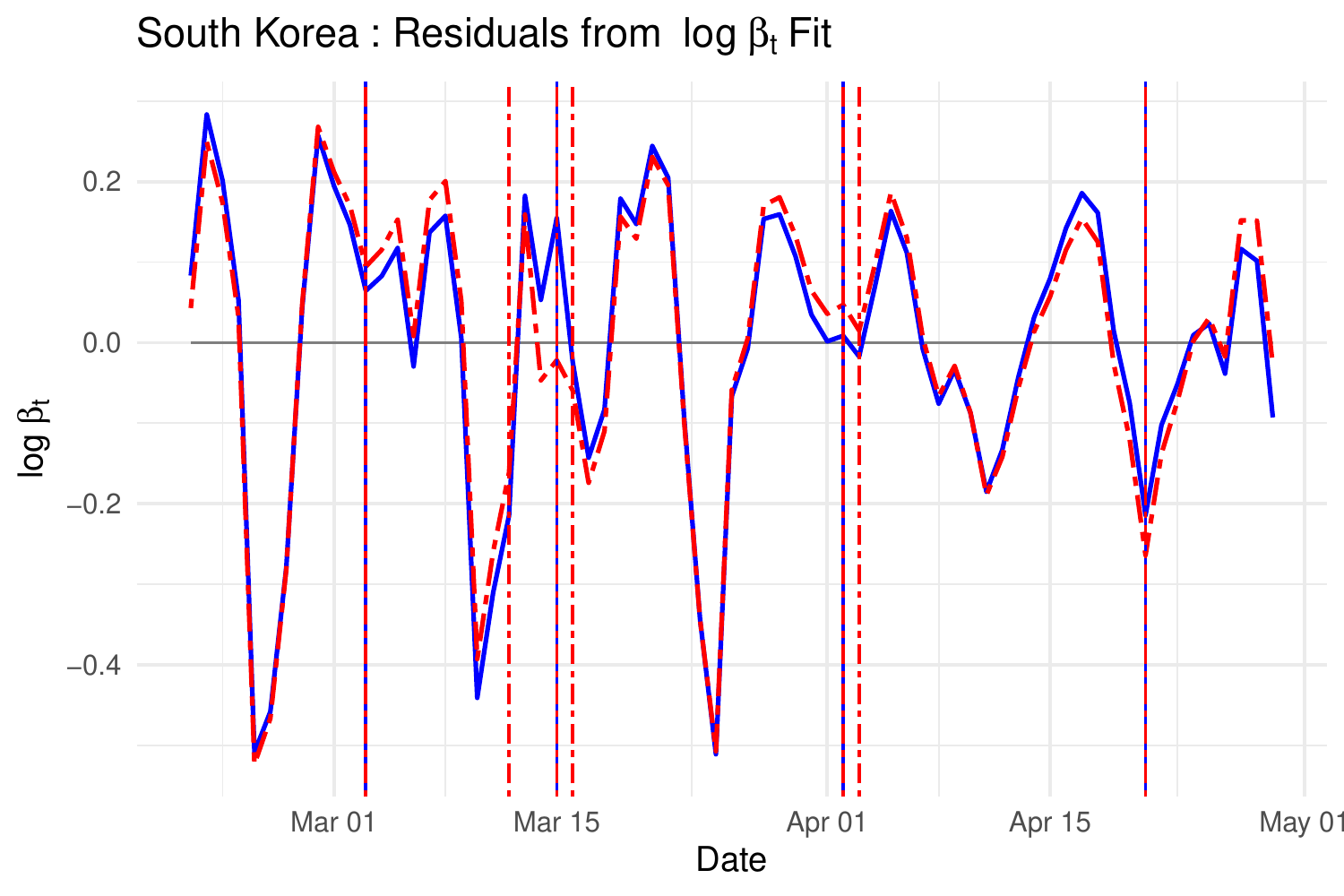}  &
\includegraphics[scale=0.5]{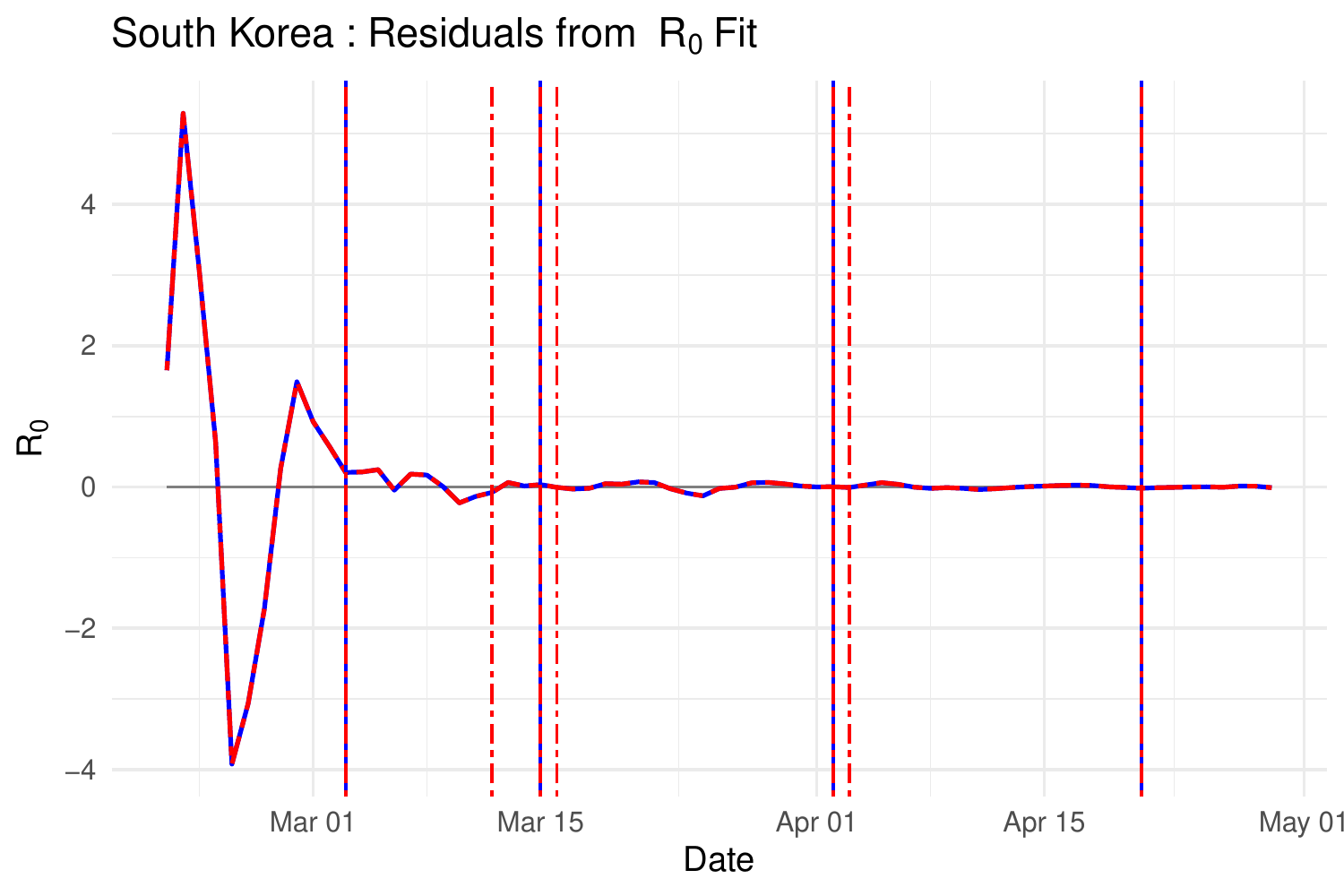}
\end{tabular}
\end{center}
\noindent
\footnotesize{Note: The Sparse HP kinks (blue) are:
March 3, March 15, April 2, and April 21.
The $\ell_1$ kinks (red) are:
March 3, March 12, March 15, March 16, April 2, April 3, and April 21.
South Korea have not imposed any nation-wide lockdown measure.}
\end{figure}

Figure~\ref{fig:Korea-filter} shows the results for South Korea. For the same reason in China, we use the data censored on April 29.
The estimated kink dates are:
March 3, March 15, April 2, and April 21. Based on them, we can classify observations into five periods:
\begin{enumerate}
\item February 21 - March 3: This period is the beginning of the coronavirus spread in South Korea.
On February 21,
Shincheonji Church of Jesus, a secretive church in South Korea was linked to a surge of infections in the country
\citep{NYtimes:timeline}.
The sharp decline of $\log \beta_t$ could be due to the fact that the number of active infections is relatively small in this period
and thus, $Y_t = \Delta C_t/ (I_{t-1} S_{t-1})$ might not be properly measured.

\item March 3 - March 15: A sharp decrease in $\log \beta_t$ in this period  corresponds to
Korean government's swift reactions to the outbreak through active  testing and contact tracing
\citep{NYtimes:Korea,ALS:NBER, kim2020estimating}, highlighted by prompt containment of
an outbreak started on March 8 at a call center in Seoul \citep{Korea:callcenter}.

\item March 15 - April 2: This period shows a modest V-turn of the contact rate in terms of the $\log \beta_t$ scale but it is
much less visible in the $R_0(t)$ scale.

\item April 2 - April 21: This period displays a further reduction of the contact rate. A remarkable event was
parliamentary elections on April 15 when 30 million people voted without triggering a new outbreak.

\item April 21 - April 29: The contact rate was flattened at a low level.
\end{enumerate}

\begin{figure}[htbp]
\begin{center}
\caption{Sparse HP and $\ell_1$ Filtering for the UK}\label{fig:UK-filter}
\vskip10pt
\begin{tabular}{cc}
\includegraphics[scale=0.5]{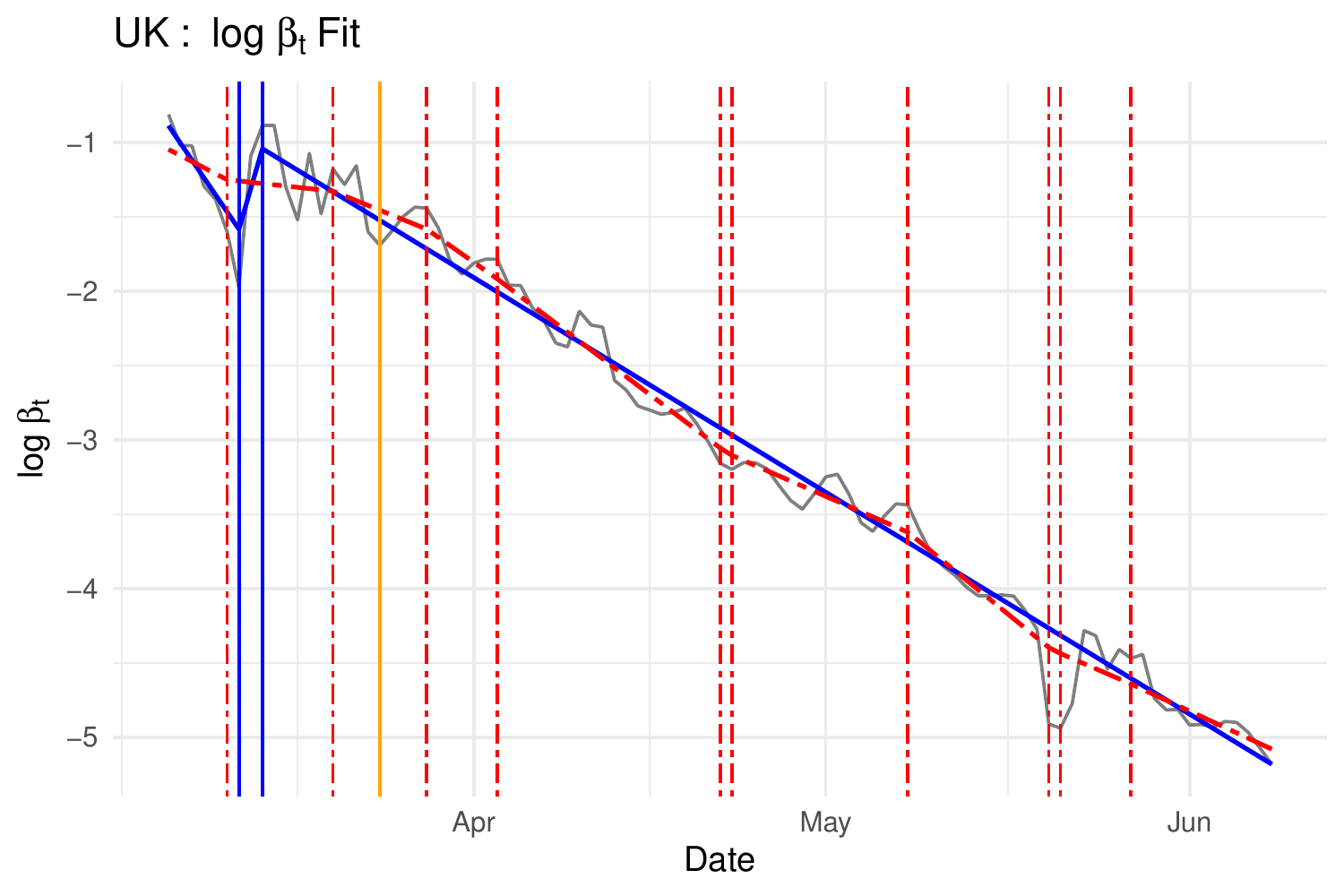} &
\includegraphics[scale=0.5]{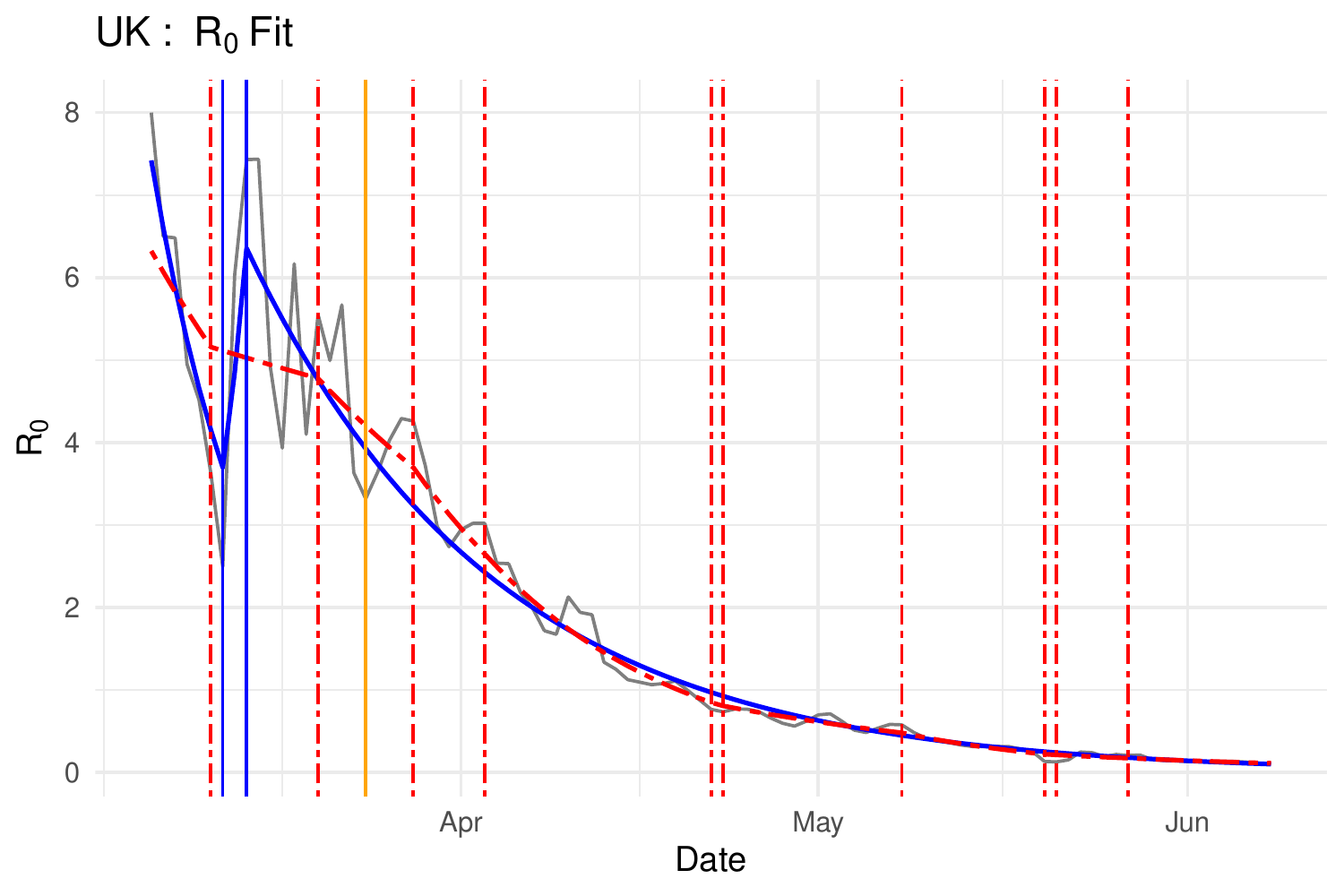} \\
\includegraphics[scale=0.5]{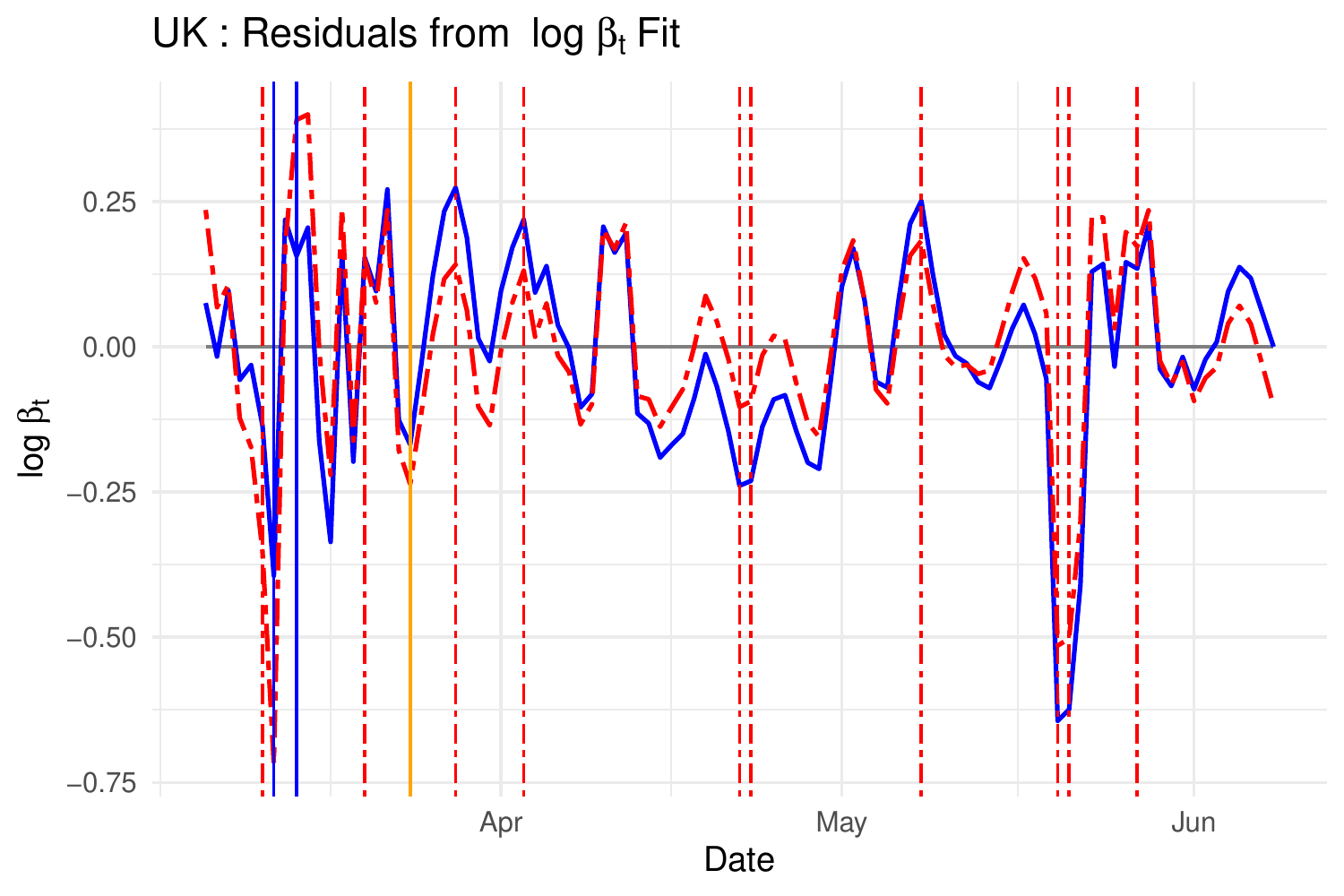}  &
\includegraphics[scale=0.5]{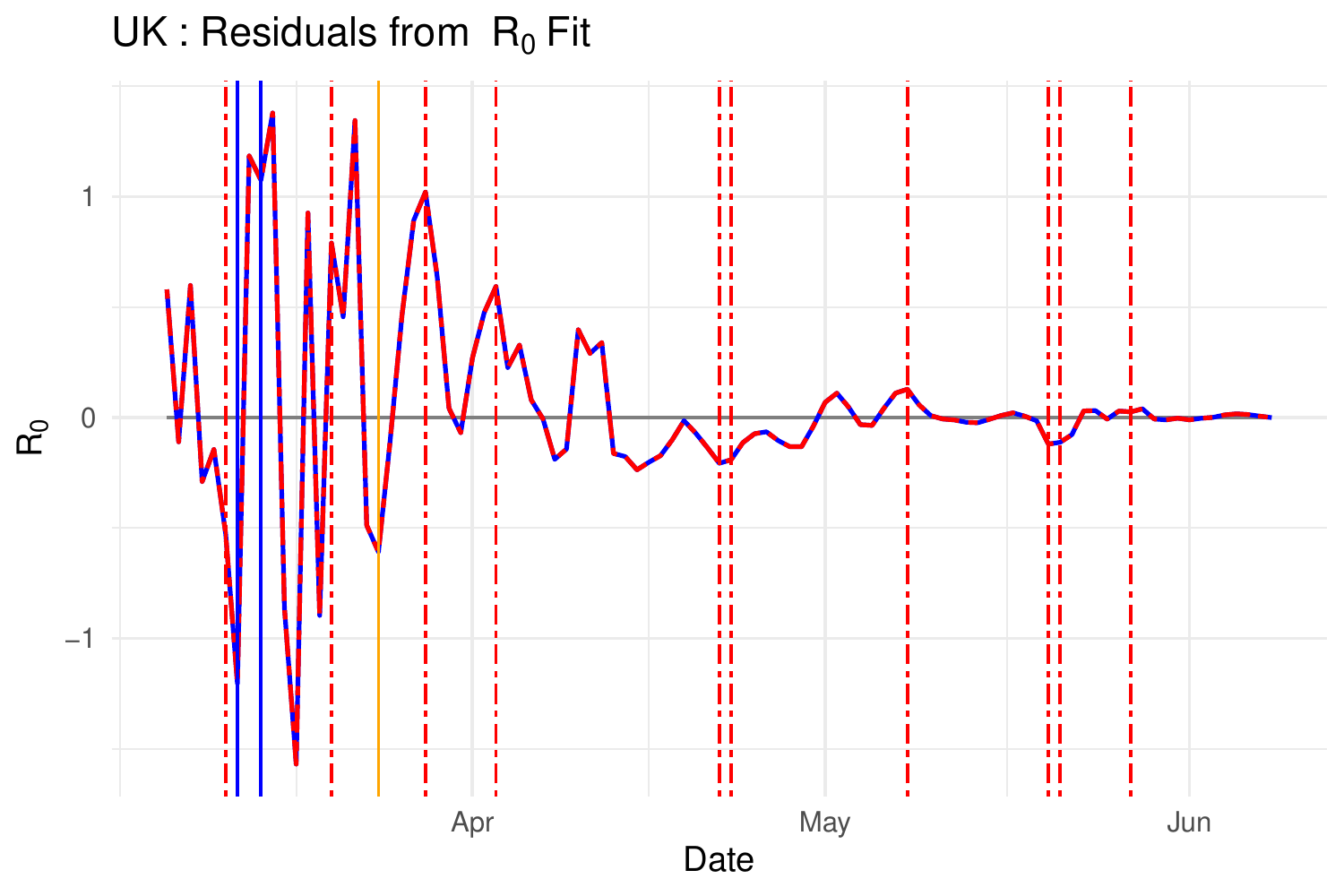}
\end{tabular}
\end{center}
\noindent \footnotesize{Note: The Sparse HP kinks (blue) are:
March 12 and March 14.
The $\ell_1$ kinks (red) are:
March 11, March 20, March 28, April 3, April 22, April 23, May 8, May 20, May 21, and May 27.
The orange vertical line denotes the lockdown date, March 24.}
\end{figure}

Figure~\ref{fig:UK-filter} shows the empirical results of the UK.
The estimated kink dates are:
March 12 and March 14.
Based on them, we can classify observations into three periods:
\begin{enumerate}
\item March 6 - March 12: This is an initial period of the epidemic in the UK. The downward trend might be due to the fact that the cumulative number of confirmed cases is relatively small and therefore, its growth rate can be easily over-estimated.
\item March 12 - March 14: This is still an early stage of the epidemic. The steep increase in the contact rate is again possibly due to the small number of the confirmed cases.
\item March 14 - June 8: This period shows a steady and constant decrease in the contact rate. The lockdown measures began in the UK on March 23 \citep{BBC:lockdown}. On May 10, the British prime minister Boris Johnson relaxed certain restrictions and announced the plan for reopening \citep{BBC:reopening} but it keeps the downward trending.
\end{enumerate}

Overall, the trend of the contact rate is quite similar to those of the US and Canada. The location of the kinks are around more in the initial periods but it shows the steady downward trending after the prime minister's lockdown announcement.
This results in a smooth curve in the $R_0(t)$ scale.
The trend estimates of the $\ell_1$ filter is almost identical to those of the sparse HP filter; however, it indicates 10 kinks, which seem overly excessive.

\subsection{A Measure of Surveillance and Policy Implications}

The sparse HP filter produces the kinks where the slope changes in the $\log \beta_t$ scale,  thereby providing a good surveillance measure for monitoring the ongoing epidemic situation. The policy responses are based on various scenarios and the contact rate is one of the most important measures that determine different developments.
As a summary statistic of the time-varying contact rate, we propose to consider the time-varying growth rate of the contact rate, which we call \emph{contact growth rates}:
\begin{align*}
\xi (t) :=  \frac{\beta_t - \beta_{t-1}}{\beta_{t-1}} \times 100.
\end{align*}
Recall that we have defined the time-varying basic reproduction number by
$R_0(t) = \beta_t/\gamma$.
Because $\gamma$ is fixed over time, we have that
\begin{align*}
\xi (t) =  \frac{R_0(t) - R_0(t-1)}{R_0(t-1)} \times 100.
\end{align*}
Therefore, $\xi(t)$ can be interpreted as the \emph{time-varying growth rate of the basic reproduction number};
it does not require the knowledge of $\gamma$ and solely depends on $\beta_t$.
Furthermore, by simple algebra,
\begin{align}\label{growth_rate:R0}
\xi (t) =  \left[ \exp(\log \bt_t - \log \bt_{t-1}) -1 \right] \times 100,
\end{align}
which implies that $\xi(t)$ will be piecewise constant if $\log \bt_t$ is piecewise linear.
This simple algebraic relationship shows that a change in the slope at the kink
in the $\log \bt_t$ scale
is translated to a break in the time-varying contact growth rates and therefore in growth rates of the time-varying basic reproduction number.
When $\xi (t)$ is a large positive number, that will be a warning signal for the policymakers.  On the contrary, if $\xi(t)$ is a big negative number, that may suggest that  policy measures imposed before are effective to reduce the contagion.

\begin{table}[]
\caption{Time-Varying Contact Growth Rates}
\label{tb:xi_t}
\begin{center}
\begin{tabular}{lccccc}
\hline
 & \multicolumn{1}{c}{US} & \multicolumn{1}{c}{Canada} & \multicolumn{1}{c}{China} & \multicolumn{1}{c}{South Korea} & \multicolumn{1}{c}{UK} \\
\hline
Period 1 & -1.55 & 7.08  & 15.04  & -15.23 & -10.96 \\
Period 2 & 7.48  & -5.02 & -12.27 & -20.34 & 31.10 \\
Period 3 & -7.67 & -2.82 & 30.23  & 4.47   & -4.70 \\
Period 4 & -3.39 &   NA  & 4.41   & -7.88  & NA \\
Period 5 & -1.04 &   NA  & -22.95 & 1.57   & NA \\
\hline
\end{tabular}
\end{center}
\noindent \footnotesize{Note: The growth rates, expressed as percentages, are
 obtained by \eqref{growth_rate:R0} using the sparse HP trend estimates.
 The contact  growth rates are also growth rates of $R_0(t)$.
The kink dates separating distinct periods are different for each country and they are reported in Sections \ref{section:US} and \ref{section:RoW}.
}
\end{table}


Table~\ref{tb:xi_t} reports the time-varying contact growth rates in the five countries that we investigate, using
the sparse HP trend estimates.
For the US, the  explosive growth rate of 7.5\% in the second period is followed by the negative growth rates of $-7.7$\%, $-3.4$\%, and $-1$\%, albeit at diminishing magnitudes.
The  trajectory of Canada is similar to that of the US.
The growth rates of China  fluctuated up and down:  it started with a high positive 15\% followed by $-12$\%; a sharp V-turn at the end of the second period (March 14) with the resulting explosive growth rate of 30\%, followed by moderate 4\% and impressive $-23$\%.
It might be the case that  the up-and-down pattern observed in China is in part due to data quality issues since China was the first country to experience the pandemic.
For South Korea, we can see the stunning drop of the growth rates culminating on March 15 (the end of the second period).
A modest positive growth rate during period 3 is offset by a larger magnitude of negative growth rate in period 4.
The UK has experienced steady---but not spectacular---negative growths over the sample period following a sharp fluctuation in mid-March. This hints the degrees of effectiveness of the UK lockdown policy.
As early pandemic epicenters, China and South Korea experienced V-turns in the time-varying growth rates of basic reproduction number.
 Canada, the UK and the US may face similar trajectories as they reopen their countries.
 Our  surveillance statistic
can be a useful indicator to monitor a new outbreak of COVID-19.
However, it will be mainly useful for a short-term projection of the contact growth rate because  it is not designed to make long-term trend predictions.

\section{Theory}\label{sec:theory}

In this section, we examine theoretical properties of the sparse HP and $\ell_1$ filters in terms of risk consistency.
Let $\| \cdot \|_0$ denote the usual $\ell_0$-(pseudo)norm, that is the number of nonzero elements,
and let $\| \cdot \|_r$ and $\| \cdot \|_{\infty}$, respectively, denote the $\ell_r$ norm for $r=1,2$ and the sup-norm.

\subsection{Risk Consistency of the Sparse HP Filter}


Define
\begin{align}\label{F:def}
\mathcal{F} = \mathcal{F}(\kappa,M) := \left\{ \bm{f}: \|  \bm{D} \bm{f} \|_{0} \leq \kappa,
\|  \bm{D} \bm{f} \|_{\infty} \leq M \right\},
\end{align}
where $M$ is defined in $\eqref{def:M}$.
For each $\bm{f} \in \mathcal{F}$, define
\begin{align*}
S( \bm{f} ) := \mathbb{E}_{\bm{y}} \left[ \frac{1}{T} (\bm{y}- \bm{f})^\top (\bm{y}- \bm{f})  \right].
\end{align*}
Let $\bm{f^*}$ denote the ideal sparse filter in the sense that
\begin{align*}
\bm{f^*} \in \text{argmin}_{\bm{f}  \in \mathcal{F}} S( \bm{f} ).
\end{align*}
Let $\bm{\widehat{f}}$ denote the sparse HP filter defined in Section~\ref{sec:SHP}. Then,
\begin{align}\label{excess-risk}
R(\bm{\widehat{f}}, \bm{f^*}) :=  S(\bm{\widehat{f}}) -  S(\bm{f^*})
\end{align}
is always nonnegative. Following the literature on empirical risk minimization, we bound the excess risk $R$ in \eqref{excess-risk}
and establish conditions under which it converges to zero.

Recall that  the sparse HP filter minimizes
\begin{align*}
Q_n(\bm{f}) := \frac{1}{T} (\bm{y}- \bm{f})^\top (\bm{y}- \bm{f})  +   \frac{\lambda}{T} \bm{f}^\top \bm{D}^\top \bm{D} \bm{f}
\end{align*}
subject to $\bm{f} \in \mathcal{F}$.

Let $S_n( \bm{f} ) := T^{-1} (\bm{y}- \bm{f})^\top (\bm{y}- \bm{f})$.
Write
\begin{align*}
R(\bm{\widehat{f}}, \bm{f^*})
&= S(\bm{\widehat{f}}) - Q_n(\bm{f^*}) + Q_n(\bm{f^*}) -  S(\bm{f^*}) \\
&\leq S(\bm{\widehat{f}}) - Q_n(\bm{\widehat{f}}) + Q_n(\bm{f^*}) -  S(\bm{f^*}) \\
&= S(\bm{\widehat{f}}) - S_n( \bm{\widehat{f}} )
- \frac{\lambda}{T} \bm{\widehat{f}}^\top \bm{D}^\top \bm{D} \bm{\widehat{f}}
 + S_n( \bm{f^*} )
+ \frac{\lambda}{T} {\bm{f^*}}^\top \bm{D}^\top \bm{D} \bm{f^*}
 -  S(\bm{f^*}) \\
&\leq 2 \sup_{\bm{f} \in \mathcal{F}} \left| S_n( \bm{f} )  - S(\bm{f})  \right|
+ 2 \frac{\lambda}{T} \sup_{\bm{f} \in \mathcal{F}} {\bm{f}}^\top \bm{D}^\top \bm{D} \bm{f}.
\end{align*}
Therefore, it suffices to bound two terms above.
For the second term, we can use \eqref{def:M} and \eqref{F:def} to bound
\begin{align*}
2 \frac{\lambda}{T} \sup_{\bm{f} \in \mathcal{F}} {\bm{f}}^\top \bm{D}^\top \bm{D} \bm{f}
&\leq \frac{2 \lambda M^2\kappa}{T}.
\end{align*}
We summarize discussions above in the following lemma.

\begin{lem}\label{lem:consistency}
	Let $\bm{\widehat{f}}$ denote the sparse HP filter. Then,
\begin{align*}
R(\bm{\widehat{f}}, \bm{f^*}) \leq 2 \sup_{\bm{f} \in \mathcal{F}} \left| S_n( \bm{f} )  - S(\bm{f})  \right| + \frac{2  \lambda\kappa}{T} \max_{t=2,\ldots,T-1} |y_{t-1} - 2 y_t + y_{t+1}|^2.
\end{align*}
\end{lem}

To derive an asymptotic result, we introduce subscripts indexed by the sample size $ T $, when necessary for clarification.
Let $ \mathcal{G}_{\kappa}  $ denote the set of every continuous and piecewise linear function whose slopes and the function itself  is bounded by $ C_1 $  and $ C_2 $, respectively, and the number of kinks is bounded by $ \kappa $.
\begin{assum}\label{key:regularity}
Assume that	$\mathcal{F}$ in \eqref{F:def} satisfies
	\begin{align}
	\mathcal{F}(\kappa,M) \subseteq \mathcal{F}_T  := \left\{ \bm{f_{T}}=(f_{T,1},...,f_{T,T}):
	f_{T,t} = f(t/T), f \in \mathcal{G}_{\kappa} \right\}.
	\end{align}
Moreover,  $ y_t = f_{T,t}^* + u_t $, where $\log \beta_t = f_{T,t}^*$, $ \bm{f^*_{T}} \in \mathcal{F}_T  $ and  $ u_t $ satisfies $ \sup_{t=1,2,\ldots} \mathbb{E} |u_t|^p <\infty $ for some $ p \geq 2 $ and Assumption \ref{assum:beta_t}.
Finally, $\lambda \kappa T^{-(1-1/p)} \rightarrow 0$ as $T \rightarrow \infty$.
\end{assum}

Then, we have the following proposition.

\begin{pro}\label{pro:consistency}
Let Assumption \ref{key:regularity} hold. Then, we have that
 as $T \rightarrow \infty$,
	\begin{align}\label{ULLN}
	\sup_{\bm{f} \in \mathcal{F}} \left| \frac{1}{T} \sum_{t=1}^T \left\{ (y_t - f_t)^2 - \mathbb{E} (y_t - f_t)^2 \right\} \right| \rightarrow_p 0
	\end{align}
	and
	\begin{align}\label{lambda:condition}
	\frac{\lambda\kappa}{T} \max_{t=2,\ldots,T-1} |y_{t-1} - 2 y_t + y_{t+1}|^2 \rightarrow_p 0.
	\end{align}
	Therefore, $R(\bm{\widehat{f}}, \bm{f^*}) \rightarrow_P 0$.
\end{pro}

Theorem~\ref{pro:consistency} establishes the consistency in terms of the excess risk $R$.
Assumption \ref{key:regularity} provides sufficient conditions for \eqref{ULLN} and \eqref{lambda:condition}.
Condition \eqref{ULLN} is a uniform law of large numbers for the class $\mathcal{F}$
and condition \eqref{lambda:condition} imposes a weak condition on $\lambda$.
Proposition~\ref{pro:consistency} follows immediately from Lemma~\ref{lem:consistency} once \eqref{ULLN} and \eqref{lambda:condition} are established.

\bigskip

\begin{proof}[Proof of Proposition \ref{pro:consistency}]
	Note that the summand in \eqref{ULLN} can be rewritten $
	(y_t^2 - \mathbb{E} y_t^2 ) -2 f_t(y_t - \mathbb{E}y_t) $.
	Then,
	$ y_t^2 - \mathbb{E} y_t^2  = u_t^2 - \sigma_u^2 + 2(f_{T,t}^* -f_t )u_t $ and $ 2 f_t(y_t - \mathbb{E}y_t) = 2 f_t u_t $.
	Furthermore,  $ T^{-1} \sum_{t=1}^T u_t^2- \sigma_u^2 =o_p(1) $ due to the law of large numbers (LLN) for a martingale difference sequence (mds).

	We now turn to $  \sup_{\bm{f} \in \mathcal{F}}\left(T^{-1} \sum_{t=1}^T f_t u_t\right)  $. The marginal convergence is straightforward since $ f_t u_t $ is an mds with bounded second moments due to the LLN for mds. Next, note that for a constant $\eta>0$
	\[
	\sup_{ |\bm{f}-\bm{f'}|_{\infty} < \eta} \left|T^{-1} \sum_{t=1}^T (f_t - f_t') u_t\right| \leq \eta \left(T^{-1} \sum_{t=1}^T |u_t| \right),   \]
	which implies the stochastic equicontinuity of the process indexed by $ \bm{f} \in \mathcal{F}_T$. Finally, recall Arzel\`{a}-Ascolli theorem, see e.g. \cite{van1996weak}, to conclude that $ \mathcal{G}_\kappa $ is totally bounded with respect to $ |\cdot|_\infty $.   Therefore,
	$$  \sup_{\bm{f} \in \mathcal{F}_T}\left(T^{-1} \sum_{t=1}^T f_t u_t\right) = o_p(1) $$
	by a generic uniform convergence theorem, e.g. \cite{andrews1992generic}.

	To show the condition \eqref{lambda:condition}, note that it is bounded by $ 16 	\frac{\lambda\kappa}{T}\max_{t=1,\ldots,T} y_t^2 $, which is in turn $ O_p(\lambda \kappa T^{-(1-1/p)}) $ due to the moment condition on $ u_t $.
\end{proof}

\subsection{Risk Consistency of the $\ell_1$ Filter}

The $\ell_1$ trend filtering (\ref{L1filter}) can be expressed as
 $$
\widetilde{\bm{f}}  :=\arg\min_{\bm{f}\in\mathbb R^T} \|\bm{y}-\bm{f}\|_2^2+\lambda \|\bm{D}\bm{f}\|_1.
 $$
 We now derive the deviation bound for $\|\widetilde{\bm{f}} -\bm{f}^*\|_2.$  First, the problem is equivalent to a regular LASSO problem as stated in Lemma~\ref{leblassm} below.

 Write  $ \bm{D}= ( \bm{D}_3 , \bm{D}_2) $  where $\bm{D}_2$  has two columns.  Additionally, write
   $$
\bm{G}_2:=  \begin{pmatrix}
 \bm{D}_3^{-1}   \\
\mathbf{0}
 \end{pmatrix},\quad \bm{g}_1:=  \begin{pmatrix}
 -\bm{D}_3^{-1}\bm{D}_2\\
  \bm{I}_2
 \end{pmatrix},
 $$
where $\bm{0}$ is $2\times (T-2)$,    $\bm{g}_1$ is $T\times 2$ and $\bm{G}_2$ is $T\times (T-2).$  Let
 $
\bm{P}_{\bm{g}_1} =  \bm{g}_1(\bm{g}_1^\top \bm{g}_1)^{-1}\bm{g}_1^\top.
 $

 \begin{lem}\label{leblassm}
We have  $ \widetilde{\bm{f}}=  \bm{y}-\widetilde{\bm{y}}+\widetilde{\bm{X}}\widehat\theta$, where $\widetilde{\bm{y}}:=(\bm{I}-\bm{P}_{\bm{g}_1})\bm{y}$, $\widetilde{\bm{X}}:=(\bm{I}-\bm{P}_{\bm{g}_1}) \bm{G}_2$ and
 $$
\widehat\theta:= \arg\min_{\theta}  \| \widetilde{\bm{y}}-   \widetilde{\bm{X}}\theta\|_2^2+ \lambda  \|  \theta\|_1.
 $$
 \end{lem}

\begin{proof}[Proof of Lemma~\ref{leblassm}]
Let $ \bm{D}_1 = (\mathbf{0} : \bm{I}_2)$ be a $2\times T$ matrix, so that
 $$
 \bar{\bm{D}} :=\begin{pmatrix}
 \bm{D} \\
 \bm{D}_1
 \end{pmatrix}
 $$
 is upper triangular and invertible.    Then,
  $
 \bm{G}:= \bar{\bm{D}}^{-1} = ( \bm{G}_2,\bm{g}_1).
 $
Then for a generic $\bm{f}\in\mathbb R^T$, we can define
    $$
\bm{ \alpha}:=\bar{\bm{D}} \bm{f}=\begin{pmatrix}
 \bm{D} \bm{f}  \\
 \bm{D}_1 \bm{f}
 \end{pmatrix}:=\begin{pmatrix}
 \theta\\
\bm{a}
 \end{pmatrix}.
 $$
   So $(\theta, \bm{a})$ also depend on $\bm{f}$ and
$
 \bm{G} \bm{\alpha}=    \bm{g}_1 \bm{a}+ \bm{G}_2\theta.
 $
 Then the problem is equivalent to:  $\widetilde{\bm{f}} =  \bm{g}_1 \widehat{\bm{a}} + \bm{G}_2\widehat \theta$, where
 $$
(\widehat{\bm{a}}, \widehat\theta):= \min_{\bm{a},\theta}  \|\bm{y}-  (\bm{g}_1\bm{a}+\bm{G}_2\theta)\|_2^2+ \lambda  \|  \theta\|_1.
 $$
 To solve the problem,  we concentrate out $\bm{a}$:  Given $\theta$, the optimal $\bm{a}$ is
 $
 (\bm{g}_1^\top \bm{g}_1)^{-1} \bm{g}_1^\top  (\bm{y}-\bm{G}_2\theta)
 $
 and the optimal $\bm{g}_1 \bm{a}$ is $\bm{P}_{\bm{g}_1} (\bm{y}- \bm{G}_2\theta)$. Substituting, so the problem becomes a regular LASSO problem:
 $$
 \min_{\theta}  \| \widetilde{\bm{y}}-   \widetilde{\bm{X}}\theta\|_2^2+ \lambda  \|  \theta\|_1.
 $$
 Finally, $\widetilde{\bm{f}} = \bm{P}_{\bm{g}_1} (\bm{y}- \bm{G}_2\widehat\theta)+ \bm{G}_2\widehat\theta= \bm{y}-\widetilde{\bm{y}}+\widetilde{\bm{X}}\widehat\theta.
$
 \end{proof}

 Next,   let  $J$ denote the indices of $t$ so that $f_{0,t-1}-f_{0,t}\neq f_{0,t}-f_{0,t+1}$ when $t\in J$;  let  $J^c$ denote the indices of $t$ so that $f_{0,t-1}-f_{0,t}= f_{0,t}-f_{0,t+1}$ when $t\in J$. Here, $\{ f_{0,t}: t=1,
 \ldots,T \}$ denote the true elements of $\bm{f}$.
 For a generic vector $\theta\in\mathbb R^{T-2}$, let $\theta_J$ and $\theta_{J^c}$ respectively be its subvectors whose elements are in $J$ and $J^c.$  No we define the restricted eigenvalue constant
 $$
 \zeta:=\inf_{\|\theta_{J^c}\|_1\leq 9\|\theta_J\|_1} \frac{ \|\frac{1}{\sqrt{T}}   \widetilde{\bm{X}}  \theta\|_2^2 }{\|\theta\|_2^2}.
 $$

 \begin{pro}\label{pro:l1:consistency}
   Let $\bm{f}^*$ denote  the true value of $\bm{f}$ and $\bm{u} := \bm{y} - \bm{f}^*$.
Suppose the event  $2.5 \|  {\bm{u}}^\top \widetilde{\bm{X}}\|_{\infty}<\lambda$ holds.  Then on this event
 \begin{align}\label{consistency:ineq}
R(\widetilde{\bm{f}},\bm{f}^*) &\leq \frac{2}{T}\bm{u}^\top \bm{P}_{\bm{g}_1}\bm{u}
+ 2\| \frac{1}{T}  \widetilde{\bm{X}}^\top  \widetilde{\bm{X}}\|_\infty \left( \frac{18\lambda}{\zeta T} \|J\|_0 \right)^2.
 \end{align}
\end{pro}

\begin{proof}[Proof of Proposition~\ref{pro:l1:consistency}]
Let  $\theta^*=\bm{D} \bm{f}^*$.  Consider the vector form of the model $\bm{y}=\bm{f}^*+\bm{u}$.    Then
  $\widetilde{\bm{y}}= \widetilde{\bm{X}}\theta^*+\widetilde{\bm{u}}$ where $\widetilde{\bm{u}}=
  (\bm{I}-\bm{P}_{\bm{g}_1})\bm{u}$. By Lemma  \ref{leblassm},
$ \widetilde{\bm{f}}=  \bm{y}-\widetilde{\bm{y}}+\widetilde{\bm{X}}\widehat\theta$, where  $$
\widehat\theta:= \arg\min_{\theta}  \| \widetilde{\bm{y}}-   \widetilde{\bm{X}}\theta\|_2^2+ \lambda  \|  \theta\|_1.
 $$
The standard  argument for the LASSO deviation bound implies,
on the event $2.5 \|  {\bm{u}}' \widetilde{\bm{X}}\|_{\infty}<\lambda$,
  $$
\|  \widehat\theta-\theta^*\|_1\leq \frac{18\lambda}{\zeta T} \|J\|_0 .
  $$
  Finally,
 $
 \widetilde{\bm{f}}-\bm{f}^*=      \bm{P}_{\bm{g}_1}\bm{u}+ \widetilde{\bm{X}}(  \widehat\theta-\theta^*)
 $ implies
 $$
R(\widetilde{\bm{f}},\bm{f}^*) = \frac{1}{T} \|\widetilde{\bm{f}}-\bm{f}^*\|_2^2
 \leq  \frac{2}{T}\bm{u}^\top \bm{P}_{\bm{g}_1}\bm{u} +2 \| \frac{1}{T}  \widetilde{\bm{X}}^\top  \widetilde{\bm{X}}\|_\infty \|  \widehat\theta-\theta^*\|_1^2.
 $$
\end{proof}

To achieve risk consistency,
$\lambda$ has to be chosen to make the second term on the right-hand side of \eqref{consistency:ineq} asymptotically small
and to ensure that the event  $2.5 \|  {\bm{u}}^\top \widetilde{\bm{X}}\|_{\infty}<\lambda$ holds with high probability.
The first term on the right-hand side of \eqref{consistency:ineq} will converge to zero under mild conditions on $\bm{u}$.
 It is reassuring that  the $\ell_1$ trend filter fits COVID-19 data  well in our empirical results.

\subsection{Risk Consistency of $\exp(\widehat{f}_t)$ and $\exp(\widetilde{f}_t)$}

In this subsection, we obtain risk consistency of  $\exp(\widehat{f}_t)$ and $\exp(\widetilde{f}_t)$. 
To do so, we first rewrite the excess risk in \eqref{excess-risk} as
\[
R(\bm{f}, \bm{f^*})=\frac{1}{T}\sum_{t=1}^T\mathbb E (f_t-f_t^*)^2,
\]
for $\bm{f} = ( f_1,..., f_T)^\top$
and 
$\bm{f^*}=(f_1^*,...,f_T^*)^\top$.
We have proved in the previous sections that 
$R(\bm{\widehat{f}}, \bm{f^*}) \rightarrow_P 0$
and
$R(\bm{\widetilde{f}}, \bm{f^*}) \rightarrow_P 0$.
Then, under the assumption that there exists a constant $C < \infty$ such that $\max_t|f_t|+\max_t|f_t^*|<C$ for all $\{f_t\}$ on the parameter space, we get, uniformly for all $f$ on the parameter space,
\[
\frac{1}{T}\sum_{t=1}^T\mathbb E (\exp(f_t)-\exp(f_t^*))^2
= \frac{1}{T}\sum_{t=1}^T\mathbb E \exp(2\tilde f_t)(f_t-f_t^*)^2
\leq \exp(2C) \frac{1}{T}\sum_{t=1}^T\mathbb E (f_t-f_t^*)^2,
\]
where in the first equality we used the mean value theorem for some $\tilde f_t\in(f_t, f_t^*).$
Therefore, 
$$
R(\exp(\bm{\widehat{f}}), \exp({\bm{f^*}}))  \leq \exp(2C) R(\bm{\widehat{f}}, \bm{f^*}) =o_P(1)
$$
and the analogous result folds for $\exp(\bm{\widetilde{f}})$.

\section{Conclusions}\label{sec:conclusions}

We have developed a novel method to estimate the time-varying COVID-19 contact rate using data on actively infected, recovered  and deceased cases. Our preferred method called the sparse HP filter has produced the kinks that are well aligned with actual events in each of five countries we have examined.
We have also proposed contact growth rates to document and monitor outbreaks.
Theoretically, we have outlined the basic properties of the sparse HP and $\ell_1$ filters in terms of risk consistency.
The next step might be to establish a theoretical result that may distinguish between the two methods by looking at
kink selection consistency.
It would also be important to develop a test for presence of kinks as well as an inference method on the location and magnitude of kinks and on contact growth rates.
 In the context of the nonparametric kernel regression of the trend function, \citet{Delgado2000} explored the distribution theory for the jump estimates but did not offer testing for the presence of a jump. Compared to the kernel smoothing approach,  it is easier to determine the number of kinks using our approach, as we have demonstrated. Furthermore, the linear trend specification is more suitable for forecasting immediate future outcomes, at least until the next kink arises. The long-term prediction is more challenging and it is beyond the scope of this paper.
Finally, it would be useful to develop a panel regression model for the contact rate  at the level of city, state or country.
These are interesting research topics for future research.


\appendix

\section*{Appendices}

\section{Under-Reporting of Positive Cases}\label{sec:under}

In Section~\ref{sec:model}, it is assumed that we observe $(C_t, R_t, D_t)$.
In this appendix, we show that our time series model in Section~\ref{sec:model} is robust to some degree of under-reporting of
positive cases.

Assume  that what we observe is only a fraction of changes in $C_t$.
This assumption reflects  the reality that a daily reported number of newly positive cases of COVID-19 is likely to be underreported.
Suppose that
we observe $\Delta c_t$ in period $t$ such that
$
\Delta c_t := \rho \Delta C_t,
$
where $0 < \rho < 1$ is unknown.
Then,
$$
 c_t = \sum_{t=1}^T \Delta c_t = \rho \sum_{t=1}^T \Delta C_t = \rho C_t,
 $$
 assuming that $c_0 = C_0 = 0$.
 In words, $\rho$ is the constant ratio between reported and true cases.
 Formally, we make the following assumption.

 \begin{assum}[Fraction Reporting]\label{assum:rho_t}
 For each $t$, we observe $(c_t, r_t, d_t)$ such that
 \begin{align*}
c_t := \rho C_t, \ \ r_t := \rho R_t \ \
\ \ \text{ and } \ \
d_t := \rho D_t,
\end{align*}
where $0 < \rho < 1$.
 \end{assum}

The two simplifying conditions in Assumption~\ref{assum:rho_t} is that (i) $\rho$ is identical among the three time series
and (ii) $\rho$ is constant over time.
In reality, a fraction of reported deaths might be higher than that of reported cases; $\rho$ might be time-varying especially in the beginning of the pandemic due to capacity constraints in testing.
However,  we believe that $\rho$ is unlikely to vary over time as much as $\beta_t$ changes over time; thus,
we take a simple approach to minimize complexity. The common $\rho$ can be thought of a broad measure of detecting COVID-19 in a community.

 Define $i_t := c_t - r_t - d_t$ and $s_t := 1 - c_t$.
Under Assumption~\ref{assum:rho_t}, the reported fraction infected at time $t$ ($i_t$) is underestimated, but
the reported fraction of the proportion that is susceptible at time $t$ ($s_t$) is overestimated.
Note that
\begin{align*}
g_t := \frac{\Delta c_{t}}{i_{t-1}}
= \frac{\rho \Delta C_{t}}{\rho I_{t-1}} = \frac{\Delta C_{t}}{I_{t-1}}.
\end{align*}
However,
\begin{align*}
s_{t-1}
= 1 - \rho_{t-1} C_{t-1}
 \neq S_{t-1}.
\end{align*}
In words, we have a  measurement error problem on $s_{t-1}$ but not on $g_t$.
It follows from  \eqref{eq:infected:stoc:coint} that the observed
$g_t$ and $s_{t-1}$ are related by
\begin{align}\label{eq:est:model}
g_t  = \beta_t s_{t-1} + v_t,
\end{align}
where
\begin{align}\label{eq:mes:error}
v_t =  \beta_t (S_{t-1} - s_{t-1})
=
 \beta_t (\rho-1) C_{t-1}.
\end{align}
The  right-hand side of \eqref{eq:mes:error} is likely to exhibit an increasing trend since $C_{t-1}$ is the cumulative fraction ever infected.
To alleviate this problem, we now divide both sides of \eqref{eq:est:model} by $c_{t-1}$, which is positive, to obtain
\begin{align}\label{main-eq-in-level}
\frac{g_t}{c_{t-1}} = \beta_t   \left[ \frac{s_{t-1}}{c_{t-1}} + \frac{\rho - 1}{\rho} \right].
\end{align}
On one hand, if $\rho = 1$, \eqref{main-eq-in-level} is identical to \eqref{eq:infected:stoc:coint}.
On other hand, if $\rho \rightarrow 0$, the term inside the brackets on the right-hand side of \eqref{main-eq-in-level}
diverges to infinity.

In the intermediate case, it depends on the relative size between
${s_{t-1}}/{c_{t-1}}$ and ${(\rho - 1)}/{\rho}$.
We now use the UK data to argue that the latter is negligible to the former.
According to the estimate by  \cite{UK:rho},
``an average of 0.25\% of the community population had COVID-19 in England at any given time between 4 May and 17 May 2020
(95\% confidence interval: 0.16\% to 0.38\%).''
In the UK data used for estimation, the changes in the number of cumulative positives between 4 May and 17 May 2020
is 0.08\% of the UK population.  Then, an estimate of $\rho = 0.08/0.25 = 0.32$, resulting in $(\rho-1)/\rho = -2.12$.
However, the sample maximum, median, minimum values of ${s_{t-1}}/{c_{t-1}}$ are
572412, 804, and 264, respectively. Therefore, the correction term $(\rho-1)/\rho$ is negligible and therefore,
\eqref{main-eq-in-level} reduces to
\begin{align}\label{main-eq-in-level-approx}
g_t  \approx \beta_t   s_{t-1},
\end{align}
which is virtually the same as \eqref{eq:infected:stoc:coint}.


\section{Sparse HP Filtering: Leave-One-Out Cross-Validation for Canada, China, South Korea and the UK}\label{appendix:SHP}

\begin{figure}[htbp]
\begin{center}
\caption{Sparse HP Filtering: LOOCV for Other Countries}\label{data-RoW-SHP-CV}
\vskip10pt
\begin{tabular}{cc}
\includegraphics[scale=0.5]{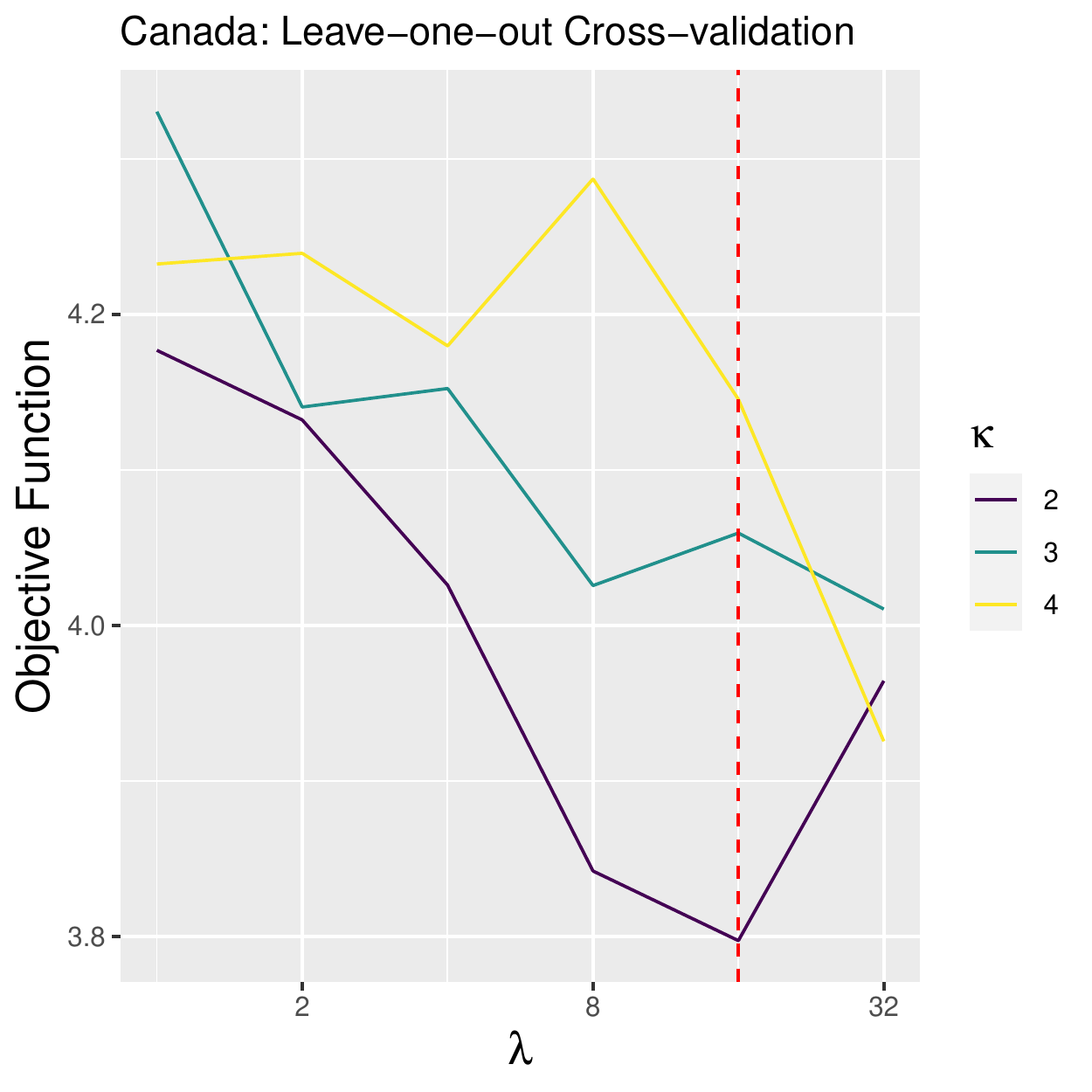} &
\includegraphics[scale=0.5]{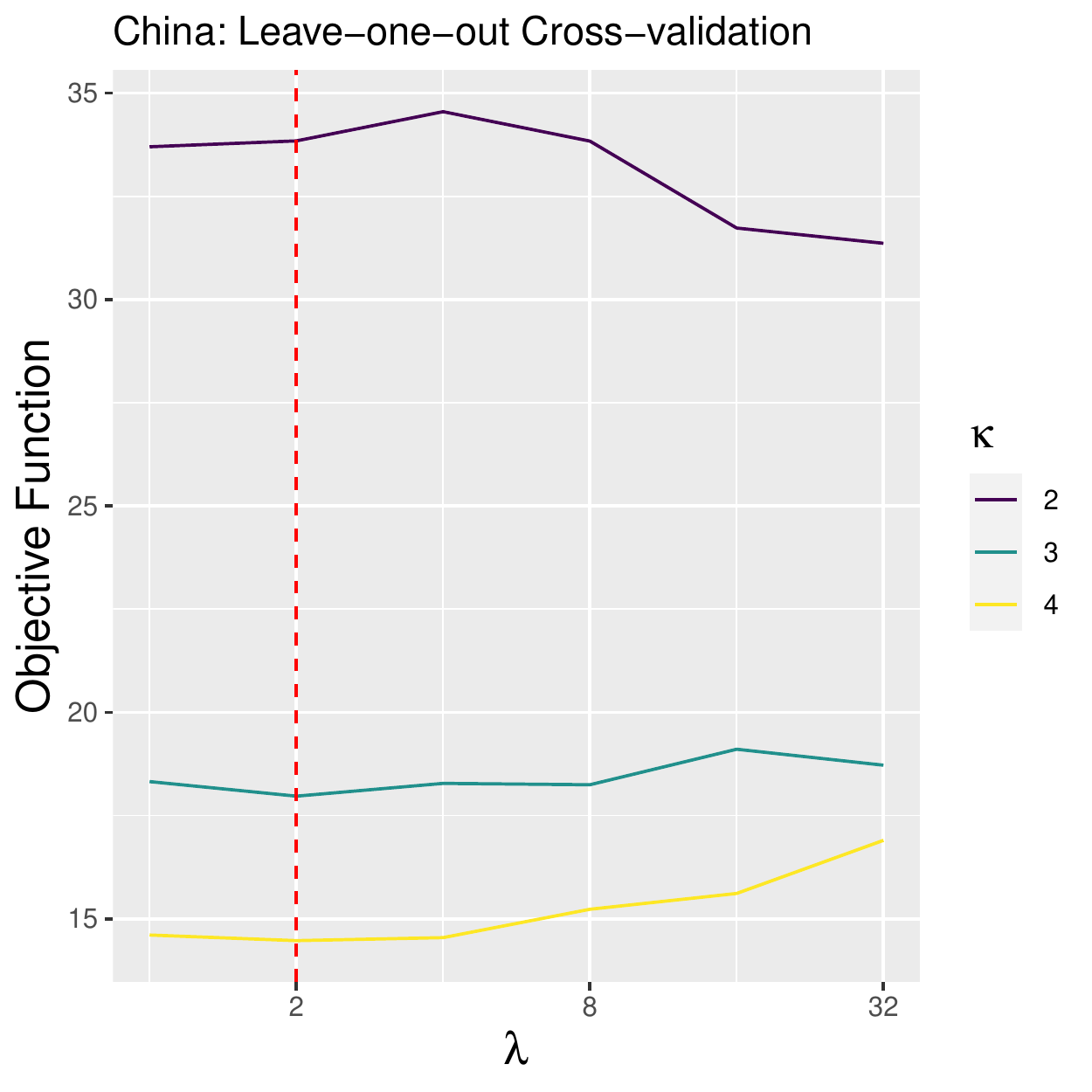} \\
\includegraphics[scale=0.5]{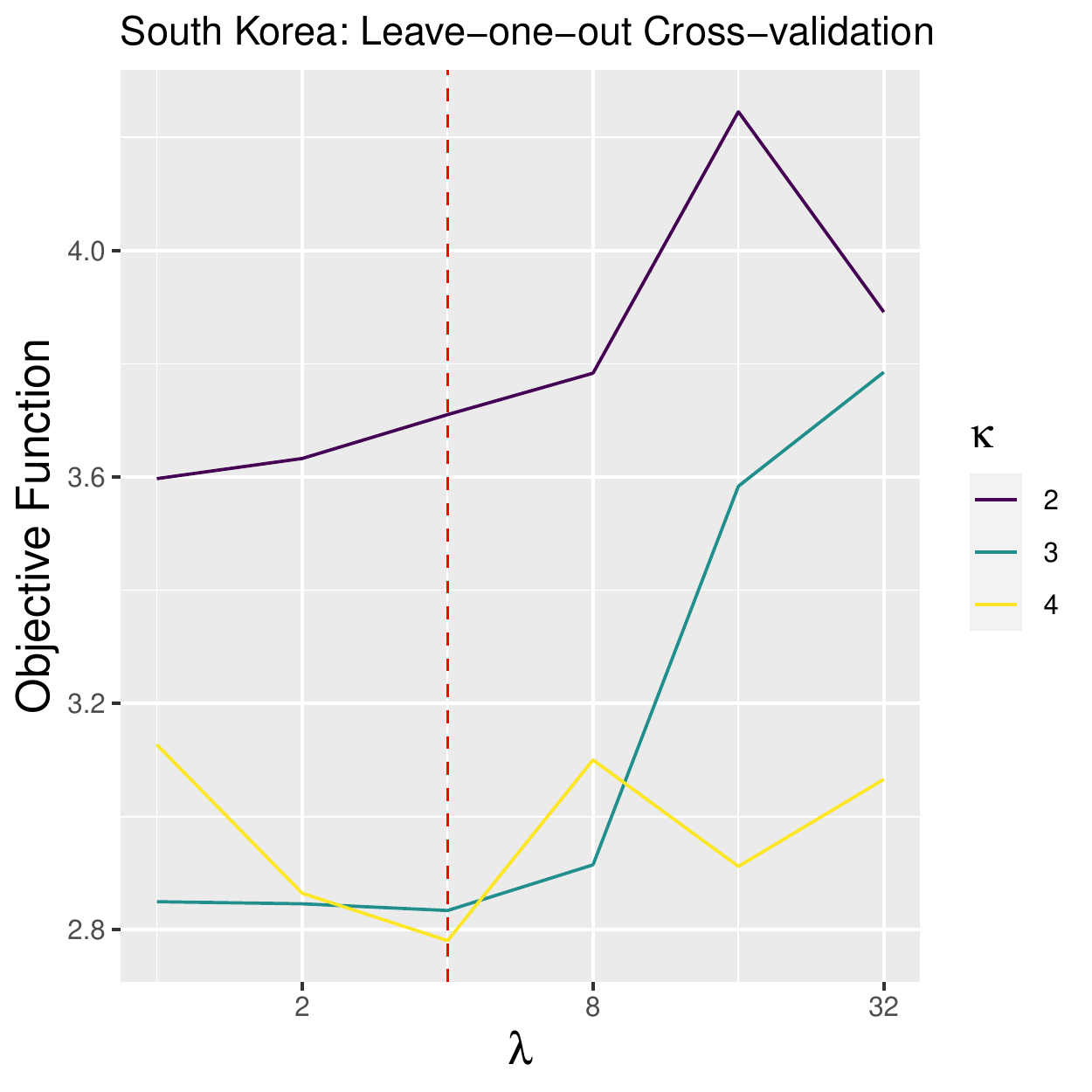} &
\includegraphics[scale=0.5]{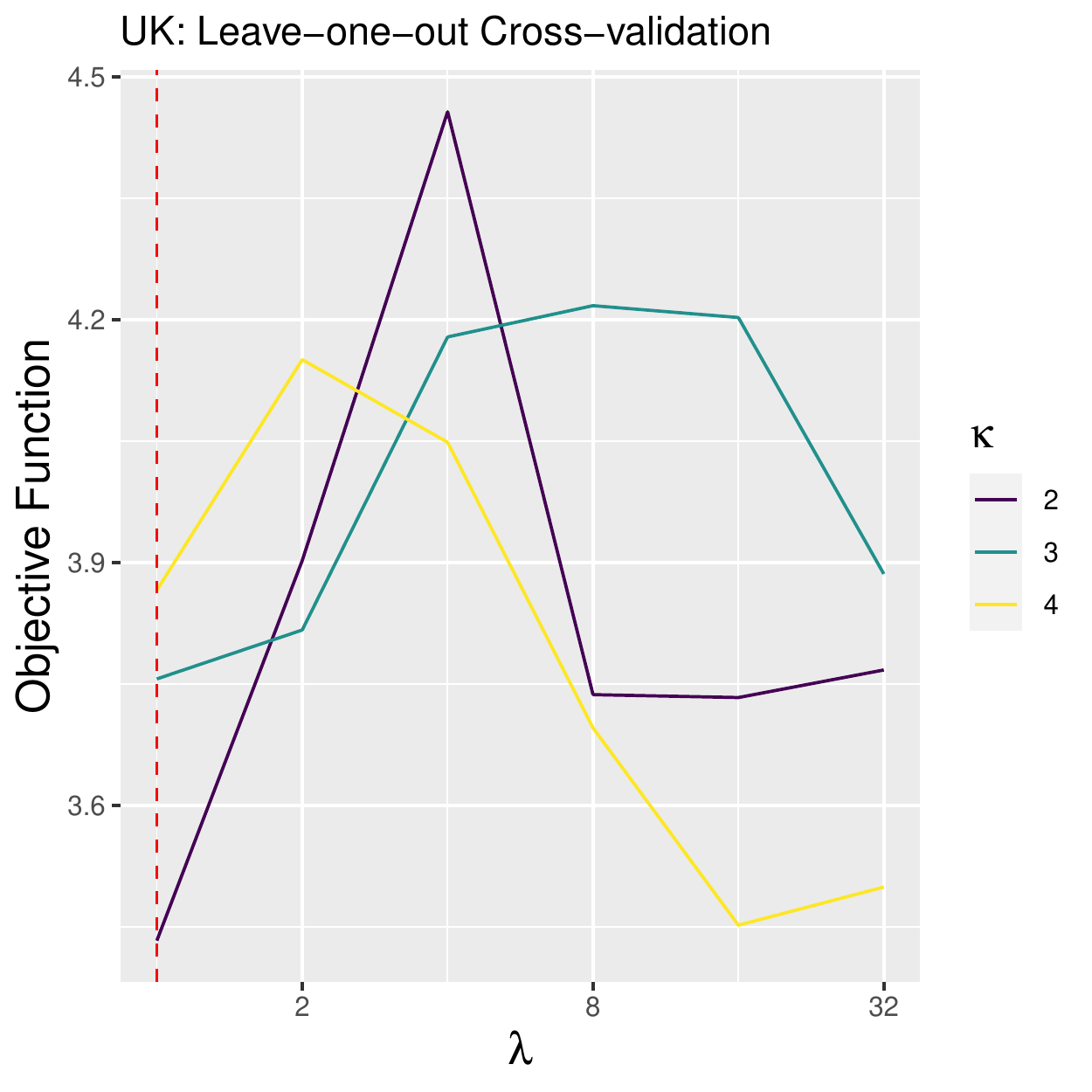} \\
\end{tabular}
\end{center}
\noindent
\footnotesize{Note: The red dashed line denotes the minimizer of the cross-validation objective function:
$(\widehat\kappa, \widehat\ld) =(2,16)$ for Canada;
$(\widehat\kappa, \widehat\ld) =(4,2)$ for China;
$(\widehat\kappa, \widehat\ld) =(4,4)$ for South Korea;
and
$(\widehat\kappa, \widehat\ld)_=(2,1)$ for the UK.
The analysis period is ended if the number of newly confirmed cases averaged over 3 days is smaller than 10:
April 26 (China) and April 29 (South Korea).
The grid points are: $\kappa\in\{2,3,4\}$  and $\ld=\{2^{0}, 2^{1}, \ldots, 2^{5}\}$.
The x-axis is represented by the $\log_2$ scale.}
\end{figure}

\clearpage

\section{$\ell_1$ Trend Filtering: Selection of $\lambda$}\label{appendix:L1}

\begin{figure}[htbp]
\begin{center}
\caption{$\ell_1$ Trend Filtering: Selection of $\lambda$}\label{data-RoW-L1-CV}
\vskip10pt
\begin{tabular}{cc}
\includegraphics[scale=0.6]{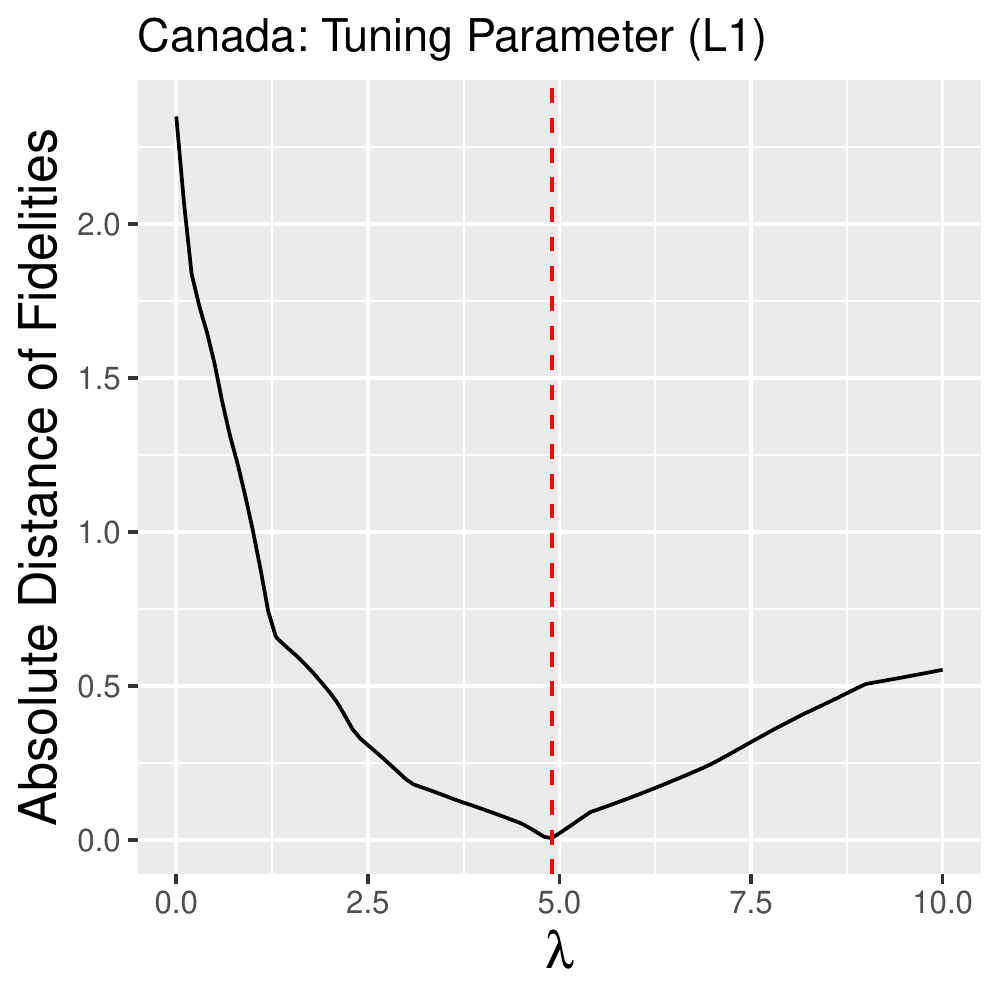} &
\includegraphics[scale=0.6]{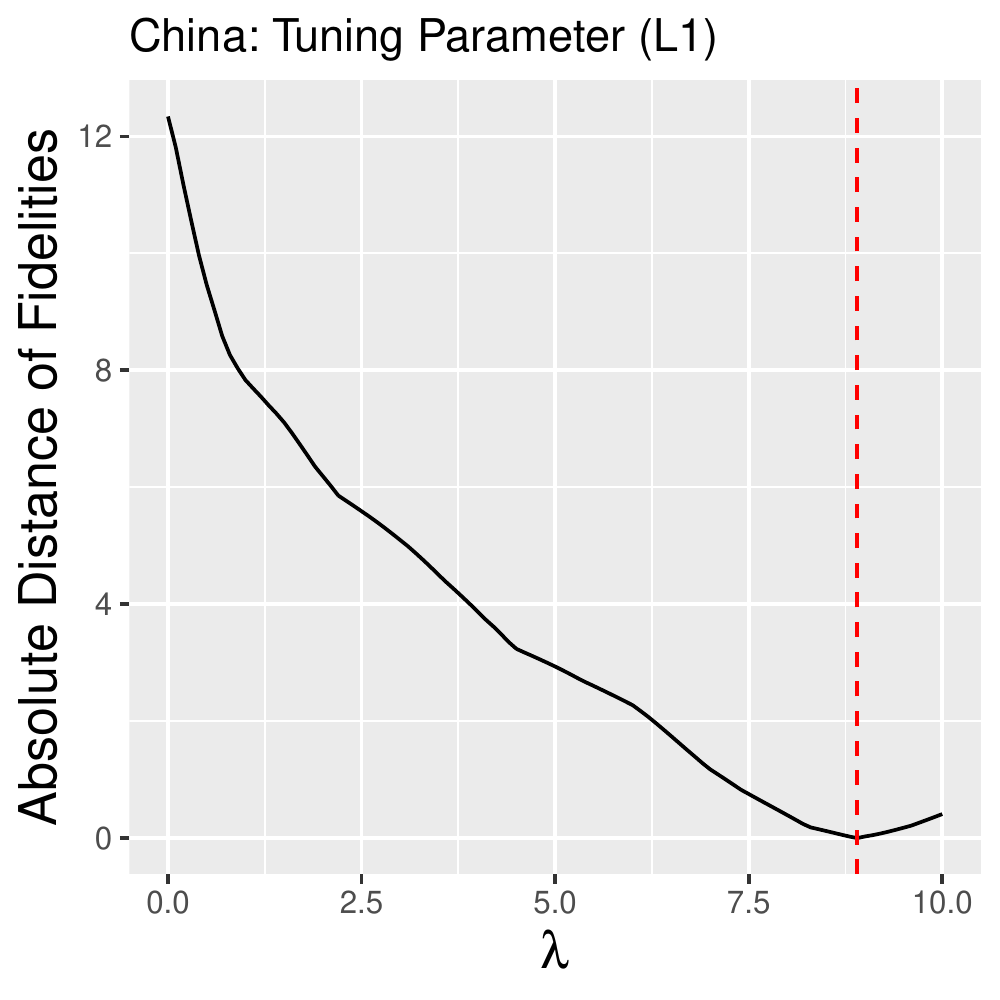} \\
\includegraphics[scale=0.6]{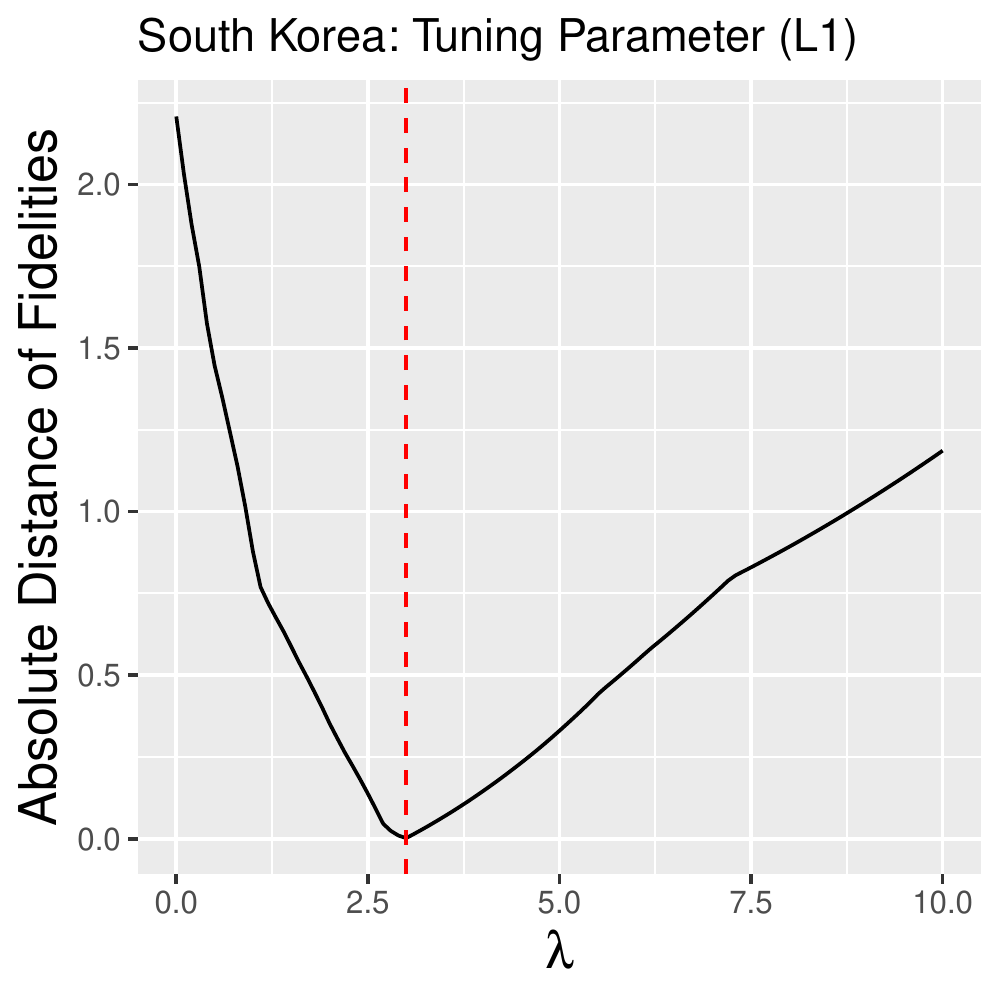} &
\includegraphics[scale=0.6]{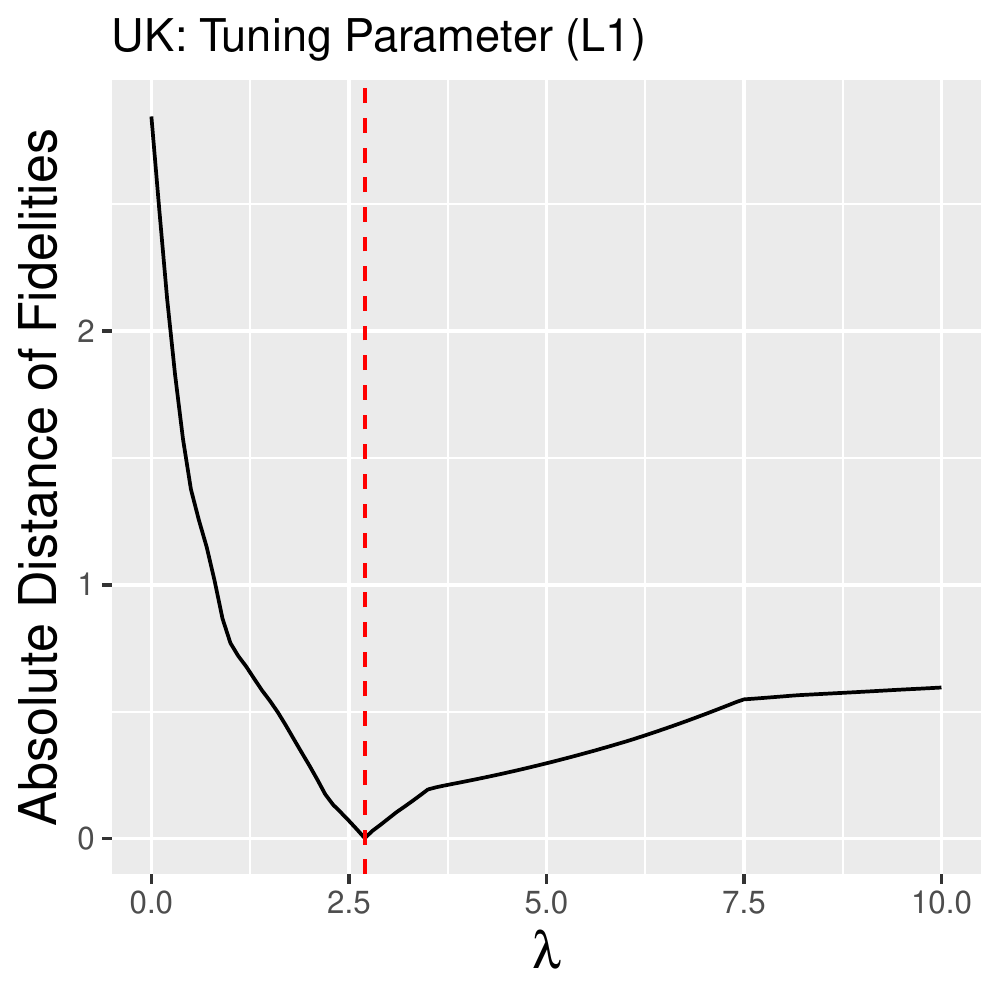} \\
\end{tabular}
\end{center}
\noindent
\footnotesize{Note: The red dashed line denotes the equalizer between the fidelity of
the sparse HP filter and that of the $\ell_1$ filter:
$\widehat\ld = 4.9$ for Canada;
$\widehat\ld = 8.9$ for China;
$\widehat\ld  = 3.0$ for South Korea;
and
$\widehat\ld = 2.7$ for the UK.
The analysis period is ended if the number of newly confirmed cases averaged over 3 days is smaller than 10:
April 26 (China) and April 29 (South Korea).}
\end{figure}

\clearpage

{\singlespacing
\bibliographystyle{chicago}
\bibliography{LLSS_bib}
}

\end{document}